\pdfoutput=1 \documentclass[aps,prd,amsmath,floats,floatfix, twocolumn,superscriptaddress,nofootinbib,showpacs,longbibliography]{revtex4-2}

\usepackage[T1]{fontenc}
\usepackage[utf8]{inputenc}
\usepackage{lmodern}

\usepackage{verbatim}

\usepackage[dvipsnames, usenames]{xcolor}
\definecolor{linkcolor}{rgb}{0.0,0.3,0.5}
\usepackage[hypertexnames=false, unicode, colorlinks=true, linkcolor=linkcolor,citecolor=linkcolor, filecolor=linkcolor, urlcolor=linkcolor, pdfusetitle]{hyperref}

\usepackage[all]{hypcap}
\usepackage{graphicx}
\usepackage{xspace}
\usepackage{amssymb}
\usepackage[normalem]{ulem} 
\usepackage{bm} 

\usepackage{microtype}

\usepackage[english]{babel}
\usepackage{blindtext}

\usepackage[caption=false]{subfig}

\usepackage{etoolbox} 
\usepackage{orcidlink}

\graphicspath{%
  {figs/}%
}

\DeclareMathAlphabet{\mathpzc}{OT1}{pzc}{m}{it}

\newcommand{\enclosepar}[1]{\left( #1 \right)}
\newcommand{\enclosebra}[1]{\left[ #1 \right]}
\newcommand{\enclosecurly}[1]{\left\{ #1 \right\}}

\newcommand{\encloseangle}[1]{\left\langle #1 \right\rangle}

\begin{document}

\title{Error quantification and comparison of binary neutron star gravitational waveforms from numerical relativity codes}

\newcommand{\Cornell}{\affiliation{Cornell Center for Astrophysics and Planetary Science, Cornell University, Ithaca, New York, NY 14853, United States of America}}
\newcommand{\Caltech}{\affiliation{Theoretical Astrophysics 350-17, California Institute of Technology, Pasadena, CA 91125, USA}}

\author{Sarah Habib\,\orcidlink{0000-0002-4725-4978}}
\email{shabib@caltech.edu}
\Caltech

\author{Elias R. Most\,\orcidlink{0000-0002-0491-1210}}
\Caltech

\author{Nils Deppe\,\orcidlink{0000-0003-4557-4115}}
\Cornell

\author{Francois Foucart\,\orcidlink{0000-0003-4617-4738}}
\affiliation{Department of Physics and Astronomy, University of New Hampshire, 9 Library Way, Durham NH 03824, USA}

\author{Mark A. Scheel\,\orcidlink{0000-0001-6656-9134}}
\Caltech

\author{Saul Teukolsky\,\orcidlink{0000-0001-9765-4526}}
\Caltech
\Cornell

\author{Michael Boyle\,\orcidlink{0000-0002-5075-5116}}
\Cornell

\author{Matthew Duez\,\orcidlink{0000-0002-0050-1783}}
\affiliation{ Department of Physics and Astronomy, Washington State University, Pullman, WA 99164, United States of America}

\author{Larry Kidder\,\orcidlink{0000-0001-5392-7342}}
\Cornell

\author{Jordan Moxon\,\orcidlink{0000-0001-9891-8677}}
\Caltech

\author{Kyle C. Nelli\,\orcidlink{0000-0003-2426-8768}}
\Caltech

\author{Harald Pfeiffer\,\orcidlink{0000-0001-9288-519X}}
\affiliation{Max Planck Institute for Gravitational Physics (Albert Einstein Institute), D-14467 Potsdam, Germany}

\author{William Throwe\,\orcidlink{0000-0001-5059-4378}}
\Cornell

\author{Nils Vu\,\orcidlink{0000-0002-5767-3949}}
\Caltech

\hypersetup{pdfauthor={Habib et al.}}

\begin{abstract}
Future gravitational wave detections of merging binary neutron star systems
have the possibility to tightly constrain the equation of state of dense
nuclear matter. In order to extract such constraints, gravitational
waveform models need to be calibrated to accurate numerical relativity
simulations of the late inspiral and merger. In this work, we take an
essential step toward classifying the error and potential systematics in
current generation numerical relativity simulations of merging binary
neutron stars. To this end, we perform a direct comparison of two codes
(\texttt{FIL}, \texttt{SpEC}), which differ in many aspects,
including the numerical methods and discretizations used and equations
solved. We find that despite these different approaches, the codes are --
within current numerical resolution bounds -- fully consistent, and broadly
comparable in cost for a given accuracy level. Our results indicate that
the error in the waveforms is primarily dominated by the hydrodynamic
evolution, consistent with earlier findings in the literature. 
We also discuss current limitations and cost estimates for
numerical relativity simulations to reach the accuracies required in the
era of next-generation gravitational wave detectors.
\end{abstract}

\maketitle

\section{Introduction}
\label{sec:introduction}

Binary neutron star (BNS) mergers are exciting gravitational wave sources.
With two events
detected~\cite{LIGOScientific:2017vwq,LIGOScientific:2020aai}, and many
more expected before the end of this decade and
beyond~\cite{Petrov:2021bqm,2020LRR....23....3A}, there are multiple
scientific opportunities to be leveraged. Apart from multi-messenger
astronomy of afterglows and gamma-ray bursts associated with BNS (see, e.g.
Ref.~\cite{Corsi:2024vvr} for a recent review), the gravitational wave (GW)
signal itself promises to provide a wealth of information on the dense
matter equation of state (EOS) (e.g.,
Refs.~\cite{Chatziioannou:2020pqz,Raithel:2019uzi}). This is because
neutron stars, unlike black holes~\cite{LeTiec:2020spy,Chia:2020yla}, can
be tidally deformed, altering the gravitational waveform in the final
orbits approaching merger~\cite{Flanagan:2007ix, Read:2013zra}. Extracting
this deformation for the first GW event of a merging BNS, GW170817, has
already led to strong constraints on the
EOS~\cite{Annala:2017llu,LIGOScientific:2018cki,Most:2018hfd,Raithel:2018ncd,De:2018uhw},
with future detectors promising to deliver extremely tight constraints,
e.g., on the radii of neutron
stars~\cite{Chatziioannou:2021tdi,Finstad:2022oni} (see e.g.,
Refs.~\cite{Raithel:2022efm,Essick:2023fso} for caveats stemming from phase
transitions, chemical equilibration
effects~\cite{Ripley:2023lsq,Counsell:2025hcv,Pnigouras:2025muo}, and mode resonances~\cite{Pratten:2021pro}). One of the limiting factors in interpreting GW signals to sufficient accuracy to extract this information are well-calibrated waveform models (see, e.g., Ref.~\cite{Dietrich:2019kaq} for current models). Exacerbating this need, currently available models are not accurate enough for next-generation detectors~\cite{Gamba:2020wgg}. Since tidal deformability imprints on the waveform cannot be fully computed from post-Newtonian theory, which breaks down near merger, they are commonly fitted from numerical relativity (NR) simulations~\cite{Dietrich:2017aum,Dietrich:2018uni,Dietrich:2019kaq}. However, NR simulations of BNS are only available in a limited number of public catalogs~\cite{Dietrich:2018phi,Gonzalez:2022mgo,Kiuchi:2017pte,Kiuchi:2019kzt}, which are not systematically sampled, as most of the simulations are targeting primarily the post-merger phase, which has different physics and accuracy requirements~\cite{Chatziioannou:2017ixj,Wijngaarden:2022sah,Criswell:2022ewn}. It is therefore important to ask what would be required to build a systematic catalog.

While NR simulations of binary black hole (BBH) mergers have been carried
out at extreme precisions already~\cite{Boyle:2019kee, Scheel:2025jct}, BNS merger
simulations so far lag substantially behind, in part because of limited
convergence and substantial errors stemming from the hydrodynamical
modeling of the stellar
material~\cite{Radice:2013hxh,Bernuzzi:2016pie,Most:2019kfe,Liebling:2020jlq}.
Different groups have implemented a number of strategies for computing GW
waves from merging BNS, especially on the hydrodynamics side, using
high-order numerical
methods~\cite{Radice:2013hxh,Bernuzzi:2016pie,Most:2019kfe,Kiuchi:2025ksk}, including
entropy based limiting~\cite{Doulis:2022vkx,Doulis:2024aew}, and
spectral~\cite{Duez:2008rb} or finite-element methods~\cite{Deppe:2024ckt,Adhikari:2025nio}. Because of their algorithmic
or implementation differences, these codes have different intrinsic errors
and computational costs. While most codes solve a version of the
BSSN~\cite{Shibata:1995we,Baumgarte:1998te} or Z4 set of
equations~\cite{Bona:2003fj,Bernuzzi:2009ex,Hilditch:2012fp,Alic:2011gg},
others use a generalized harmonic set of
variables~\cite{Pretorius:2004jg,Lindblom:2005qh}. Initial conditions for
the simulations need to be computed numerically, e.g., by solving the
extended conformally thin sandwich (XCTS) equations~\cite{Pfeiffer:2002iy}.
This is commonly done using
spectral~\cite{Gourgoulhon:2000nn,Tichy:2009yr,Tacik:2015tja,Papenfort:2021hod}
or finite-difference~\cite{East:2012zn,Uryu:2011ky} discretizations. Given
all these various parts it seems important to ask whether using any such
combination of different methods and codes lead to comparable errors or
systematic differences when used under production settings. However, only a
limited number of direct code comparisons have been carried
out~\cite{Espino:2022mtb,Hamilton:2024ziw,Neuweiler:2024jae,Kuan:2025bzu}. In light of
the above question of building a waveform catalog it seems, however,
imperative to have a well-defined error budget, as has been established in
the case of BBH simulations~\cite{Ajith:2012az}.

In this work, we present a direct code comparison between two different
NR codes (\texttt{FIL}~\cite{Most:2019kfe,Etienne:2015cea} and
\texttt{SpEC}~\cite{Scheel:2006gg,Scheel:2008rj,Szilagyi:2009qz,Szilagyi:2014fna})
for BNS merger simulations. These use different numerical algorithms,
formulations of the Einstein equations, and independent initial-data codes, such
as the \texttt{FUKA}~\cite{Papenfort:2021hod,Grandclement:2009ju} and
\texttt{SpELLs}~\cite{Tacik:2015tja} libraries. As such, the comparison of codes
is as different as currently possible with codes in the community, and allows
for a faithful assessment of intrinsic consistency and error budgets of current
generation numerical waveforms. We outline the currently (limited) accuracy of
typical production-level simulations and discuss potential requirements for
future simulation work.

This paper is structured as follows. In Sec.~\ref{sec:methods} we present
the setup for the comparison. The main results are shown in Sec.~\ref{sec:results}, before concluding in Sec.~\ref{sec:conclusion}.
This work uses a unit convention of $G=c=1$.

\section{Methods}
\label{sec:methods}

For this comparison we adopt an equal mass BNS system with negligible neutron star spin and a total mass of $M= 1.350 M_\odot$ at infinite separation, and a baryon mass of $M_b= 1.4958\, M_\odot$ per NS. We adopt an initial separation of $47.67\,\rm km$. The initial data are prescribed using the extended conformal thin-sandwich (XCTS) formulation~\cite{Pfeiffer:2002iy}. For more details see, e.g., Refs.~\cite{Tichy:2009yr,Tacik:2015tja}. We further adopt eccentricity-reducing initial parameters following the parametrization of Ref.~\cite{Buonanno:2010yk}. Specifically, we use 
\begin{align}
  \dot{a} &= -8.09518351\times 10^{-5}\,, \label{eqn:dota_ecc}\\
  M_\odot^2 \Omega &= 0.008017218957\,, \label{eqn:Omega_ecc}
\end{align}
where $\dot{a}d/2$ is the initial radial velocity of the stars, $d$ the binary separation, and $\Omega$ their initial angular velocity.

The choice of equation of state (EOS) dictates tidal effects and contributes to
observable features in the GW signal, particularly during and after late inspiral. We use the SLy~\cite{Douchin:2001sv} spectral EOS
implemented in Ref.~\cite{Foucart:2019yzo}.
At the time of writing,
the SLy$\Gamma2$ model is within current EOS constraints from BNS
observations~\cite{LIGOScientific:2018cki, LIGOScientific:2018hze}. A spectral
EOS is represented as a set of basis functions with coefficients~\cite{Lindblom:2010bb}. 
In the EOS representation we use~\cite{Foucart:2019yzo}, the pressure $P$ is related to the rest-mass density $\rho$, via
\begin{align}\label{eqn:EOS}
    P(x,T) = P_0 \exp\left(\Gamma_0 x + \Theta (x) \left[\gamma_2 \frac{x^3}{3} + \gamma_3 \frac{x^4}{4}\right]\right) + \rho T\,, 
\end{align}
where $T$ is the temperature, $x = \log (\rho/\rho_0)$, $P_0=3.3625\times10^{-7}$,
$\gamma_2 = 0.4029$, $\gamma_3=-0.1008$, $\rho_0=1.0118\times10^{-4}$, $\Gamma_0=2$, and
$\Theta(x)$ is the Heaviside function.
This representation is chosen to minimize loss of convergence due to non-smoothness in the equation of state~\cite{Foucart:2019yzo,Raithel:2022san}.

{Initial data is computed independently using the \texttt{FUKA}
\cite{Papenfort:2021hod} and \texttt{SPELLS} \cite{Tacik:2015tja} frameworks.
Both codes solve the same XCTS equations on a spectral multigrid domain,
but differ in the details of the implementation, especially in the handling of
the EOS. In \texttt{SPELLS}, Eq. \eqref{eqn:EOS} is used directly, whereas
in \texttt{FUKA} it is first tabulated using a log-linear sampled grid of
about 1,000 points. While we use the same eccentricity reduction parameters
in both cases, the choice of resolution is choosen independently and
according to typical simulation configurations for the evolution codes
\cite{Papenfort:2021hod}. While systematic comparisons of different initial
data codes have been carried out \cite{Tsokaros:2016eik}, 
we find that despite using different initial data solvers, the two codes
produce results consistent within estimated errors from the time evolution
and waveform extraction. We thus expect the impact of using different
initial data solvers to be a subdominant source of error.
}



%
\subsection{Evolution codes}

We use the standard $3+1$ decomposition in numerical relativity, in which the spacetime metric $g_{\mu\nu}$ takes the form
\begin{align}
    ds^2 = -\alpha^2 dt^2 + \gamma_{ij}\enclosepar{dx^i + \beta^i dt}\enclosepar{dx^j + \beta^j dt}
    \label{eq:GR}
\end{align}
where $\alpha$ is the lapse, $\beta^i$ is the shift, and $\gamma_{ij}$ is the spatial metric. 


All tested codes solve the general relativistic (magneto-)hydrodynamics
(GR(M)HD) system of equations in flux-balanced conservation
form~\cite{Duez:2005sf},
\begin{align}
  \partial_t \bold{U} + \partial_i \bold{F}^i(\bold{U})
  = \bold{S}(\bold{U}).
  \label{eq:HD}
\end{align}
Here $\bold{U}$ is the state vector of conserved variables to be evolved, $\bold{F}^i(\bold{U})$ are the fluxes, and $\bold{S}(\bold{U})$ are source terms. 
More details on the discretizations and the codes are provided in the
following sections below. All simulations were performed on comparable hardware, i.e.~AMD Epyc CPUs with $\sim 2.5$ GHz clock speed.

\subsubsection{\texttt{FIL}}
\label{sec:fil}
The \texttt{Frankfurt/IllinoisGRMHD (FIL)} code is based on the Einstein
Toolkit infrastructure~\cite{Loffler:2011ay}. It implements the GRMHD
equations in 3+1 form~\cite{Duez:2005sf}, which are solved using a
fourth-order accurate version of the conservative finite-difference ECHO
scheme~\cite{DelZanna:2007pk}. Reconstruction to cell interfaces uses
WENO-Z~\cite{borges2008improved}, with fluxes $\mathcal{F}$ being computed
using a HLLE Riemann solver~\cite{einfeldt1988godunov}. We additionally
limit the fluxes using an approximate second-order a priori
positivity-preserving limiter based on the density~\cite{Radice:2013xpa},
combined with third order Runge-Kutta time integration. The resulting limited fluxes are then corrected using a DER4 corrector~\cite{DelZanna:2007pk} to achieve overall higher order. In detail, we compute fluxes at cell interfaces, $i+1/2$,
\begin{align}
    F_{i+1/2} = \frac{13}{12} \mathcal{F}_{i+1/2}-\frac{1}{24}\left(\mathcal{F}_{i+3/2} - \mathcal{F}_{i-1/2}\right)\,.
\end{align}
Primitive inversion is carried out using the scheme of Ref.~\cite{Kastaun:2020uxr}, with a fall-back for purely hydrodynamical flows~\cite{Galeazzi:2013mia}. We also use an entropy-based backup solver. Equation of state handling is provided using \texttt{FIL}'s microphysics infrastructure, which offers two ways of handling the spectral EOS used here. First, the EOS routine has been implemented using direct numerical integration outlined in Ref.~\cite{Foucart:2019yzo}. However, we have found it more convenient to simply tabulate the spectral EOS using a uniformly sampled table with 1,000 grid points, which is then interpolated linearly in logarithmic quantities.

The Einstein equations are solved using the Z4c
formalism~\cite{Bernuzzi:2009ex,Hilditch:2012fp} in moving puncture
gauge~\cite{Alcubierre:2002kk,Campanelli:2005dd,Baker:2005vv}. Specifically, we adopt
\begin{align}
    &\partial_t \alpha - \beta^i \partial_i \alpha = - 2 \alpha K\,,\\
    &\partial_t \beta^k - \beta^i \partial_i \beta^k = \frac{3}{4} B^k\,,\\
    &\partial_t B^k - \beta^i \partial_i B^k = \left(\partial_t - \beta^i\partial_i\right) \tilde{\Gamma}^k - \eta B^k\,,
\end{align}
where $K$ is the trace of extrinsic curvature, $\tilde{\Gamma}^k$ and $B^k$ are the variables for the Gamma-driver, and $\eta M_\odot=0.2$. We adopt Z4c damping parameters of $\kappa=0.04$. All damping parameters have a roll-off with inverse distance that sets in at a radius of $ R/M_\odot = 64$.

The resulting set of equations is then solved using a strong-stability preserving third-order Runge-Kutta scheme~\cite{gottlieb2001strong}.

\texttt{FIL} uses a domain of $7$ uniform-resolution AMR grids centered on each NS, with a total domain extent of $ 2,048\, M_\odot$.

The initial data configuration is computed using the \texttt{FUKA} code~\cite{Papenfort:2021hod}. \texttt{FUKA} uses the \texttt{KADATH} spectral solver library~\cite{Grandclement:2009ju}. The equation of state is handled using the same log-linear table used for the evolution.

\subsubsection{SpEC}
\label{sec:spec}
\texttt{SpEC} evolves the gravitational and hydrodynamic systems on two separate
spatial grids~\cite{Duez:2008rb}. The Einstein equations are evolved on a
pseudospectral grid; the hydrodynamics equations are evolved on a
finite-difference grid. Evolution is done with a third-order Runge-Kutta time
stepper. The time step on the pseudospectral grid is chosen adaptively to reach
a target tolerance varying with the chosen grid resolution. The time step on the
finite-difference grid is allowed to be larger: up to $\Delta t = \Delta x/4$,
with $\Delta x$ the minimum grid spacing. At the end of a time step on the
finite-difference grid, the metric and its derivatives are interpolated onto the
finite-difference grid by first refining the spectral data using spectral
interpolation, then using third-order interpolation onto the finite-difference
grid. The primitive fluid variables are interpolated onto the pseudospectral
grid using monotonicity-preserving polynomial interpolation. Data at
intermediate times is obtained using linear interpolation (higher-order
interpolation is possible but has no practical impact on the accuracy of the
simulations~\cite{Knight:2023kqw}). More details on the time stepping methods
can be found in Ref.~\cite{Knight:2023kqw}, while grid-to-grid interpolation is
described in Ref.~\cite{Duez:2008rb}.

The pseudospectral grid during inspiral is composed of a set of spherical shells centered on each NS, balls around the NS interior regions, distorted cubes connecting the spherical regions to the wave zone, and spherical shells centered on the center of mass of the binary in the wave zone. \texttt{SpEC} employs adaptive mesh refinement (AMR) for the pseudospectral grid, adjusting resolution based on the errors estimated from the coefficients of the spectral expansion for each evolved variable. The target truncation error is scaled as $\Delta x_{FD}^{5}$, with details in Ref.~\cite{Foucart:2012vn}. The finite-difference grid uses a constant resolution grid during inspiral, and fixed mesh refinement after merger. In both cases, the finest level of the computational domain is divided into blocks of $\sim 6\,{\rm km^3}$ that are only evolved if matter is present in the region of space that they cover. More specifically, if all grid cells in a block have densities below $6\times10^9\,{\rm g/cm^3}$, it is removed from the computational domain. If matter with density above $10^{10}\,{\rm g/cm^3}$ approaches within 3 grid cells of a removed block, that block is added and evolved once more. The threshold densities are decreased far away from the center of the binary, as described in Ref.~\cite{Foucart:2012vn}. As the \texttt{SpEC} grid is constructed so that the centers of the compact objects are fixed on the grid during inspiral, the effective grid resolution increases as the neutron stars spiral in. Whenever the grid spacing is reduced by $20\%$, we interpolate onto a new finite-difference grid with the original grid resolution. Once the BNS approaches merger and the NSs deform, spherical pseudospectral domains around each NS no longer characterize symmetry in the binary, so both grids are restructured. When the maximum density on the pseudospectral grid grows to $3\%$ higher than its initial value, the grid switches to a ball centered at the coordinate center of mass of the binary system, surrounded by spherical shells covering the wave zone. After contact, the finite-difference grid is made of four nested cubes. The finest resolution grid covers a $40\,{\rm km^3}$ region around the center of mass of the system. Each coarser level of refinement is twice as large in each dimension.

To keep the center of mass of the neutron stars fixed on the grid during inspiral, \texttt{SpEC} uses a time-dependent map between grid coordinates and ``inertial'' coordinates. That map includes both a global rescaling of the coordinates and a rotation around the polar axis. The control system used to evolve the scaling factor and rotation angle is described in more detail in Ref.~\cite{Hemberger:2012jz}.

The spacetime metric is evolved using the first-order GH formulation~\cite{Lindblom:2005qh}, in which the coordinates $\bold{x}$ obey the wave equation
\begin{align}
  g_{\mu\nu} \nabla^\sigma \nabla_\sigma x^\nu = H_\mu(\bold{x}, g_{\mu\nu})
\end{align}
where $H_\mu(\bold{x}, g_{\mu\nu})$ is an arbitrary gauge source function, which in \texttt{SpEC} is set to the harmonic gauge $H_\mu=0$. This gives a system of equations and several constraint equations for $g_{\mu\nu}$, $\Phi_{i\mu\nu} \equiv \partial_i g_{\mu\nu}$, and $\Pi_{\mu\nu} \equiv n^{\gamma} \partial_{\gamma}g_{\mu\nu}$, where $n^\gamma$ is the unit normal vector to the spatial slice. Constraints are not enforced during numerical evolution and need to be monitored. \texttt{SpEC} damps the constraints $\mathcal{C}_\mu = \Gamma_\mu + H_\mu$ and $\mathcal{C}_{i\mu\nu} = \Phi_{i\mu\nu} - \partial_ig_{\mu\nu}$, where $\Gamma_\mu=\Gamma^\nu_{\nu\mu}$ is the contracted Christoffel symbol and $\Phi_{i\mu\nu}$ the first-order reduction variable in the GH system. The constraint damping parameters are chosen the same as in ref.~\cite{Foucart:2012vn}. In addition to constraint damping, violations are controlled by constraint-preserving boundary conditions enforced at the outer boundary of the computational domain as extra terms in the evolved variables~\cite{Lindblom:2005qh, Rinne:2007ui}.

The hydrodynamics evolution uses numerical fluxes at cell faces that are computed from the characteristic fluxes and characteristic variables reconstructed on cell faces using the fifth-order accurate MP5 reconstructor, using the methods introduced in Ref.~\cite{Radice:2013xpa}. A density floor of $\rho_{\rm floor}=6\times10^4\,{\rm g/cm^3}$ is imposed, below which densities are reset to $\rho_{\rm floor}$. Below $\rho_{\rm atm}\approx 1\times10^6\,{\rm g/cm^3}$, we also apply atmosphere corrections limiting the velocity of the fluid in the corotating frame to $< 0.0001c$ and the pressure to $P(\rho,T)<1.01P(\rho,0)$. We also require $T\geq 0$. Finally, corrections to the conservative variables are performed whenever the evolution reaches values of the conserved variables that are close to becoming unphysical (i.e.~to no longer correspond to physical values of the primitive variables), as described in Ref.~\cite{Foucart:2012vn}.

{\texttt{SpEC} uses the built-in \texttt{SPELLS} initial data framework
\cite{Tacik:2015tja}.}


\begin{table*}[t]
    \begin{tabular}{| c || c | c | c | c | c | c |}
        \hline
        \textbf{Run name} & \textbf{Code} & \textbf{Grid spacing $\bm{[M_\odot]}$} & \textbf{Grid spacing $\bm{[\mathrm{m}]}$} & \textbf{Computational cost to merger [core-hours]} \\
        \hline
        \hline
        SpEC Lev1 & \texttt{SpEC} & 0.17667 & 261.5 & 16,855 (25,491) \\
        SpEC Lev2 & \texttt{SpEC} & 0.14134 & 209.2 & 44,476 (93,118) \\
        SpEC Lev3 & \texttt{SpEC} & 0.11307 & 167.3 & 109,759 (187,331) \\
        \hline
        FIL Lev0 & \texttt{FIL} & 0.22857 & 338.3 & 46,080 \\
        FIL Lev1 & \texttt{FIL} & 0.178 & 263.4 & 71,349 \\
        FIL Lev2 & \texttt{FIL} & 0.1379 & 204.1 & 245,760 \\
        \hline
    \end{tabular}
    \caption{The computational cost and grid spacing of each run studied. For \texttt{SpEC} we give the computational cost to merger as the time, and costs in parentheses include post-merger runtime. For \texttt{FIL} we give the computational cost to merger as the time the maximum amplitude in the gravitational waveform is computed at a distance of $600\ M_\odot$. In \texttt{SpEC} the post-merger is significantly more expensive than the inspiral because of the much larger finite-difference grid.
    }
    \label{tab:run_table}
\end{table*}

\subsection{Waveform Extraction}
\label{sec:extraction}
For a given mode of the GW strain $h_{lm}$, the waveform amplitude $A$ and
phase $\phi$ are defined from the strain as,
\begin{align}
    rh_{lm} = A_{lm}e^{-i\phi_{lm}}\,.
\end{align}
Sign conventions are not consistent among all codes (see
Ref.~\cite{Boyle:2019kee}, Appendix C for a discussion). For the following
results, we use the sign convention,
\begin{align}
    h = h_{+} -ih_\times\,.
\end{align}
For our analysis, we primarily focus on the $l=2,m=2$ mode.

We consider two $l=2,m=2$ waveforms to be equivalent if they differ by only by an overall phase shift (equivalent to a rotation in the $x$-$y$ plane) and/or an overall time offset. So when comparing two different waveforms, we first align them by explicitly time-shifting and phase-shifting one of them. This must be done carefully because differences between waveforms are very sensitive to alignment. A typical alignment method is to time-shift and phase-shift so that the two waveforms reach their maximum amplitude at the same time and phase. However, we find that this results in phase differences being dominated by errors during the merger and post-merger, and we are most interested in errors during inspiral. So instead, we align waveforms by choosing the optimal time and phase shift to minimize the difference in amplitude and phase of the $l=2,m=2$ strain mode over a chosen time window during the inspiral; this is a two-dimensional root-finding problem. The time window chosen for alignment, typically around $[700\ M_\odot,\ 2000\ M_\odot]$, must start sufficiently late such that junk radiation or CCE junk is not included in the minimization.

Asymptotic GW quantities can be extracted upon completion of a numerical simulation. The three primary methods for achieving this are using Nakano extrapolation (NE), finite-radius extrapolation (FRE), and Cauchy-Characteristic evolution (CCE). In the following results we apply these different methods and assess their contribution to overall simulation error.

\subsubsection{Nakano extrapolation}
\label{sec:Nakano ext}

\texttt{FIL} extracts the Weyl scalar $\Psi_4$ at coordinate spheres of finite
radii $r$. We use the perturbative formula proposed as Eq.~3 in
Ref.~\cite{Lousto:2010qx} (see also Ref. \cite{Fontbute:2025ixd}),
\begin{align}
  \lim_{r \to \infty} r \Psi_4^{lm} &=
  \enclosepar{1 - \frac{2M}{\bar{r}}} \times \\
  & \enclosebra{\bar{r}
    \Psi_4^{lm} -
    \frac{(l-1)(l+2)}{2\bar{r}}
    \int \bar{r} \Psi_4^{lm} \ dt}
\end{align}
for a given extraction radius $\bar{r}$. We use this formula to obtain $\Psi_4$ at future null infinity for \texttt{FIL}.

The $\Psi_4$ spherical harmonic waveform can then be integrated to obtain the modes of the gravitational wave (GW) strain $h$, which are given as
\begin{align}
  h_{lm} = -\int^{t}_{-\infty} \int^{t'}_{-\infty} \Psi_4^{lm} \ dt'' \ dt'\,,
\end{align}
where the sign is appropriately chosen for each code. However, simply integrating $\Psi_4$ in time tends to cause a spurious nonlinear drift in the strain~\cite{Reisswig:2010di}. The standard solution is to compute the strain using fixed frequency integration (FFI)~\cite{Reisswig:2010di}, i.e., integration of $\Psi_4$ in the frequency domain while fixing the contribution from frequencies below a chosen cutoff. The cutoff frequency is a free parameter that must be fine tuned and chosen with care to avoid damping physical frequencies or amplifying unphysical features. Where necessary we use a cutoff near the orbital frequency, the lowest physical frequency present in the system, as suggested in Ref.~\cite{Reisswig:2010di}. All \texttt{FIL} waveforms presented are computed using the NE GW extraction method.

\subsubsection{Finite radius extrapolation}
\label{sec:ext}

\texttt{SpEC} uses FRE to extract GW data. Specifically, in \texttt{SpEC}, $\Psi_4$ and $h$ are computed on a series of concentric spherical shells approximately evenly spaced in $1/r$ and centered on the binary. The waveforms are then extrapolated to future null infinity by fitting a series in powers of $1/r$ to the data on the concentric shells~\cite{Boyle:2009vi}. Less care is taken in using the ``best'' tetrad when computing $\Psi_4$ and $h$ on the shells since the errors from tetrad differences will extrapolate away. We emphasize that in addition to extracting $\Psi_4$ directly, \texttt{SpEC} \textit{directly} and \textit{independently} extracts the GW strain $h$ using the Sarbach and Tiglio formulation~\cite{Sarbach:2001qq} of the Regge-Wheeler and Zerilli equations~\cite{Regge:1957td, Zerilli:1970se} with implementation details described in Refs.~\cite{Rinne:2008vn, Boyle:2019kee, Boyle:2007ft}. Unless otherwise stated, we use the extrapolated waveforms for \texttt{SpEC} in our analysis.


\subsubsection{Cauchy-Characteristic Evolution\label{sec:cce}}

Cauchy-characteristic evolution (CCE)~\cite{1996PhRvD..54.6153B, Bishop:2016lgv, 2016CQGra..33v5007H, 2020PhRvD.102b4004B, 2020PhRvD.102d4052M, Moxon:2021gbv} is a waveform extraction method where Einstein's equations are solved out to future null infinity, $\mathcal{I}^+$. Metric data on a worldtube at some finite radius is written to disk during the Cauchy evolution, i.e., during the GH simulation in \texttt{SpEC}. This data is then used as a boundary condition to the characteristic evolution that evolves the outgoing gravitational radiation to $\mathcal{I}^+$. CCE is able to capture physical effects like gravitational wave memory that are not present in extrapolated waveforms~\cite{Mitman:2020pbt,Mitman:2020bjf}. By solving the Einstein equations to $\mathcal{I}^+$, CCE circumvents extrapolation errors inherent in FRE waveforms. Thus, CCE will generally produce more realistic waveforms than any extrapolation procedure. Despite these advantages, CCE is not completely parameter-free since initial data on the initial null slice needs to be chosen and different radii for the worldtube can be used. An additional complication that CCE presents is that the waveforms are output in an arbitrary Bondi-Metzner-Sachs (BMS) frame. This frame is not generally going to match the one that the extrapolated waveforms are in. We choose to always map the waveforms to their own superrest frame~\cite{Mitman:2020bjf}.

We use several worldtube radii for CCE, and find that the resulting waveforms differ slightly. \texttt{SpEC} uses radii of $\{129, 496, 863, 1230\} \ M_\odot$. All CCE waveforms are extracted using the \texttt{SpECTRE} CCE module~\cite{Moxon:2021gbv}. For each code, we choose the extraction radius
with the smallest $\Psi_2$ constraint.

We use \texttt{SpEC} to study the difference between extrapolated and CCE
waveforms for BNS simulations. Since some codes such as \texttt{SpECTRE} do not output any finite radius or extrapolated waveforms, it is crucial to understand if systematic difference between the extrapolated and CCE waveforms are a dominant source of discrepancy between different codes. 
\section{Results}
\label{sec:results}

In this section, we perform several side-by-side comparisons of BNS simulations with the same initial parameters run by different codes. Table~\ref{tab:run_table} describes the runs examined in this section. We analyze \texttt{SpEC} and \texttt{FIL} at three different finite-difference resolutions each, allowing for convergence tests. Each level increases the resolution by about 25\%. We restrict our analysis to inspiral and merger in the following.

\subsection{Waveform Properties}
\label{sec:waveform}
\begin{figure}[h]
    \centering
    \includegraphics[width=\columnwidth]{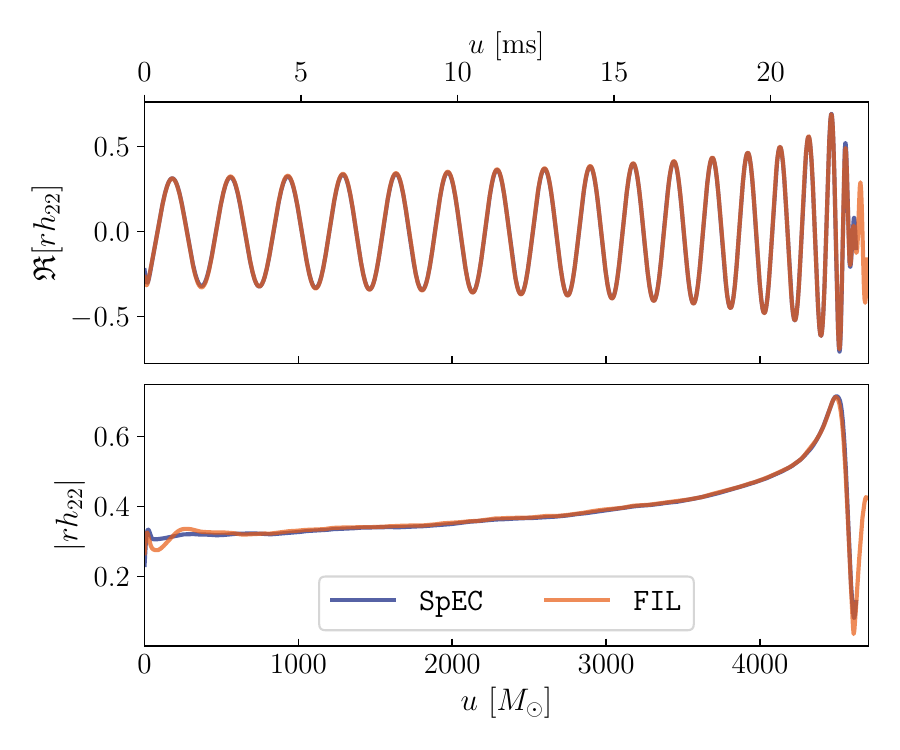}
    \caption{Real part of the $l=2, m=2$ strain mode
      $\mathfrak{R}(rh_{22})$ and amplitude of the $l=2, m=2$ strain mode
      $|rh_{22}|$ for the highest resolution runs available from each code (SpEC
      Lev3, and FIL Lev2). Runs are aligned to SpEC Lev3 with a time and phase
      shift. All waveforms generally show close agreement during inspiral and
      reach merger at similar times, having similar cycle evolutions, peak
      amplitudes, and inspiral lengths.}
    \label{fig:strain_and_phase_best}
\end{figure}

Here we compare waveform features during the inspiral. Figure~\ref{fig:strain_and_phase_best} shows the real part of the GW strain $rh_{22}$ and the strain amplitude $|rh_{22}|$ for the highest-resolution runs available from each code. Waveforms in Fig.~\ref{fig:strain_and_phase_best} have been individually aligned to SpEC Lev3 by time and phase shifts as described in Sec.~\ref{sec:extraction}.

In general, all codes show remarkable agreement by eye in frequency and phase evolution until merger. The fact that we start out with very good agreement already implies that we should be able to meaningfully compare the different codes quantitatively. Since the initial configuration is constructed without gravitational wave content from previous orbits being present in the initial domain, {\it junk radiation}, spurious waveform content sourced from the relaxation of initial data in the beginning of evolution, and CCE junk are present (see initial oscillations in the amplitude), but do not significantly impact the observed code agreement.

\begin{figure*}
  \centering
    \includegraphics[width=1.8\columnwidth]{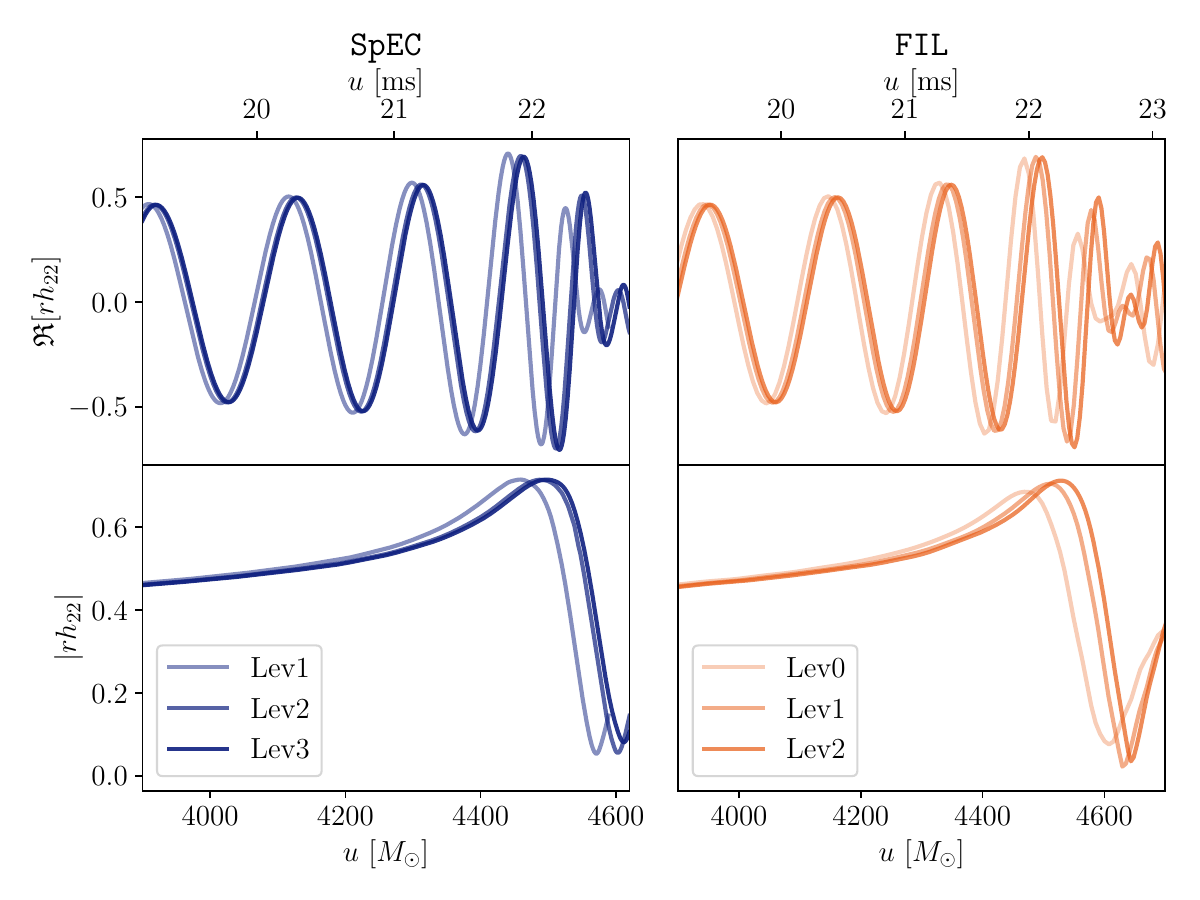}
    \caption{Real part of the $l=2, m=2$ strain mode $\mathfrak{R}(r h_{22})$
      and amplitude of the $l=2, m=2$ strain mode $|rh_{22}|$ for all
      resolutions available from each code. The waveforms are not aligned in
      time. Darker curves correspond to higher resolution runs. Inspiral is
      consistent between different resolutions, especially in \texttt{SpEC} and
      \texttt{FIL}. Runs with higher resolutions coalesce at later times because
      of decreased numerical dissipation.}
    \label{fig:strain_and_phase_levs}
\end{figure*}

Figure~\ref{fig:strain_and_phase_levs} provides a zoom-in on the time of merger for all resolutions. Runs in this figure are unaligned and are instead matched in time to the retarded time at infinity, $u$, as best estimated for each code.
Since numerical resolution strongly affects (spurious) dissipation of orbital energy, we expect a monotonic increase in the time of merger with increasing resolution. In all codes, higher resolution runs do reach merger at later times, by as much as $\sim 60 \ M_\odot$. Note that the coalescence time in both \texttt{SpEC} and \texttt{FIL} appears to converge asymptotically with increasing grid resolution. As such, the waveforms we use should have a sufficiently high fidelity for a code comparison, which we provide in quantitative terms in the following.

\subsection{Error Analysis}
\label{sec:error}
There are several possible sources of error in numerical-relativity waveforms. One source is roundoff error, originating from the finite precision of values represented by computer hardware. More important and usually dominant is truncation error, which comes from approximating continuous values as discrete. In numerical waveforms, truncation error includes contributions from discretization in space and time. Additional error sources come from extrapolation of waveform quantities to future null infinity. While CCE does not have extrapolation errors, it is still subject to truncation errors in the discretization of the Einstein equations on the characteristic domain, and systematic errors from choosing the worldtube radius and initial data on the initial null slice. Improving the accuracy of numerical waveforms requires understanding which of these error sources dominate.

In the following sections, we show several measures of error using the phase of the $l=2,m=2$ mode of the strain. In Sec.~\ref{sec:convergence} we perform self-convergence tests with \texttt{SpEC} and \texttt{FIL} to check the self-consistency of measured errors with expected convergence rates. In Sec.~\ref{sec:extraction_error} we measure the impact of waveform extraction in overall phase error, and in Sec.~\ref{sec:richardson} we extrapolate continuum solutions from \texttt{SpEC} and \texttt{FIL} waveforms to assess systematic agreement between the two codes. Lastly, in Sec.~\ref{sec:implications} we discuss what our findings imply for target accuracy requirements with respect to next-generation gravitational wave observatories.

\subsubsection{Phase Error Convergence}
\label{sec:convergence}
Convergent behavior in phase is a particularly important benchmark for numerical waveforms since the phase encodes various information about the system, including tidal deformability~\cite{Flanagan:2007ix}. In this section, we perform self-convergence tests using \texttt{SpEC} and \texttt{FIL} waveforms to determine whether phase error scales predictably with grid resolution within each code.

Suppose we have a numerical waveform evolved at three uniform spatial resolutions $\Delta x_H < \Delta x_M < \Delta x_L$. If truncation error from a FD scheme is the dominant error source,
then for some exact solution $f(x)$ we can write the expected solution obtained, $\tilde{f}(\Delta x)$, as
\begin{align}
    \tilde{f}(x) = f(\Delta x) + C\Delta x^p
    \label{eq:error_expansion}
\end{align}
where $C \Delta x^p$ is the contribution from truncation error. The variable $p$ is the convergence order, which depends on and can be computed for
a given choice of numerical FD scheme. Thus we expect that
\begin{align}
    \frac{\tilde{f}(\Delta x_M) - \tilde{f}(\Delta x_H)}{\tilde{f}(\Delta x_L) - \tilde{f}(\Delta x_M)} = \frac{\Delta x^p_M - \Delta x^p_H}{\Delta x^p_L - \Delta x^p_M}
    \label{eq:rescale}
\end{align}
We calculate phase errors as the difference in time series at different resolutions, e.g.~$\tilde{f}(\Delta x_M) - \tilde{f}(\Delta x_H)$. 

A numerical result is said to be convergent if the equality in Eq.~\eqref{eq:rescale} holds for the expected $p$, meaning the error scales predictably with a limiting factor of the numerical accuracy, e.g.~grid resolution. Convergence indicates that truncation error is the dominant error source and that implementations of the contributing numerical schemes are correct. Quantifying this error can be straightforward in numerical codes that have a fixed discretization, as in \texttt{FIL}, whereas in hybrid schemes, as in \texttt{SpEC}, in which the spectral and FD schemes have different orders, error cannot easily be analyzed in this way unless, e.g., the FD error dominates.

\begin{figure*}
    \centering
    \includegraphics[width=1.8\columnwidth]{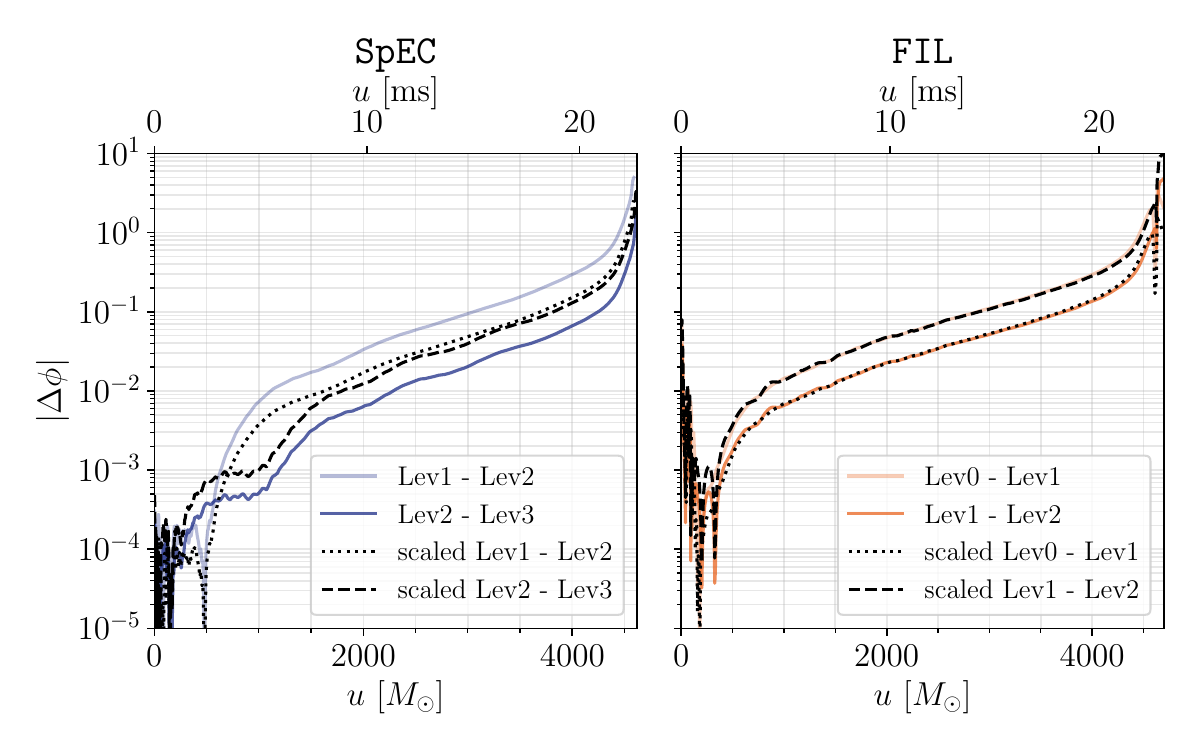}
    \caption{Phase difference over time between runs at varying resolutions of
      the same code, with \texttt{SpEC} in the left panel and \texttt{FIL} in
      the right panel. Waveforms are unaligned as in
      Fig.~\ref{fig:strain_and_phase_levs}. Dashed lines show error curves
      rescaled by an assumed convergence factor -- both \texttt{SpEC} errors and
      \texttt{FIL} errors are scaled by third order. The rescaled errors match
      the measured errors if the code conforms to the assumed convergence order;
      otherwise the error is not well-described by a single convergence
      order. \texttt{SpEC} is generally expected to be limited to fifth-order
      convergence but instead scales with seventh order, indicating that in
      reality \texttt{SpEC} has no clear convergence order in this
      regime. \texttt{FIL} is consistent with the expected and previously shown
      third-order convergence during most of inspiral.
    }
    \label{fig:phase_convergence}
\end{figure*}

In Fig.~\ref{fig:phase_convergence}, we show the phase difference $\Delta \phi$ in radians over time between consecutive resolutions given three runs for \texttt{SpEC} and \texttt{FIL}. Each difference is calculated with the higher resolution time series interpolated onto the time grid of the lower resolution time series. Waveforms are matched by estimated retarded time without alignment, similar to those shown in Fig.~\ref{fig:strain_and_phase_levs}. Colored lines show the phase differences between two waveforms of differing resolution, and dashed lines show the phase differences rescaled by the expected factor from Eq.~\eqref{eq:rescale} assuming some convergence order.

Errors in dashed lines are scaled from the higher resolution error and ideally should match the lower resolution error, and vice versa for dotted lines. \texttt{SpEC} errors are rescaled assuming third-order convergence ($p=3$). General convergence in \texttt{SpEC} is difficult to assess because of its coupled evolution grids and the competing effects of several numerical schemes. However, the rescaled errors do not conform to the measured errors, indicating that the error is not entirely dominated by the hydrodynamic sector, at least for the lowest resolution simulation available. It is likely that the multiple schemes in \texttt{SpEC} interact differently at different grid resolutions (e.g.~such that errors between SpEC Lev2 and SpEC Lev3 cancel). For different setups, \texttt{SpEC} has previously shown convergence during inspiral for BHNS waveforms~\cite{Duez:2008rb}. 

\texttt{FIL} errors are rescaled assuming the expected third order ($p=3$). The rescaled errors show that \texttt{FIL} is consistent with third-order convergence during inspiral as demonstrated previously~\cite{Most:2019kfe}. Although the FD scheme in \texttt{FIL} is formally fourth-order convergent, accuracy is limited either by the third-order time stepper or the third-order fallback reconstruction in the WENO-Z algorithm.

Overall, we confirm that phase errors systematically decrease with resolution for \texttt{SpEC} and \texttt{FIL}, and find that overall phase errors are small, on the order of $\mathcal{O}(10^{-2})$.

\subsubsection{Waveform Extraction Error}
\label{sec:extraction_error}
In this section, we assess the error contribution from waveform extraction. It
has previously been established that extraction errors can dominate in the early
inspiral of BNS simulations (e.g, Ref.~\cite{Bernuzzi:2016pie}). In BBH
simulations, errors in waveform extraction have been well investigated (e.g.,
Refs.~\cite{Chu:2015kft,Boyle:2019kee}). BBH waveform extraction has undergone
many recent advances with the development of CCE~\cite{Moxon:2021gbv,
  2020PhRvD.102d4052M}. Unlike FRE, CCE evolves the spacetime from a chosen
timelike worldtube out to future null infinity and is capable of resolving
gravitational wave memory~\cite{Pollney:2010hs}. Since CCE is generally
considered to be the best mathematically well-motivated extraction method, we
evaluate extraction error by comparing between CCE and FRE waveforms in
\texttt{SpEC}, effectively treating CCE as a reference point. We prefer this
method over comparing different FRE waveforms as it gives a measure of total FRE
error. It has been shown that for BBH, CCE is more accurate than FRE in that it
obeys the Bondi constraints better and matches better to PN at early
times~\cite{Mitman:2020bjf,Mitman:2021xkq,Mitman:2022kwt}, and it has been shown
that the numerical truncation error associated with CCE is much smaller than the
numerical truncation error of the spectral BBH Cauchy
evolution~\cite{Moxon:2021gbv,2020PhRvD.102d4052M}. Given that even for spectral
BBH simulations the FRE error is smaller than truncation
error~\cite{Boyle:2019kee}, we expect truncation error (in particular due to
hydrodynamics) and not wave extraction error to be dominant for current BNS
waveforms. Note, however, that some aspects of CCE are under active
investigation, and beyond the scope of this work.

\begin{figure}[h]
    \centering
    \includegraphics[width=\columnwidth]{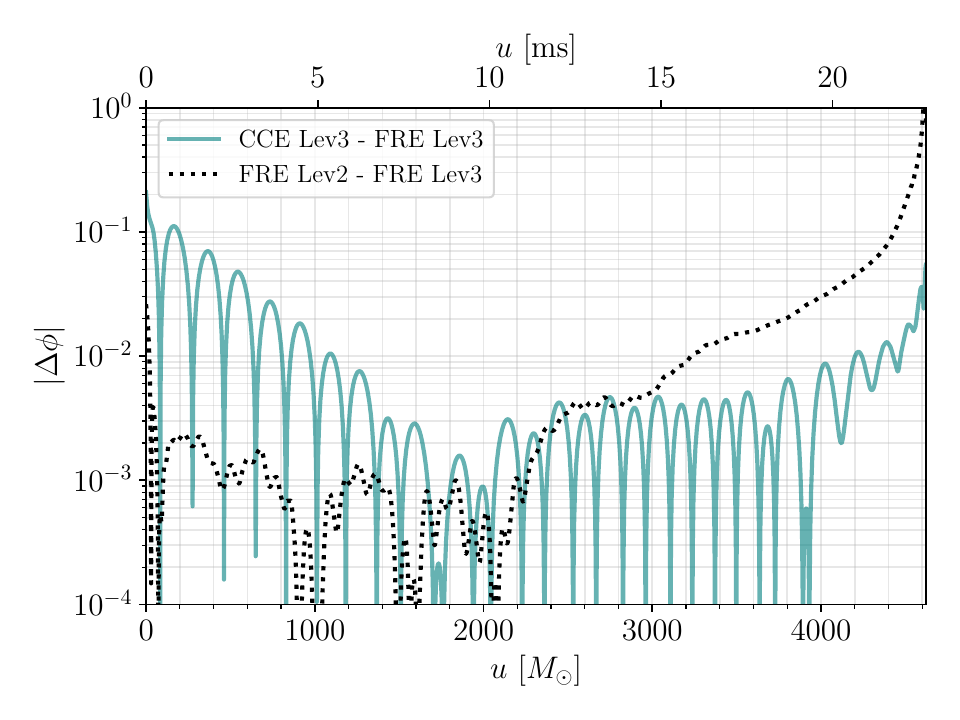}
    \caption{Phase difference between FRE and CCE waveforms at the same grid resolution, with the CCE waveform extracted using a worldtube radius of $R = 496 \ M_\odot$. The black dashed line shows relative truncation error between SpEC Lev2 and SpEC Lev3 extracted with FRE. Extraction error is comparable to or lower than truncation error throughout inspiral. This trend holds for all examined CCE worldtube radii, although there is no clear trend in error with respect to worldtube radius.}
    \label{fig:spec_cce}
\end{figure}

Figure~\ref{fig:spec_cce} shows the phase difference between CCE and FRE waveforms at the same FD grid resolution. The shown CCE waveform is extracted using a worldtube radius of $R = 496 \ M_\odot$. The CCE waveform has been aligned with a time and phase shift to SpEC Lev3 as extrapolated with FRE. The black dashed line shows the phase difference (i.e.~the truncation error, similar to those shown in Fig.~\ref{fig:phase_convergence}) between SpEC Lev2 and SpEC Lev3, both extrapolated with FRE, for comparison of truncation error with extraction error. In general, truncation error is either larger than or comparable to the phase difference between CCE and FRE waveforms through inspiral and is clearly larger than extraction error at merger, indicating that truncation error tends to dominate over extraction error, as expected.

It turns out that the choice of CCE worldtube radius makes a small difference in
the CCE waveforms, for reasons that are not fully understood and beyond the
scope of this paper. In theory, the best choice of worldtube radius minimizes
Bondi constraint values. For Fig.~\ref{fig:spec_cce}, we choose the worldtube
radius that minimizes the Bondi constraint
\begin{equation}
  \label{eq:Psi2 time bondi constraint}
  \partial_t \Psi_2=\eth\Psi_3 + \sigma\Psi_4,
\end{equation}
where $\Psi_2$, $\Psi_3$, and $\Psi_4$ are the Weyl scalars. We minimize this
constraint because $\Psi_2$ is used in BMS frame fixing.
 However, extraction error is
subdominant relative to truncation error at all examined worldtube radii, so our
findings are not significantly impacted by the choice of radius.

\subsubsection{Phase Error Comparison}
\label{sec:richardson}
In this section, our goal is to directly compare the phase errors of the different codes presented in this paper. This is important because it allows us to draw conclusions about potential systematic differences between the codes (e.g., small differences even on the initial data level will lead to slightly different physical parameters of the system). It also allows us to estimate the computational cost per level of accuracy, which we do in the next section.

One way of facilitating a comparison is to designate a reference solution and compare to it. This is easiest if a clear convergence order of a code can be established, as it allows us to extrapolate the numerical solutions to their continuum limit (at zero grid spacing), {so-called Richardson extrapolation}. Given a numerical solution $\tilde{f}(\Delta x)$ computed at two different grid resolutions $\Delta x_1,\ \Delta x_2$ and known convergence order $p$, the solution in the limit of zero grid spacing can be estimated from Eq.~\eqref{eq:error_expansion} as
\begin{align}
  f(x) & \approx \tilde{f}(\Delta x_1) + \frac{\tilde{f}(\Delta x_1) -
         \tilde{f}(\Delta x_2)}{(\Delta x_2 / \Delta x_1)^p - 1}
         \label{eq:richardson}
\end{align}
assuming $C$ is independent of grid resolution (which it should be).

\texttt{FIL} results in a systematic third-order convergent solution, as shown in Sec.~\ref{sec:convergence} (see also Ref.~\cite{Most:2019kfe}). We therefore adopt the extrapolated \texttt{FIL} solution as our reference for the purpose of this comparison and compute phase differences for all waveforms. The Richardson extrapolation assumes third-order convergence in \texttt{FIL} (as verified in Sec.~\ref{sec:convergence}) and uses the highest two resolution runs available, FIL Lev1 and FIL Lev2. In particular, the $l=2,\ m=2$ strain mode is extrapolated, and the reference phase is computed from the Richardson extrapolated strain as a derived quantity.

\begin{figure*}
    \centering
    \includegraphics[width=1.8\columnwidth]{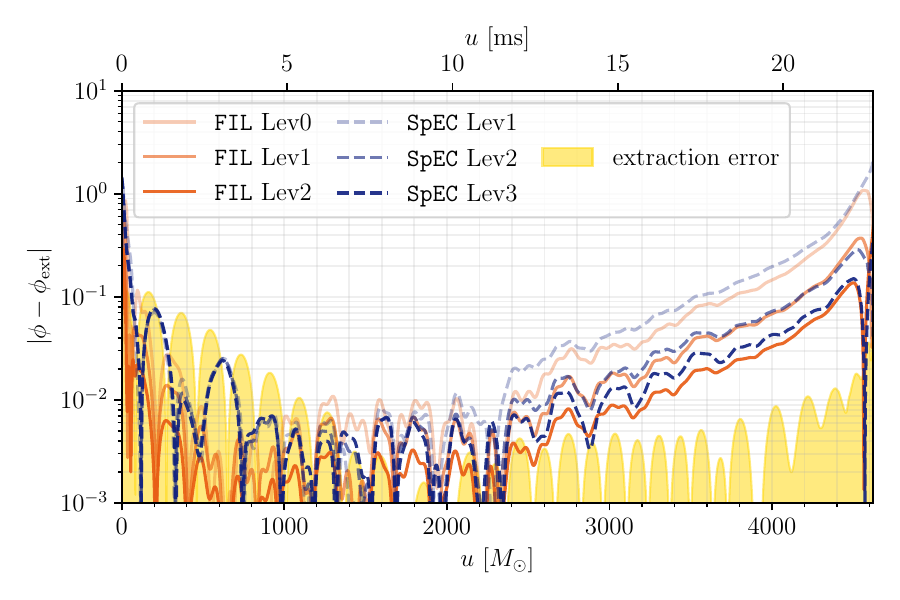}
    \caption{Phase difference over time in radians relative to the waveform Richardson-extrapolated from \texttt{FIL} Lev1 and \texttt{FIL} Lev2 for all runs. The top panel shows phase errors from different codes together (with \texttt{SpEC} error in dashed lines and \texttt{FIL} error in solid lines), and the bottom panels show the errors measured for each code separately. The shaded region shows extraction error as measured from \texttt{SpEC} in Sec.~\ref{sec:extraction_error}, using the CCE radius $R = 496 \ M_\odot$. \texttt{SpEC} and \texttt{FIL} show similar magnitudes in error, indicating that there is no resolvable systematic difference between the two codes at the chosen grid resolutions. Both \texttt{SpEC} and \texttt{FIL} appear to approach 0 error with increasing resolution. Additionally, phase differences are consistently larger than extraction error, indicating that evolution error dominates throughout inspiral. Low phase error in the range of roughly $[700 \ M_\odot, \ 2000 \ M_\odot]$ is a consequence of waveform alignment, which minimizes amplitude and phase differences within the chosen time window.}
    \label{fig:richardson_phase}
\end{figure*}

Figure~\ref{fig:richardson_phase} shows the phase error relative to the reference phase for all runs. Each run has independently been aligned to the Richardson extrapolated waveform using the process outlined in Sec.~\ref{sec:extraction}. The phase differences oscillate around zero at early times because of
the choice of alignment window.

The behavior of the errors in each code as grid resolution changes indicates whether there is a systematic phase discrepancy between the two codes. If \texttt{SpEC} converged to a significantly different phase evolution from \texttt{FIL}, the difference in the \texttt{SpEC} phase from the Richardson extrapolated \texttt{FIL} phase would asymptotically approach some large value as grid resolution increases. In Fig.~\ref{fig:richardson_phase} we find that \texttt{SpEC} and \texttt{FIL} approach similar magnitudes of phase error with increasing grid resolution, and \texttt{SpEC} appears to approach 0 with increasing resolution instead of leveling off at a finite value. The errors in SpEC Lev3 and FIL Lev2 (the highest resolution runs from each code) differ from each other on the scale of $\mathcal{O}(10^{-2})$ through inspiral. Given that, as shown in Sec.~\ref{sec:convergence}, truncation error is also generally of order $\mathcal{O}(10^{-2})$, we conclude that no systematic difference in phase between \texttt{SpEC} and \texttt{FIL} is resolvable from truncation error at these resolutions. Figure~\ref{fig:richardson_phase} indicates a rough upper bound on truncation error in phase for \texttt{FIL}, which is on the order of $\mathcal{O}(10^{-1})$ for the resolutions we adopt.

\subsection{Implications}
\label{sec:implications}
In order to keep up with the accuracy demands of next-generation gravitational wave detectors, numerical waveforms need to reach mismatch errors much smaller than what is currently achieved~\cite{Purrer:2019jcp,Gamba:2020wgg}. In this section we discuss the numerical resolution and computational cost required for such an improvement.

A way of estimating accuracy and cost requirements is through waveform mismatches. Mismatch is a standard tool in gravitational waveform analysis for not only error measurement but also matching waveform templates to signal data. Our definition of mismatch, which we summarize here, is the same as in Ref.~\cite{Boyle:2019kee}.

Given two complex waveform strains $h_1$ and $h_2$, we define their inner product $\encloseangle{h_1 | h_2}$ as
\begin{align}
    \encloseangle{h_1 | h_2} = \int^{+\infty}_{-\infty} \frac{\tilde{h}_1(f){\tilde{h}_2}^*(f)}{S_n(f)} \ df
\end{align}
where $\tilde{h}_1(f)$ and $\tilde{h}_2(f)$ are the strains in the frequency domain, ${\tilde{h}_2}^*(f)$ is the complex conjugate of $h_2(f)$, and $S_n(f)$ is noise power spectral density. $S_n(f)$ is set by detector precision, and we use the Advanced LIGO noise curve~\cite{LIGO-T1800042-v5} for the following calculations. Then, we define mismatch $\mathcal{M}(h_1, h_2)$ as
\begin{align}
    \mathcal{M}(h_1, h_2) = 1 - \max_{\delta \phi,\ \delta t} \mathfrak{R}\enclosebra{\frac{\encloseangle{h_1 | h_2}}{\sqrt{\encloseangle{h_1 | h_1} \encloseangle{h_2 | h_2}}}}
\end{align}
in which $h_1$ and $h_2$ are aligned with phase and time shifts $\delta \phi,\ \delta t$ that minimize mismatch. We use the alignment procedure described in Sec.~\ref{sec:extraction} to align waveforms for the purposes of mismatch analysis. We compute mismatches using only the $l=2,m=2$ strain modes, and we use the extent of each numerical waveform from the end of junk radiation to the peak amplitude.

We compare mismatches to the faithfulness criterion defined in~\cite{Flanagan:1997kp,Damour:2010zb} and perform an analysis similar to that done in~\cite{Doulis:2022vkx}. For faithful GW parameter estimation, a numerical waveform $h$ should satisfy
\begin{align}
    \mathcal{M}(h, h_\mathrm{ref}) < \frac{N}{2\tilde{\rho}^2}
    \label{eq:threshold}
\end{align}
where $\tilde{\rho}$ is the desired signal-to-noise ratio (SNR). Formally, $N \leq 1$, but in practice it is sometimes set to the number of intrinsic parameters in order to obtain a less strict threshold value~\cite{Chatziioannou:2017tdw}, so we use both $N=1$ and $N=6$. We use the Richardson extrapolated strain computed in Sec.~\ref{sec:richardson} as $h_\mathrm{ref}$. To determine what grid spacing would be necessary to reach a given mismatch threshold, we can extrapolate mismatch to arbitary resolution by fitting computed mismatches to a power law ($\mathcal{M} \propto \Delta x^{-p}$, where $p$ is convergence order). For the following analysis, we assume a convergence order of $p=3$ for both codes, which is likely an underestimate for \texttt{SpEC} and consistent with the previously shown results in Sec.~\ref{sec:convergence} for \texttt{FIL}.

\begin{figure*}
    \centering
    \includegraphics[width=2\columnwidth]{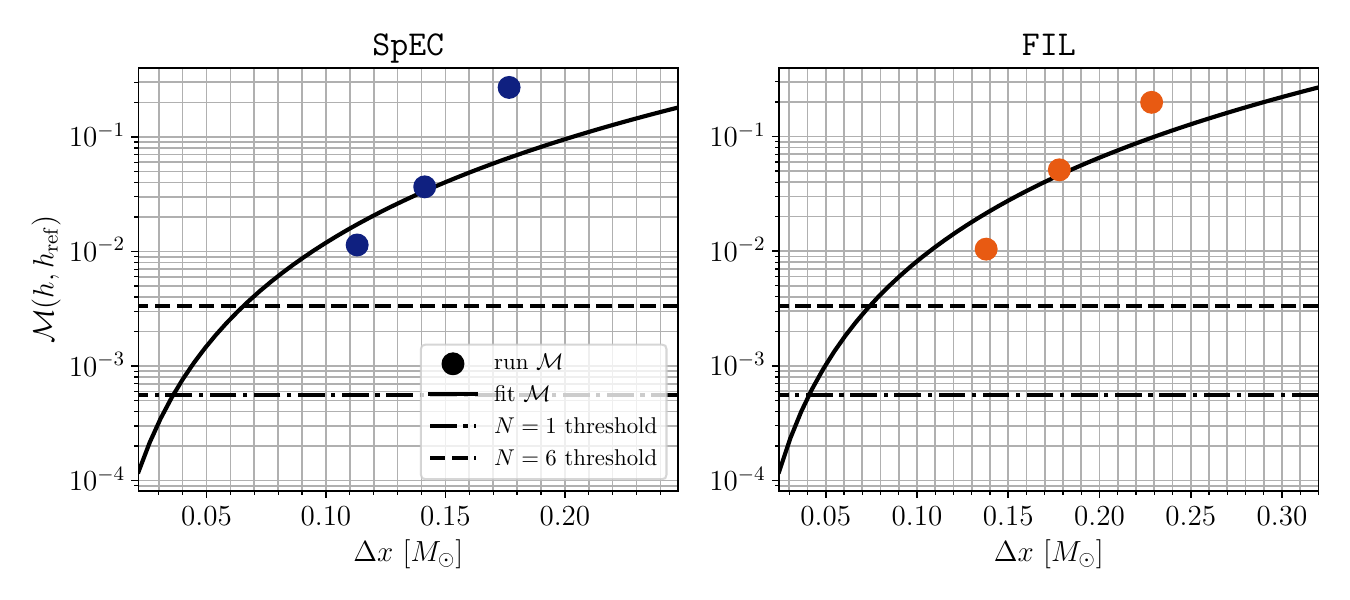}
    \caption{Computed mismatches between the Richardson extrapolated \texttt{FIL} waveform and presented numerical BNS waveforms from \texttt{SpEC} and \texttt{FIL} of varying grid separation (grid separation increases from left to right in each subplot). Solid lines denote the extrapolated mismatch as a function of grid resolution, as determined by the power law $\mathcal{M} \propto \Delta x^{-p}$. For both codes, we assume $p=3$. Mismatch extrapolations use the two data points of lower grid separation in each subplot. Dashed lines mark the threshold mismatch values, defined by Eq.~\eqref{eq:threshold}.}
    \label{fig:faithfulness}
\end{figure*}

\begin{figure}[h]
    \centering
    \includegraphics[width=\columnwidth]{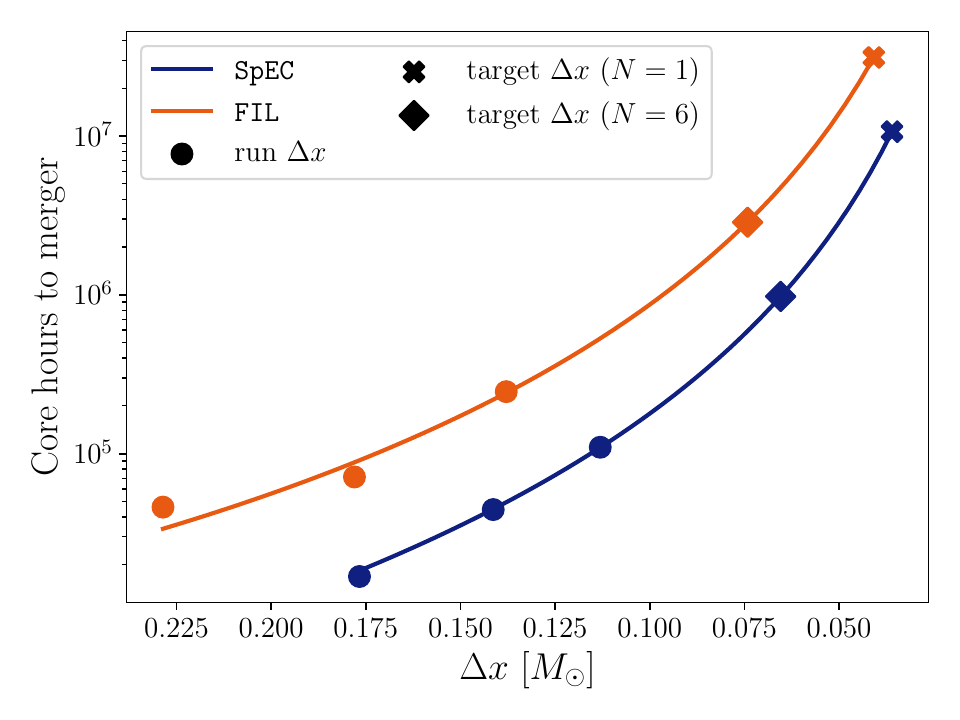}
    \caption{The computational cost in core hours for a BNS waveform through merger in relation to the finite difference grid separation $\Delta x$. Dots mark compute times of runs presented in this paper, crosses mark the projected costs given the grid separations required to reach the $N=1$ mismatch threshold shown in Fig.~\ref{fig:faithfulness}, and diamonds mark the respective costs for the $N=6$ threshold. Curves are costs extrapolated from known run costs assuming cost scales as $\mathcal{O}(\Delta x^4)$. For either \texttt{SpEC} or \texttt{FIL}, a BNS run with high enough grid resolution to reach the target mismatch would require on the order of $1$ million to $10$ million core hours.}
    \label{fig:compute cost}
\end{figure}

Figure~\ref{fig:faithfulness} shows mismatches versus grid separation for
reference mismatches of the various numerical BNS runs from \texttt{SpEC} and
\texttt{FIL}, extrapolated mismatches for each code as solid curves, and the
distinguishability thresholds with $\tilde{\rho}=30$ and $N=\enclosecurly{1,6}$
as dashed lines. Figure~\ref{fig:compute cost} then shows total compute time in
core hours until merger as a function of grid separation. The mismatches
$\mathcal{M}(h, h_\mathrm{ref})$ of SpEC Lev2 and SpEC Lev3 are 0.0336 and
0.0114 respectively. Extrapolating from these mismatches, we find that, assuming
third order error convergence, a \texttt{SpEC} run must have a grid spacing of
$\sim 0.036 \ M_\odot$ in order to reach the $N=1$ threshold, i.e.~a $3\times$
resolution increase and an $81\times$ cost increase. For a more conservative
cost estimate, we can treat the convergence order as a fitting parameter instead
of fixing it. This gives a resulting convergence order of $p = 5.2$, which is
plausible for \texttt{SpEC}, and the target resolution becomes $\sim 0.063 \
M_\odot$, leading to a $\sim 9\times$ cost increase. For FIL Lev2 and FIL Lev1,
we find mismatches $\mathcal{M}(h, h_\mathrm{ref})$ of 0.0104 and 0.0512. The
SNR 30 threshold resolution assuming third order convergence is then $\sim 0.041
\ M_\odot$, resulting in a $\sim 134\times$ cost increase. Note that these cost
increases also assume the current simulation length is sufficient for accurate
hybridization with semi-analytic models like post-Newtonian to produce very long
waveforms. If this is not true, then the computational cost would increase
significantly beyond the estimates here.

\section{Conclusion\label{sec:conclusion}}

Next-generation gravitational wave detectors offer exciting prospects for probing dense matter with BNS~\cite{Chatziioannou:2021tdi,Finstad:2022oni}. In order to extract this information from the GW signal, highly accurate models for inspiral gravitational wave emission will be needed~\cite{Gamba:2020wgg}. Crucially, these rely on calibration to NR simulations~\cite{Dietrich:2019kaq}, which in turn will have similar accuracy requirements.

In this work, we investigated potential systematics in gravitational
waveforms computed with current numerical relativity simulations. We have
done so by performing a direct code comparison of two NR codes (\texttt{FIL}, \texttt{SpEC}) that are very different in both the equations they solve and the numerical schemes they use. We have presented a detailed BNS error analysis that covers GW phase error, convergence behavior, and extraction error across multiple resolutions of BNS runs with the same initial data parameters. By quantifying error in a BNS waveform observable and considering the relative impacts of evolution error and extraction error, this analysis contributes to an area of limited study in the current literature.

Reassuringly, we find that within currently used numerical resolutions the codes
systematically agree, and are convergent.
We further show that waveform extraction errors are at present negligible when
compared to a more accurate and well-defined CCE approach. Also, \texttt{SpEC}
uses pseudo-spectral methods for the wavezone. In this region, BNS simulations
are the same as BBH simulations for these codes, and BBH simulations with
\texttt{SpEC} have substantially smaller errors than BNS (e.g.,
Refs.~\cite{Boyle:2019kee,Lovelace:2024wra}). These considerations likely imply
that the waveform error is currently limited by the hydrodynamic evolution. 

Using the current computational cost, which for a given accuracy we find to be
roughly comparable among all codes within a factor of a few (see
Table~\ref{tab:run_table}), we predict that resolutions three to four times
higher than what are used now may be needed to built a catalog to meet the
requirements of next-generation facilities (see also Ref.~\cite{Doulis:2022vkx}
for similar conclusions). Additional improvements on the hydrodynamics
side~\cite{Doulis:2022vkx,Kiuchi:2022ubj,Radice:2013hxh,Bernuzzi:2016pie,Most:2019kfe},
as well as the use of new hardware acceleration~\cite{Fields:2024pob} or
different numerical
approaches~\cite{Fambri:2018udk,Hebert:2018xbk,Deppe:2021bhi,Deppe:2024ckt,Adhikari:2025nio}
may be necessary to overcome present limitations.

As a starting point, this paper only considers one binary configuration
with equal mass components (see, e.g., Ref.~\cite{Jan:2023raq} for
potential challenges in BBH waveforms). Additionally, our analysis is
confined to the $l=2, \ m=2$ waveform mode. Higher-order modes have even
greater accuracy requirements in order to be numerically resolved, but are likely
important for unequal mass systems. Future work will be needed to
address these issues.

\begin{acknowledgments}
We would like to thank Keefe Mitman, Yoonsoo Kim, and Nicholas Rui for helpful
discussions. This material is based upon work supported by the National Science
Foundation under Grants No.~PHY-2309211, No.~PHY-2309231, and No.~OAC-2209656 at
Caltech and No.~PHY-2407742, No.~PHY-2207342, and No.~OAC-2209655 at Cornell.
ERM acknowledge support by the National Science Foundation under grants No.
PHY-2309210 and OAC-2103680. ERM also acknowledges the use of Delta at the
National Center for Supercomputing Applications (NCSA) through allocation
PHY210074 from the Advanced Cyberinfrastructure Coordination Ecosystem: Services
\& Support (ACCESS) program, which is supported by National Science Foundation
grants \#2138259, \#2138286, \#2138307, \#2137603, and \#2138296. Additional
simulations were performed on the NSF Frontera supercomputer under grant
AST21006. Finally, ERM acknowledges support through DOE NERSC supercomputer
Perlmutter under grant m4575, which uses resources of the National Energy
Research Scientific Computing Center, a DOE Office of Science User Facility
supported by the Office of Science of the U.S. Department of Energy under
Contract No. DE-AC02-05CH11231 using NERSC award NP-ERCAP0028480. FF gratefully
acknowledges support from the Department of Energy, Office of Science, Office of
Nuclear Physics, under contract number DE-SC0020435 and DE-SC0025023, and from
NASA through grant 80NSSC22K0719. MD gratefully acknowledges support from the
NSF through grant PHY-2407726. Any opinions, findings, and conclusions or
recommendations expressed in this material are those of the author(s) and do not
necessarily reflect the views of the National Science Foundation. This work was
supported by the Sherman Fairchild Foundation at Caltech and Cornell.
\end{acknowledgments}

\section*{References}

\bibliography{References}

\begin{thebibliography}{129}%
\makeatletter
\providecommand \@ifxundefined [1]{%
 \@ifx{#1\undefined}
}%
\providecommand \@ifnum [1]{%
 \ifnum #1\expandafter \@firstoftwo
 \else \expandafter \@secondoftwo
 \fi
}%
\providecommand \@ifx [1]{%
 \ifx #1\expandafter \@firstoftwo
 \else \expandafter \@secondoftwo
 \fi
}%
\providecommand \natexlab [1]{#1}%
\providecommand \enquote  [1]{``#1''}%
\providecommand \bibnamefont  [1]{#1}%
\providecommand \bibfnamefont [1]{#1}%
\providecommand \citenamefont [1]{#1}%
\providecommand \href@noop [0]{\@secondoftwo}%
\providecommand \href [0]{\begingroup \@sanitize@url \@href}%
\providecommand \@href[1]{\@@startlink{#1}\@@href}%
\providecommand \@@href[1]{\endgroup#1\@@endlink}%
\providecommand \@sanitize@url [0]{\catcode `\\12\catcode `\$12\catcode
  `\&12\catcode `\#12\catcode `\^12\catcode `\_12\catcode `\%12\relax}%
\providecommand \@@startlink[1]{}%
\providecommand \@@endlink[0]{}%
\providecommand \url  [0]{\begingroup\@sanitize@url \@url }%
\providecommand \@url [1]{\endgroup\@href {#1}{\urlprefix }}%
\providecommand \urlprefix  [0]{URL }%
\providecommand \Eprint [0]{\href }%
\providecommand \doibase [0]{https://doi.org/}%
\providecommand \selectlanguage [0]{\@gobble}%
\providecommand \bibinfo  [0]{\@secondoftwo}%
\providecommand \bibfield  [0]{\@secondoftwo}%
\providecommand \translation [1]{[#1]}%
\providecommand \BibitemOpen [0]{}%
\providecommand \bibitemStop [0]{}%
\providecommand \bibitemNoStop [0]{.\EOS\space}%
\providecommand \EOS [0]{\spacefactor3000\relax}%
\providecommand \BibitemShut  [1]{\csname bibitem#1\endcsname}%
\let\auto@bib@innerbib\@empty
\bibitem [{\citenamefont {Abbott}\ \emph {et~al.}(2017)\citenamefont {Abbott}
  \emph {et~al.}}]{LIGOScientific:2017vwq}%
  \BibitemOpen
  \bibfield  {author} {\bibinfo {author} {\bibfnamefont {B.~P.}\ \bibnamefont
  {Abbott}} \emph {et~al.} (\bibinfo {collaboration} {LIGO Scientific,
  Virgo}),\ }\bibfield  {title} {\bibinfo {title} {{GW170817: Observation of
  Gravitational Waves from a Binary Neutron Star Inspiral}},\ }\href
  {https://doi.org/10.1103/PhysRevLett.119.161101} {\bibfield  {journal}
  {\bibinfo  {journal} {Phys. Rev. Lett.}\ }\textbf {\bibinfo {volume} {119}},\
  \bibinfo {pages} {161101} (\bibinfo {year} {2017})},\ \Eprint
  {https://arxiv.org/abs/1710.05832} {arXiv:1710.05832 [gr-qc]} \BibitemShut
  {NoStop}%
\bibitem [{\citenamefont {Abbott}\ \emph {et~al.}(2020)\citenamefont {Abbott}
  \emph {et~al.}}]{LIGOScientific:2020aai}%
  \BibitemOpen
  \bibfield  {author} {\bibinfo {author} {\bibfnamefont {B.~P.}\ \bibnamefont
  {Abbott}} \emph {et~al.} (\bibinfo {collaboration} {LIGO Scientific,
  Virgo}),\ }\bibfield  {title} {\bibinfo {title} {{GW190425: Observation of a
  Compact Binary Coalescence with Total Mass $\sim 3.4 M_{\odot}$}},\ }\href
  {https://doi.org/10.3847/2041-8213/ab75f5} {\bibfield  {journal} {\bibinfo
  {journal} {Astrophys. J. Lett.}\ }\textbf {\bibinfo {volume} {892}},\
  \bibinfo {pages} {L3} (\bibinfo {year} {2020})},\ \Eprint
  {https://arxiv.org/abs/2001.01761} {arXiv:2001.01761 [astro-ph.HE]}
  \BibitemShut {NoStop}%
\bibitem [{\citenamefont {Petrov}\ \emph {et~al.}(2022)\citenamefont {Petrov},
  \citenamefont {Singer}, \citenamefont {Coughlin}, \citenamefont {Kumar},
  \citenamefont {Almualla}, \citenamefont {Anand}, \citenamefont {Bulla},
  \citenamefont {Dietrich}, \citenamefont {Foucart},\ and\ \citenamefont
  {Guessoum}}]{Petrov:2021bqm}%
  \BibitemOpen
  \bibfield  {author} {\bibinfo {author} {\bibfnamefont {P.}~\bibnamefont
  {Petrov}}, \bibinfo {author} {\bibfnamefont {L.~P.}\ \bibnamefont {Singer}},
  \bibinfo {author} {\bibfnamefont {M.~W.}\ \bibnamefont {Coughlin}}, \bibinfo
  {author} {\bibfnamefont {V.}~\bibnamefont {Kumar}}, \bibinfo {author}
  {\bibfnamefont {M.}~\bibnamefont {Almualla}}, \bibinfo {author}
  {\bibfnamefont {S.}~\bibnamefont {Anand}}, \bibinfo {author} {\bibfnamefont
  {M.}~\bibnamefont {Bulla}}, \bibinfo {author} {\bibfnamefont
  {T.}~\bibnamefont {Dietrich}}, \bibinfo {author} {\bibfnamefont
  {F.}~\bibnamefont {Foucart}},\ and\ \bibinfo {author} {\bibfnamefont
  {N.}~\bibnamefont {Guessoum}},\ }\bibfield  {title} {\bibinfo {title}
  {{Data-driven Expectations for Electromagnetic Counterpart Searches Based on
  LIGO/Virgo Public Alerts}},\ }\href
  {https://doi.org/10.3847/1538-4357/ac366d} {\bibfield  {journal} {\bibinfo
  {journal} {Astrophys. J.}\ }\textbf {\bibinfo {volume} {924}},\ \bibinfo
  {pages} {54} (\bibinfo {year} {2022})},\ \Eprint
  {https://arxiv.org/abs/2108.07277} {arXiv:2108.07277 [astro-ph.HE]}
  \BibitemShut {NoStop}%
\bibitem [{\citenamefont {{Abbott}}\ \emph {et~al.}(2020)\citenamefont
  {{Abbott}}, \citenamefont {{Abbott}}, \citenamefont {{Abbott}}, \citenamefont
  {{Abraham}}, \citenamefont {{Acernese}}, \citenamefont {{Ackley}},
  \citenamefont {{Adams}}, \citenamefont {{Adya}}, \citenamefont {{Affeldt}},
  \citenamefont {{Agathos}}, \citenamefont {{Agatsuma}}, \citenamefont
  {{Aggarwal}}, \citenamefont {{Aguiar}}, \citenamefont {{Aiello}},
  \citenamefont {{Ain}}, \citenamefont {{Ajith}}, \citenamefont {{Akutsu}},
  \citenamefont {{Allen}}, \citenamefont {{Allocca}}, \citenamefont {{Aloy}},
  \citenamefont {{Altin}}, \citenamefont {{Amato}}, \citenamefont {{Ananyeva}},
  \citenamefont {{Anderson}}, \citenamefont {{Anderson}}, \citenamefont
  {{Ando}}, \citenamefont {{Angelova}}, \citenamefont {{Antier}}, \citenamefont
  {{Appert}}, \citenamefont {{Arai}}, \citenamefont {{Arai}}, \citenamefont
  {{Arai}}, \citenamefont {{Araki}}, \citenamefont {{Araya}}, \citenamefont
  {{Araya}}, \citenamefont {{Areeda}}, \citenamefont {{Ar{\`e}ne}},
  \citenamefont {{Aritomi}}, \citenamefont {{Arnaud}}, \citenamefont {{Arun}},
  \citenamefont {{Ascenzi}}, \citenamefont {{Ashton}}, \citenamefont {{Aso}},
  \citenamefont {{Aston}}, \citenamefont {{Astone}}, \citenamefont {{Aubin}},
  \citenamefont {{Aufmuth}}, \citenamefont {{Aultoneal}}, \citenamefont
  {{Austin}}, \citenamefont {{Avendano}}, \citenamefont {{Avila-Alvarez}},
  \citenamefont {{Babak}}, \citenamefont {{Bacon}}, \citenamefont
  {{Badaracco}}, \citenamefont {{Bader}}, \citenamefont {{Bae}}, \citenamefont
  {{Bae}}, \citenamefont {{Baiotti}}, \citenamefont {{Bajpai}}, \citenamefont
  {{Baker}}, \citenamefont {{Baldaccini}}, \citenamefont {{Ballardin}},
  \citenamefont {{Ballmer}}, \citenamefont {{Banagiri}}, \citenamefont
  {{Barayoga}}, \citenamefont {{Barclay}}, \citenamefont {{Barish}},
  \citenamefont {{Barker}}, \citenamefont {{Barkett}}, \citenamefont
  {{Barnum}}, \citenamefont {{Barone}}, \citenamefont {{Barr}}, \citenamefont
  {{Barsotti}}, \citenamefont {{Barsuglia}}, \citenamefont {{Barta}},
  \citenamefont {{Bartlett}}, \citenamefont {{Barton}}, \citenamefont
  {{Bartos}}, \citenamefont {{Bassiri}}, \citenamefont {{Basti}}, \citenamefont
  {{Bawaj}}, \citenamefont {{Bayley}}, \citenamefont {{Bazzan}}, \citenamefont
  {{B{\'e}csy}}, \citenamefont {{Bejger}}, \citenamefont {{Belahcene}},
  \citenamefont {{Bell}}, \citenamefont {{Beniwal}}, \citenamefont {{Berger}},
  \citenamefont {{Bergmann}}, \citenamefont {{Bernuzzi}}, \citenamefont
  {{Bero}}, \citenamefont {{Berry}}, \citenamefont {{Bersanetti}},
  \citenamefont {{Bertolini}}, \citenamefont {{Betzwieser}}, \citenamefont
  {{Bhandare}}, \citenamefont {{Bidler}}, \citenamefont {{Bilenko}},
  \citenamefont {{Bilgili}}, \citenamefont {{Billingsley}}, \citenamefont
  {{Birch}}, \citenamefont {{Birney}}, \citenamefont {{Birnholtz}},
  \citenamefont {{Biscans}}, \citenamefont {{Biscoveanu}}, \citenamefont
  {{Bisht}}, \citenamefont {{Bitossi}}, \citenamefont {{Bizouard}},
  \citenamefont {{Blackburn}}, \citenamefont {{Blair}}, \citenamefont
  {{Blair}}, \citenamefont {{Blair}}, \citenamefont {{Bloemen}}, \citenamefont
  {{Bode}}, \citenamefont {{Boer}}, \citenamefont {{Boetzel}}, \citenamefont
  {{Bogaert}}, \citenamefont {{Bondu}}, \citenamefont {{Bonilla}},
  \citenamefont {{Bonnand}}, \citenamefont {{Booker}}, \citenamefont {{Boom}},
  \citenamefont {{Booth}}, \citenamefont {{Bork}}, \citenamefont {{Boschi}},
  \citenamefont {{Bose}}, \citenamefont {{Bossie}}, \citenamefont
  {{Bossilkov}}, \citenamefont {{Bosveld}}, \citenamefont {{Bouffanais}},
  \citenamefont {{Bozzi}}, \citenamefont {{Bradaschia}}, \citenamefont
  {{Brady}}, \citenamefont {{Bramley}}, \citenamefont {{Branchesi}},
  \citenamefont {{Brau}}, \citenamefont {{Briant}}, \citenamefont {{Briggs}},
  \citenamefont {{Brighenti}}, \citenamefont {{Brillet}}, \citenamefont
  {{Brinkmann}}, \citenamefont {{Brisson}}, \citenamefont {{Brockill}},
  \citenamefont {{Brooks}}, \citenamefont {{Brown}}, \citenamefont {{Brown}},
  \citenamefont {{Brunett}}, \citenamefont {{Buikema}}, \citenamefont
  {{Bulik}}, \citenamefont {{Bulten}}, \citenamefont {{Buonanno}},
  \citenamefont {{Buskulic}}, \citenamefont {{Buy}}, \citenamefont {{Byer}},
  \citenamefont {{Cabero}}, \citenamefont {{Cadonati}}, \citenamefont
  {{Cagnoli}}, \citenamefont {{Cahillane}}, \citenamefont {{Bustillo}},
  \citenamefont {{Callister}}, \citenamefont {{Calloni}}, \citenamefont
  {{Camp}}, \citenamefont {{Campbell}}, \citenamefont {{Canepa}}, \citenamefont
  {{Cannon}}, \citenamefont {{Cannon}}, \citenamefont {{Cao}}, \citenamefont
  {{Cao}}, \citenamefont {{Capocasa}}, \citenamefont {{Carbognani}},
  \citenamefont {{Caride}}, \citenamefont {{Carney}}, \citenamefont
  {{Carullo}}, \citenamefont {{Diaz}}, \citenamefont {{Casentini}},
  \citenamefont {{Caudill}}, \citenamefont {{Cavagli{\`a}}}, \citenamefont
  {{Cavalier}}, \citenamefont {{Cavalieri}}, \citenamefont {{Cella}},
  \citenamefont {{Cerd{\'a}-Dur{\'a}n}}, \citenamefont {{Cerretani}},
  \citenamefont {{Cesarini}}, \citenamefont {{Chaibi}}, \citenamefont
  {{Chakravarti}}, \citenamefont {{Chamberlin}}, \citenamefont {{Chan}},
  \citenamefont {{Chan}}, \citenamefont {{Chao}}, \citenamefont {{Charlton}},
  \citenamefont {{Chase}}, \citenamefont {{Chassande-Mottin}}, \citenamefont
  {{Chatterjee}}, \citenamefont {{Chaturvedi}}, \citenamefont
  {{Chatziioannou}}, \citenamefont {{Cheeseboro}}, \citenamefont {{Chen}},
  \citenamefont {{Chen}},\ and\ \citenamefont {{Chen}}}]{2020LRR....23....3A}%
  \BibitemOpen
  \bibfield  {author} {\bibinfo {author} {\bibfnamefont {B.~P.}\ \bibnamefont
  {{Abbott}}}, \bibinfo {author} {\bibfnamefont {R.}~\bibnamefont {{Abbott}}},
  \bibinfo {author} {\bibfnamefont {T.~D.}\ \bibnamefont {{Abbott}}}, \bibinfo
  {author} {\bibfnamefont {S.}~\bibnamefont {{Abraham}}}, \bibinfo {author}
  {\bibfnamefont {F.}~\bibnamefont {{Acernese}}}, \bibinfo {author}
  {\bibfnamefont {K.}~\bibnamefont {{Ackley}}}, \bibinfo {author}
  {\bibfnamefont {C.}~\bibnamefont {{Adams}}}, \bibinfo {author} {\bibfnamefont
  {V.~B.}\ \bibnamefont {{Adya}}}, \bibinfo {author} {\bibfnamefont
  {C.}~\bibnamefont {{Affeldt}}}, \bibinfo {author} {\bibfnamefont
  {M.}~\bibnamefont {{Agathos}}}, \bibinfo {author} {\bibfnamefont
  {K.}~\bibnamefont {{Agatsuma}}}, \bibinfo {author} {\bibfnamefont
  {N.}~\bibnamefont {{Aggarwal}}}, \bibinfo {author} {\bibfnamefont {O.~D.}\
  \bibnamefont {{Aguiar}}}, \bibinfo {author} {\bibfnamefont {L.}~\bibnamefont
  {{Aiello}}}, \bibinfo {author} {\bibfnamefont {A.}~\bibnamefont {{Ain}}},
  \bibinfo {author} {\bibfnamefont {P.}~\bibnamefont {{Ajith}}}, \bibinfo
  {author} {\bibfnamefont {T.}~\bibnamefont {{Akutsu}}}, \bibinfo {author}
  {\bibfnamefont {G.}~\bibnamefont {{Allen}}}, \bibinfo {author} {\bibfnamefont
  {A.}~\bibnamefont {{Allocca}}}, \bibinfo {author} {\bibfnamefont {M.~A.}\
  \bibnamefont {{Aloy}}}, \bibinfo {author} {\bibfnamefont {P.~A.}\
  \bibnamefont {{Altin}}}, \bibinfo {author} {\bibfnamefont {A.}~\bibnamefont
  {{Amato}}}, \bibinfo {author} {\bibfnamefont {A.}~\bibnamefont {{Ananyeva}}},
  \bibinfo {author} {\bibfnamefont {S.~B.}\ \bibnamefont {{Anderson}}},
  \bibinfo {author} {\bibfnamefont {W.~G.}\ \bibnamefont {{Anderson}}},
  \bibinfo {author} {\bibfnamefont {M.}~\bibnamefont {{Ando}}}, \bibinfo
  {author} {\bibfnamefont {S.~V.}\ \bibnamefont {{Angelova}}}, \bibinfo
  {author} {\bibfnamefont {S.}~\bibnamefont {{Antier}}}, \bibinfo {author}
  {\bibfnamefont {S.}~\bibnamefont {{Appert}}}, \bibinfo {author}
  {\bibfnamefont {K.}~\bibnamefont {{Arai}}}, \bibinfo {author} {\bibfnamefont
  {K.}~\bibnamefont {{Arai}}}, \bibinfo {author} {\bibfnamefont
  {Y.}~\bibnamefont {{Arai}}}, \bibinfo {author} {\bibfnamefont
  {S.}~\bibnamefont {{Araki}}}, \bibinfo {author} {\bibfnamefont
  {A.}~\bibnamefont {{Araya}}}, \bibinfo {author} {\bibfnamefont {M.~C.}\
  \bibnamefont {{Araya}}}, \bibinfo {author} {\bibfnamefont {J.~S.}\
  \bibnamefont {{Areeda}}}, \bibinfo {author} {\bibfnamefont {M.}~\bibnamefont
  {{Ar{\`e}ne}}}, \bibinfo {author} {\bibfnamefont {N.}~\bibnamefont
  {{Aritomi}}}, \bibinfo {author} {\bibfnamefont {N.}~\bibnamefont {{Arnaud}}},
  \bibinfo {author} {\bibfnamefont {K.~G.}\ \bibnamefont {{Arun}}}, \bibinfo
  {author} {\bibfnamefont {S.}~\bibnamefont {{Ascenzi}}}, \bibinfo {author}
  {\bibfnamefont {G.}~\bibnamefont {{Ashton}}}, \bibinfo {author}
  {\bibfnamefont {Y.}~\bibnamefont {{Aso}}}, \bibinfo {author} {\bibfnamefont
  {S.~M.}\ \bibnamefont {{Aston}}}, \bibinfo {author} {\bibfnamefont
  {P.}~\bibnamefont {{Astone}}}, \bibinfo {author} {\bibfnamefont
  {F.}~\bibnamefont {{Aubin}}}, \bibinfo {author} {\bibfnamefont
  {P.}~\bibnamefont {{Aufmuth}}}, \bibinfo {author} {\bibfnamefont
  {K.}~\bibnamefont {{Aultoneal}}}, \bibinfo {author} {\bibfnamefont
  {C.}~\bibnamefont {{Austin}}}, \bibinfo {author} {\bibfnamefont
  {V.}~\bibnamefont {{Avendano}}}, \bibinfo {author} {\bibfnamefont
  {A.}~\bibnamefont {{Avila-Alvarez}}}, \bibinfo {author} {\bibfnamefont
  {S.}~\bibnamefont {{Babak}}}, \bibinfo {author} {\bibfnamefont
  {P.}~\bibnamefont {{Bacon}}}, \bibinfo {author} {\bibfnamefont
  {F.}~\bibnamefont {{Badaracco}}}, \bibinfo {author} {\bibfnamefont
  {M.~K.~M.}\ \bibnamefont {{Bader}}}, \bibinfo {author} {\bibfnamefont
  {S.~W.}\ \bibnamefont {{Bae}}}, \bibinfo {author} {\bibfnamefont {Y.~B.}\
  \bibnamefont {{Bae}}}, \bibinfo {author} {\bibfnamefont {L.}~\bibnamefont
  {{Baiotti}}}, \bibinfo {author} {\bibfnamefont {R.}~\bibnamefont {{Bajpai}}},
  \bibinfo {author} {\bibfnamefont {P.~T.}\ \bibnamefont {{Baker}}}, \bibinfo
  {author} {\bibfnamefont {F.}~\bibnamefont {{Baldaccini}}}, \bibinfo {author}
  {\bibfnamefont {G.}~\bibnamefont {{Ballardin}}}, \bibinfo {author}
  {\bibfnamefont {S.~W.}\ \bibnamefont {{Ballmer}}}, \bibinfo {author}
  {\bibfnamefont {S.}~\bibnamefont {{Banagiri}}}, \bibinfo {author}
  {\bibfnamefont {J.~C.}\ \bibnamefont {{Barayoga}}}, \bibinfo {author}
  {\bibfnamefont {S.~E.}\ \bibnamefont {{Barclay}}}, \bibinfo {author}
  {\bibfnamefont {B.~C.}\ \bibnamefont {{Barish}}}, \bibinfo {author}
  {\bibfnamefont {D.}~\bibnamefont {{Barker}}}, \bibinfo {author}
  {\bibfnamefont {K.}~\bibnamefont {{Barkett}}}, \bibinfo {author}
  {\bibfnamefont {S.}~\bibnamefont {{Barnum}}}, \bibinfo {author}
  {\bibfnamefont {F.}~\bibnamefont {{Barone}}}, \bibinfo {author}
  {\bibfnamefont {B.}~\bibnamefont {{Barr}}}, \bibinfo {author} {\bibfnamefont
  {L.}~\bibnamefont {{Barsotti}}}, \bibinfo {author} {\bibfnamefont
  {M.}~\bibnamefont {{Barsuglia}}}, \bibinfo {author} {\bibfnamefont
  {D.}~\bibnamefont {{Barta}}}, \bibinfo {author} {\bibfnamefont
  {J.}~\bibnamefont {{Bartlett}}}, \bibinfo {author} {\bibfnamefont {M.~A.}\
  \bibnamefont {{Barton}}}, \bibinfo {author} {\bibfnamefont {I.}~\bibnamefont
  {{Bartos}}}, \bibinfo {author} {\bibfnamefont {R.}~\bibnamefont {{Bassiri}}},
  \bibinfo {author} {\bibfnamefont {A.}~\bibnamefont {{Basti}}}, \bibinfo
  {author} {\bibfnamefont {M.}~\bibnamefont {{Bawaj}}}, \bibinfo {author}
  {\bibfnamefont {J.~C.}\ \bibnamefont {{Bayley}}}, \bibinfo {author}
  {\bibfnamefont {M.}~\bibnamefont {{Bazzan}}}, \bibinfo {author}
  {\bibfnamefont {B.}~\bibnamefont {{B{\'e}csy}}}, \bibinfo {author}
  {\bibfnamefont {M.}~\bibnamefont {{Bejger}}}, \bibinfo {author}
  {\bibfnamefont {I.}~\bibnamefont {{Belahcene}}}, \bibinfo {author}
  {\bibfnamefont {A.~S.}\ \bibnamefont {{Bell}}}, \bibinfo {author}
  {\bibfnamefont {D.}~\bibnamefont {{Beniwal}}}, \bibinfo {author}
  {\bibfnamefont {B.~K.}\ \bibnamefont {{Berger}}}, \bibinfo {author}
  {\bibfnamefont {G.}~\bibnamefont {{Bergmann}}}, \bibinfo {author}
  {\bibfnamefont {S.}~\bibnamefont {{Bernuzzi}}}, \bibinfo {author}
  {\bibfnamefont {J.~J.}\ \bibnamefont {{Bero}}}, \bibinfo {author}
  {\bibfnamefont {C.~P.~L.}\ \bibnamefont {{Berry}}}, \bibinfo {author}
  {\bibfnamefont {D.}~\bibnamefont {{Bersanetti}}}, \bibinfo {author}
  {\bibfnamefont {A.}~\bibnamefont {{Bertolini}}}, \bibinfo {author}
  {\bibfnamefont {J.}~\bibnamefont {{Betzwieser}}}, \bibinfo {author}
  {\bibfnamefont {R.}~\bibnamefont {{Bhandare}}}, \bibinfo {author}
  {\bibfnamefont {J.}~\bibnamefont {{Bidler}}}, \bibinfo {author}
  {\bibfnamefont {I.~A.}\ \bibnamefont {{Bilenko}}}, \bibinfo {author}
  {\bibfnamefont {S.~A.}\ \bibnamefont {{Bilgili}}}, \bibinfo {author}
  {\bibfnamefont {G.}~\bibnamefont {{Billingsley}}}, \bibinfo {author}
  {\bibfnamefont {J.}~\bibnamefont {{Birch}}}, \bibinfo {author} {\bibfnamefont
  {R.}~\bibnamefont {{Birney}}}, \bibinfo {author} {\bibfnamefont
  {O.}~\bibnamefont {{Birnholtz}}}, \bibinfo {author} {\bibfnamefont
  {S.}~\bibnamefont {{Biscans}}}, \bibinfo {author} {\bibfnamefont
  {S.}~\bibnamefont {{Biscoveanu}}}, \bibinfo {author} {\bibfnamefont
  {A.}~\bibnamefont {{Bisht}}}, \bibinfo {author} {\bibfnamefont
  {M.}~\bibnamefont {{Bitossi}}}, \bibinfo {author} {\bibfnamefont {M.~A.}\
  \bibnamefont {{Bizouard}}}, \bibinfo {author} {\bibfnamefont {J.~K.}\
  \bibnamefont {{Blackburn}}}, \bibinfo {author} {\bibfnamefont {C.~D.}\
  \bibnamefont {{Blair}}}, \bibinfo {author} {\bibfnamefont {D.~G.}\
  \bibnamefont {{Blair}}}, \bibinfo {author} {\bibfnamefont {R.~M.}\
  \bibnamefont {{Blair}}}, \bibinfo {author} {\bibfnamefont {S.}~\bibnamefont
  {{Bloemen}}}, \bibinfo {author} {\bibfnamefont {N.}~\bibnamefont {{Bode}}},
  \bibinfo {author} {\bibfnamefont {M.}~\bibnamefont {{Boer}}}, \bibinfo
  {author} {\bibfnamefont {Y.}~\bibnamefont {{Boetzel}}}, \bibinfo {author}
  {\bibfnamefont {G.}~\bibnamefont {{Bogaert}}}, \bibinfo {author}
  {\bibfnamefont {F.}~\bibnamefont {{Bondu}}}, \bibinfo {author} {\bibfnamefont
  {E.}~\bibnamefont {{Bonilla}}}, \bibinfo {author} {\bibfnamefont
  {R.}~\bibnamefont {{Bonnand}}}, \bibinfo {author} {\bibfnamefont
  {P.}~\bibnamefont {{Booker}}}, \bibinfo {author} {\bibfnamefont {B.~A.}\
  \bibnamefont {{Boom}}}, \bibinfo {author} {\bibfnamefont {C.~D.}\
  \bibnamefont {{Booth}}}, \bibinfo {author} {\bibfnamefont {R.}~\bibnamefont
  {{Bork}}}, \bibinfo {author} {\bibfnamefont {V.}~\bibnamefont {{Boschi}}},
  \bibinfo {author} {\bibfnamefont {S.}~\bibnamefont {{Bose}}}, \bibinfo
  {author} {\bibfnamefont {K.}~\bibnamefont {{Bossie}}}, \bibinfo {author}
  {\bibfnamefont {V.}~\bibnamefont {{Bossilkov}}}, \bibinfo {author}
  {\bibfnamefont {J.}~\bibnamefont {{Bosveld}}}, \bibinfo {author}
  {\bibfnamefont {Y.}~\bibnamefont {{Bouffanais}}}, \bibinfo {author}
  {\bibfnamefont {A.}~\bibnamefont {{Bozzi}}}, \bibinfo {author} {\bibfnamefont
  {C.}~\bibnamefont {{Bradaschia}}}, \bibinfo {author} {\bibfnamefont {P.~R.}\
  \bibnamefont {{Brady}}}, \bibinfo {author} {\bibfnamefont {A.}~\bibnamefont
  {{Bramley}}}, \bibinfo {author} {\bibfnamefont {M.}~\bibnamefont
  {{Branchesi}}}, \bibinfo {author} {\bibfnamefont {J.~E.}\ \bibnamefont
  {{Brau}}}, \bibinfo {author} {\bibfnamefont {T.}~\bibnamefont {{Briant}}},
  \bibinfo {author} {\bibfnamefont {J.~H.}\ \bibnamefont {{Briggs}}}, \bibinfo
  {author} {\bibfnamefont {F.}~\bibnamefont {{Brighenti}}}, \bibinfo {author}
  {\bibfnamefont {A.}~\bibnamefont {{Brillet}}}, \bibinfo {author}
  {\bibfnamefont {M.}~\bibnamefont {{Brinkmann}}}, \bibinfo {author}
  {\bibfnamefont {V.}~\bibnamefont {{Brisson}}}, \bibinfo {author}
  {\bibfnamefont {P.}~\bibnamefont {{Brockill}}}, \bibinfo {author}
  {\bibfnamefont {A.~F.}\ \bibnamefont {{Brooks}}}, \bibinfo {author}
  {\bibfnamefont {D.~A.}\ \bibnamefont {{Brown}}}, \bibinfo {author}
  {\bibfnamefont {D.~D.}\ \bibnamefont {{Brown}}}, \bibinfo {author}
  {\bibfnamefont {S.}~\bibnamefont {{Brunett}}}, \bibinfo {author}
  {\bibfnamefont {A.}~\bibnamefont {{Buikema}}}, \bibinfo {author}
  {\bibfnamefont {T.}~\bibnamefont {{Bulik}}}, \bibinfo {author} {\bibfnamefont
  {H.~J.}\ \bibnamefont {{Bulten}}}, \bibinfo {author} {\bibfnamefont
  {A.}~\bibnamefont {{Buonanno}}}, \bibinfo {author} {\bibfnamefont
  {D.}~\bibnamefont {{Buskulic}}}, \bibinfo {author} {\bibfnamefont
  {C.}~\bibnamefont {{Buy}}}, \bibinfo {author} {\bibfnamefont {R.~L.}\
  \bibnamefont {{Byer}}}, \bibinfo {author} {\bibfnamefont {M.}~\bibnamefont
  {{Cabero}}}, \bibinfo {author} {\bibfnamefont {L.}~\bibnamefont
  {{Cadonati}}}, \bibinfo {author} {\bibfnamefont {G.}~\bibnamefont
  {{Cagnoli}}}, \bibinfo {author} {\bibfnamefont {C.}~\bibnamefont
  {{Cahillane}}}, \bibinfo {author} {\bibfnamefont {J.~C.}\ \bibnamefont
  {{Bustillo}}}, \bibinfo {author} {\bibfnamefont {T.~A.}\ \bibnamefont
  {{Callister}}}, \bibinfo {author} {\bibfnamefont {E.}~\bibnamefont
  {{Calloni}}}, \bibinfo {author} {\bibfnamefont {J.~B.}\ \bibnamefont
  {{Camp}}}, \bibinfo {author} {\bibfnamefont {W.~A.}\ \bibnamefont
  {{Campbell}}}, \bibinfo {author} {\bibfnamefont {M.}~\bibnamefont
  {{Canepa}}}, \bibinfo {author} {\bibfnamefont {K.}~\bibnamefont {{Cannon}}},
  \bibinfo {author} {\bibfnamefont {K.~C.}\ \bibnamefont {{Cannon}}}, \bibinfo
  {author} {\bibfnamefont {H.}~\bibnamefont {{Cao}}}, \bibinfo {author}
  {\bibfnamefont {J.}~\bibnamefont {{Cao}}}, \bibinfo {author} {\bibfnamefont
  {E.}~\bibnamefont {{Capocasa}}}, \bibinfo {author} {\bibfnamefont
  {F.}~\bibnamefont {{Carbognani}}}, \bibinfo {author} {\bibfnamefont
  {S.}~\bibnamefont {{Caride}}}, \bibinfo {author} {\bibfnamefont {M.~F.}\
  \bibnamefont {{Carney}}}, \bibinfo {author} {\bibfnamefont {G.}~\bibnamefont
  {{Carullo}}}, \bibinfo {author} {\bibfnamefont {J.~C.}\ \bibnamefont
  {{Diaz}}}, \bibinfo {author} {\bibfnamefont {C.}~\bibnamefont {{Casentini}}},
  \bibinfo {author} {\bibfnamefont {S.}~\bibnamefont {{Caudill}}}, \bibinfo
  {author} {\bibfnamefont {M.}~\bibnamefont {{Cavagli{\`a}}}}, \bibinfo
  {author} {\bibfnamefont {F.}~\bibnamefont {{Cavalier}}}, \bibinfo {author}
  {\bibfnamefont {R.}~\bibnamefont {{Cavalieri}}}, \bibinfo {author}
  {\bibfnamefont {G.}~\bibnamefont {{Cella}}}, \bibinfo {author} {\bibfnamefont
  {P.}~\bibnamefont {{Cerd{\'a}-Dur{\'a}n}}}, \bibinfo {author} {\bibfnamefont
  {G.}~\bibnamefont {{Cerretani}}}, \bibinfo {author} {\bibfnamefont
  {E.}~\bibnamefont {{Cesarini}}}, \bibinfo {author} {\bibfnamefont
  {O.}~\bibnamefont {{Chaibi}}}, \bibinfo {author} {\bibfnamefont
  {K.}~\bibnamefont {{Chakravarti}}}, \bibinfo {author} {\bibfnamefont {S.~J.}\
  \bibnamefont {{Chamberlin}}}, \bibinfo {author} {\bibfnamefont
  {M.}~\bibnamefont {{Chan}}}, \bibinfo {author} {\bibfnamefont {M.~L.}\
  \bibnamefont {{Chan}}}, \bibinfo {author} {\bibfnamefont {S.}~\bibnamefont
  {{Chao}}}, \bibinfo {author} {\bibfnamefont {P.}~\bibnamefont {{Charlton}}},
  \bibinfo {author} {\bibfnamefont {E.~A.}\ \bibnamefont {{Chase}}}, \bibinfo
  {author} {\bibfnamefont {E.}~\bibnamefont {{Chassande-Mottin}}}, \bibinfo
  {author} {\bibfnamefont {D.}~\bibnamefont {{Chatterjee}}}, \bibinfo {author}
  {\bibfnamefont {M.}~\bibnamefont {{Chaturvedi}}}, \bibinfo {author}
  {\bibfnamefont {K.}~\bibnamefont {{Chatziioannou}}}, \bibinfo {author}
  {\bibfnamefont {B.~D.}\ \bibnamefont {{Cheeseboro}}}, \bibinfo {author}
  {\bibfnamefont {C.~S.}\ \bibnamefont {{Chen}}}, \bibinfo {author}
  {\bibfnamefont {H.~Y.}\ \bibnamefont {{Chen}}},\ and\ \bibinfo {author}
  {\bibfnamefont {K.~H.}\ \bibnamefont {{Chen}}},\ }\bibfield  {title}
  {\bibinfo {title} {{Prospects for observing and localizing gravitational-wave
  transients with Advanced LIGO, Advanced Virgo and KAGRA}},\ }\href
  {https://doi.org/10.1007/s41114-020-00026-9} {\bibfield  {journal} {\bibinfo
  {journal} {Living Reviews in Relativity}\ }\textbf {\bibinfo {volume} {23}},\
  \bibinfo {eid} {3} (\bibinfo {year} {2020})}\BibitemShut {NoStop}%
\bibitem [{\citenamefont {Corsi}\ \emph {et~al.}(2024)\citenamefont {Corsi}
  \emph {et~al.}}]{Corsi:2024vvr}%
  \BibitemOpen
  \bibfield  {author} {\bibinfo {author} {\bibfnamefont {A.}~\bibnamefont
  {Corsi}} \emph {et~al.},\ }\bibfield  {title} {\bibinfo {title}
  {{Multi-messenger astrophysics of black holes and neutron stars as probed by
  ground-based gravitational wave detectors: from present to future}},\ }\href
  {https://doi.org/10.3389/fspas.2024.1386748} {\bibfield  {journal} {\bibinfo
  {journal} {Front. Astron. Space Sci.}\ }\textbf {\bibinfo {volume} {11}},\
  \bibinfo {pages} {1386748} (\bibinfo {year} {2024})},\ \Eprint
  {https://arxiv.org/abs/2402.13445} {arXiv:2402.13445 [astro-ph.HE]}
  \BibitemShut {NoStop}%
\bibitem [{\citenamefont {Chatziioannou}(2020)}]{Chatziioannou:2020pqz}%
  \BibitemOpen
  \bibfield  {author} {\bibinfo {author} {\bibfnamefont {K.}~\bibnamefont
  {Chatziioannou}},\ }\bibfield  {title} {\bibinfo {title} {{Neutron star tidal
  deformability and equation of state constraints}},\ }\href
  {https://doi.org/10.1007/s10714-020-02754-3} {\bibfield  {journal} {\bibinfo
  {journal} {Gen. Rel. Grav.}\ }\textbf {\bibinfo {volume} {52}},\ \bibinfo
  {pages} {109} (\bibinfo {year} {2020})},\ \Eprint
  {https://arxiv.org/abs/2006.03168} {arXiv:2006.03168 [gr-qc]} \BibitemShut
  {NoStop}%
\bibitem [{\citenamefont {Raithel}(2019)}]{Raithel:2019uzi}%
  \BibitemOpen
  \bibfield  {author} {\bibinfo {author} {\bibfnamefont {C.~A.}\ \bibnamefont
  {Raithel}},\ }\bibfield  {title} {\bibinfo {title} {{Constraints on the
  Neutron Star Equation of State from GW170817}},\ }\href
  {https://doi.org/10.1140/epja/i2019-12759-5} {\bibfield  {journal} {\bibinfo
  {journal} {Eur. Phys. J. A}\ }\textbf {\bibinfo {volume} {55}},\ \bibinfo
  {pages} {80} (\bibinfo {year} {2019})},\ \Eprint
  {https://arxiv.org/abs/1904.10002} {arXiv:1904.10002 [astro-ph.HE]}
  \BibitemShut {NoStop}%
\bibitem [{\citenamefont {Le~Tiec}\ and\ \citenamefont
  {Casals}(2021)}]{LeTiec:2020spy}%
  \BibitemOpen
  \bibfield  {author} {\bibinfo {author} {\bibfnamefont {A.}~\bibnamefont
  {Le~Tiec}}\ and\ \bibinfo {author} {\bibfnamefont {M.}~\bibnamefont
  {Casals}},\ }\bibfield  {title} {\bibinfo {title} {{Spinning Black Holes Fall
  in Love}},\ }\href {https://doi.org/10.1103/PhysRevLett.126.131102}
  {\bibfield  {journal} {\bibinfo  {journal} {Phys. Rev. Lett.}\ }\textbf
  {\bibinfo {volume} {126}},\ \bibinfo {pages} {131102} (\bibinfo {year}
  {2021})},\ \Eprint {https://arxiv.org/abs/2007.00214} {arXiv:2007.00214
  [gr-qc]} \BibitemShut {NoStop}%
\bibitem [{\citenamefont {Chia}(2021)}]{Chia:2020yla}%
  \BibitemOpen
  \bibfield  {author} {\bibinfo {author} {\bibfnamefont {H.~S.}\ \bibnamefont
  {Chia}},\ }\bibfield  {title} {\bibinfo {title} {{Tidal deformation and
  dissipation of rotating black holes}},\ }\href
  {https://doi.org/10.1103/PhysRevD.104.024013} {\bibfield  {journal} {\bibinfo
   {journal} {Phys. Rev. D}\ }\textbf {\bibinfo {volume} {104}},\ \bibinfo
  {pages} {024013} (\bibinfo {year} {2021})},\ \Eprint
  {https://arxiv.org/abs/2010.07300} {arXiv:2010.07300 [gr-qc]} \BibitemShut
  {NoStop}%
\bibitem [{\citenamefont {Flanagan}\ and\ \citenamefont
  {Hinderer}(2008)}]{Flanagan:2007ix}%
  \BibitemOpen
  \bibfield  {author} {\bibinfo {author} {\bibfnamefont {E.~E.}\ \bibnamefont
  {Flanagan}}\ and\ \bibinfo {author} {\bibfnamefont {T.}~\bibnamefont
  {Hinderer}},\ }\bibfield  {title} {\bibinfo {title} {{Constraining neutron
  star tidal Love numbers with gravitational wave detectors}},\ }\href
  {https://doi.org/10.1103/PhysRevD.77.021502} {\bibfield  {journal} {\bibinfo
  {journal} {Phys. Rev. D}\ }\textbf {\bibinfo {volume} {77}},\ \bibinfo
  {pages} {021502} (\bibinfo {year} {2008})},\ \Eprint
  {https://arxiv.org/abs/0709.1915} {arXiv:0709.1915 [astro-ph]} \BibitemShut
  {NoStop}%
\bibitem [{\citenamefont {Read}\ \emph {et~al.}(2013)\citenamefont {Read},
  \citenamefont {Baiotti}, \citenamefont {Creighton}, \citenamefont {Friedman},
  \citenamefont {Giacomazzo}, \citenamefont {Kyutoku}, \citenamefont
  {Markakis}, \citenamefont {Rezzolla}, \citenamefont {Shibata},\ and\
  \citenamefont {Taniguchi}}]{Read:2013zra}%
  \BibitemOpen
  \bibfield  {author} {\bibinfo {author} {\bibfnamefont {J.~S.}\ \bibnamefont
  {Read}}, \bibinfo {author} {\bibfnamefont {L.}~\bibnamefont {Baiotti}},
  \bibinfo {author} {\bibfnamefont {J.~D.~E.}\ \bibnamefont {Creighton}},
  \bibinfo {author} {\bibfnamefont {J.~L.}\ \bibnamefont {Friedman}}, \bibinfo
  {author} {\bibfnamefont {B.}~\bibnamefont {Giacomazzo}}, \bibinfo {author}
  {\bibfnamefont {K.}~\bibnamefont {Kyutoku}}, \bibinfo {author} {\bibfnamefont
  {C.}~\bibnamefont {Markakis}}, \bibinfo {author} {\bibfnamefont
  {L.}~\bibnamefont {Rezzolla}}, \bibinfo {author} {\bibfnamefont
  {M.}~\bibnamefont {Shibata}},\ and\ \bibinfo {author} {\bibfnamefont
  {K.}~\bibnamefont {Taniguchi}},\ }\bibfield  {title} {\bibinfo {title}
  {{Matter effects on binary neutron star waveforms}},\ }\href
  {https://doi.org/10.1103/PhysRevD.88.044042} {\bibfield  {journal} {\bibinfo
  {journal} {Phys. Rev. D}\ }\textbf {\bibinfo {volume} {88}},\ \bibinfo
  {pages} {044042} (\bibinfo {year} {2013})},\ \Eprint
  {https://arxiv.org/abs/1306.4065} {arXiv:1306.4065 [gr-qc]} \BibitemShut
  {NoStop}%
\bibitem [{\citenamefont {Annala}\ \emph {et~al.}(2018)\citenamefont {Annala},
  \citenamefont {Gorda}, \citenamefont {Kurkela},\ and\ \citenamefont
  {Vuorinen}}]{Annala:2017llu}%
  \BibitemOpen
  \bibfield  {author} {\bibinfo {author} {\bibfnamefont {E.}~\bibnamefont
  {Annala}}, \bibinfo {author} {\bibfnamefont {T.}~\bibnamefont {Gorda}},
  \bibinfo {author} {\bibfnamefont {A.}~\bibnamefont {Kurkela}},\ and\ \bibinfo
  {author} {\bibfnamefont {A.}~\bibnamefont {Vuorinen}},\ }\bibfield  {title}
  {\bibinfo {title} {{Gravitational-wave constraints on the neutron-star-matter
  Equation of State}},\ }\href {https://doi.org/10.1103/PhysRevLett.120.172703}
  {\bibfield  {journal} {\bibinfo  {journal} {Phys. Rev. Lett.}\ }\textbf
  {\bibinfo {volume} {120}},\ \bibinfo {pages} {172703} (\bibinfo {year}
  {2018})},\ \Eprint {https://arxiv.org/abs/1711.02644} {arXiv:1711.02644
  [astro-ph.HE]} \BibitemShut {NoStop}%
\bibitem [{\citenamefont {Abbott}\ \emph {et~al.}(2018)\citenamefont {Abbott}
  \emph {et~al.}}]{LIGOScientific:2018cki}%
  \BibitemOpen
  \bibfield  {author} {\bibinfo {author} {\bibfnamefont {B.~P.}\ \bibnamefont
  {Abbott}} \emph {et~al.} (\bibinfo {collaboration} {LIGO Scientific,
  Virgo}),\ }\bibfield  {title} {\bibinfo {title} {{GW170817: Measurements of
  neutron star radii and equation of state}},\ }\href
  {https://doi.org/10.1103/PhysRevLett.121.161101} {\bibfield  {journal}
  {\bibinfo  {journal} {Phys. Rev. Lett.}\ }\textbf {\bibinfo {volume} {121}},\
  \bibinfo {pages} {161101} (\bibinfo {year} {2018})},\ \Eprint
  {https://arxiv.org/abs/1805.11581} {arXiv:1805.11581 [gr-qc]} \BibitemShut
  {NoStop}%
\bibitem [{\citenamefont {Most}\ \emph {et~al.}(2018)\citenamefont {Most},
  \citenamefont {Weih}, \citenamefont {Rezzolla},\ and\ \citenamefont
  {Schaffner-Bielich}}]{Most:2018hfd}%
  \BibitemOpen
  \bibfield  {author} {\bibinfo {author} {\bibfnamefont {E.~R.}\ \bibnamefont
  {Most}}, \bibinfo {author} {\bibfnamefont {L.~R.}\ \bibnamefont {Weih}},
  \bibinfo {author} {\bibfnamefont {L.}~\bibnamefont {Rezzolla}},\ and\
  \bibinfo {author} {\bibfnamefont {J.}~\bibnamefont {Schaffner-Bielich}},\
  }\bibfield  {title} {\bibinfo {title} {{New constraints on radii and tidal
  deformabilities of neutron stars from GW170817}},\ }\href
  {https://doi.org/10.1103/PhysRevLett.120.261103} {\bibfield  {journal}
  {\bibinfo  {journal} {Phys. Rev. Lett.}\ }\textbf {\bibinfo {volume} {120}},\
  \bibinfo {pages} {261103} (\bibinfo {year} {2018})},\ \Eprint
  {https://arxiv.org/abs/1803.00549} {arXiv:1803.00549 [gr-qc]} \BibitemShut
  {NoStop}%
\bibitem [{\citenamefont {Raithel}\ \emph {et~al.}(2018)\citenamefont
  {Raithel}, \citenamefont {\"Ozel},\ and\ \citenamefont
  {Psaltis}}]{Raithel:2018ncd}%
  \BibitemOpen
  \bibfield  {author} {\bibinfo {author} {\bibfnamefont {C.}~\bibnamefont
  {Raithel}}, \bibinfo {author} {\bibfnamefont {F.}~\bibnamefont {\"Ozel}},\
  and\ \bibinfo {author} {\bibfnamefont {D.}~\bibnamefont {Psaltis}},\
  }\bibfield  {title} {\bibinfo {title} {{Tidal deformability from GW170817 as
  a direct probe of the neutron star radius}},\ }\href
  {https://doi.org/10.3847/2041-8213/aabcbf} {\bibfield  {journal} {\bibinfo
  {journal} {Astrophys. J. Lett.}\ }\textbf {\bibinfo {volume} {857}},\
  \bibinfo {pages} {L23} (\bibinfo {year} {2018})},\ \Eprint
  {https://arxiv.org/abs/1803.07687} {arXiv:1803.07687 [astro-ph.HE]}
  \BibitemShut {NoStop}%
\bibitem [{\citenamefont {De}\ \emph {et~al.}(2018)\citenamefont {De},
  \citenamefont {Finstad}, \citenamefont {Lattimer}, \citenamefont {Brown},
  \citenamefont {Berger},\ and\ \citenamefont {Biwer}}]{De:2018uhw}%
  \BibitemOpen
  \bibfield  {author} {\bibinfo {author} {\bibfnamefont {S.}~\bibnamefont
  {De}}, \bibinfo {author} {\bibfnamefont {D.}~\bibnamefont {Finstad}},
  \bibinfo {author} {\bibfnamefont {J.~M.}\ \bibnamefont {Lattimer}}, \bibinfo
  {author} {\bibfnamefont {D.~A.}\ \bibnamefont {Brown}}, \bibinfo {author}
  {\bibfnamefont {E.}~\bibnamefont {Berger}},\ and\ \bibinfo {author}
  {\bibfnamefont {C.~M.}\ \bibnamefont {Biwer}},\ }\bibfield  {title} {\bibinfo
  {title} {{Tidal Deformabilities and Radii of Neutron Stars from the
  Observation of GW170817}},\ }\href
  {https://doi.org/10.1103/PhysRevLett.121.091102} {\bibfield  {journal}
  {\bibinfo  {journal} {Phys. Rev. Lett.}\ }\textbf {\bibinfo {volume} {121}},\
  \bibinfo {pages} {091102} (\bibinfo {year} {2018})},\ \bibinfo {note}
  {[Erratum: Phys.Rev.Lett. 121, 259902 (2018)]},\ \Eprint
  {https://arxiv.org/abs/1804.08583} {arXiv:1804.08583 [astro-ph.HE]}
  \BibitemShut {NoStop}%
\bibitem [{\citenamefont {Chatziioannou}(2022)}]{Chatziioannou:2021tdi}%
  \BibitemOpen
  \bibfield  {author} {\bibinfo {author} {\bibfnamefont {K.}~\bibnamefont
  {Chatziioannou}},\ }\bibfield  {title} {\bibinfo {title} {{Uncertainty limits
  on neutron star radius measurements with gravitational waves}},\ }\href
  {https://doi.org/10.1103/PhysRevD.105.084021} {\bibfield  {journal} {\bibinfo
   {journal} {Phys. Rev. D}\ }\textbf {\bibinfo {volume} {105}},\ \bibinfo
  {pages} {084021} (\bibinfo {year} {2022})},\ \Eprint
  {https://arxiv.org/abs/2108.12368} {arXiv:2108.12368 [gr-qc]} \BibitemShut
  {NoStop}%
\bibitem [{\citenamefont {Finstad}\ \emph {et~al.}(2023)\citenamefont
  {Finstad}, \citenamefont {White},\ and\ \citenamefont
  {Brown}}]{Finstad:2022oni}%
  \BibitemOpen
  \bibfield  {author} {\bibinfo {author} {\bibfnamefont {D.}~\bibnamefont
  {Finstad}}, \bibinfo {author} {\bibfnamefont {L.~V.}\ \bibnamefont {White}},\
  and\ \bibinfo {author} {\bibfnamefont {D.~A.}\ \bibnamefont {Brown}},\
  }\bibfield  {title} {\bibinfo {title} {{Prospects for a Precise Equation of
  State Measurement from Advanced LIGO and Cosmic Explorer}},\ }\href
  {https://doi.org/10.3847/1538-4357/acf12f} {\bibfield  {journal} {\bibinfo
  {journal} {Astrophys. J.}\ }\textbf {\bibinfo {volume} {955}},\ \bibinfo
  {pages} {45} (\bibinfo {year} {2023})},\ \Eprint
  {https://arxiv.org/abs/2211.01396} {arXiv:2211.01396 [astro-ph.HE]}
  \BibitemShut {NoStop}%
\bibitem [{\citenamefont {Raithel}\ and\ \citenamefont
  {Most}(2023)}]{Raithel:2022efm}%
  \BibitemOpen
  \bibfield  {author} {\bibinfo {author} {\bibfnamefont {C.~A.}\ \bibnamefont
  {Raithel}}\ and\ \bibinfo {author} {\bibfnamefont {E.~R.}\ \bibnamefont
  {Most}},\ }\bibfield  {title} {\bibinfo {title} {{Degeneracy in the Inference
  of Phase Transitions in the Neutron Star Equation of State from Gravitational
  Wave Data}},\ }\href {https://doi.org/10.1103/PhysRevLett.130.201403}
  {\bibfield  {journal} {\bibinfo  {journal} {Phys. Rev. Lett.}\ }\textbf
  {\bibinfo {volume} {130}},\ \bibinfo {pages} {201403} (\bibinfo {year}
  {2023})},\ \Eprint {https://arxiv.org/abs/2208.04294} {arXiv:2208.04294
  [astro-ph.HE]} \BibitemShut {NoStop}%
\bibitem [{\citenamefont {Essick}\ \emph {et~al.}(2023)\citenamefont {Essick},
  \citenamefont {Legred}, \citenamefont {Chatziioannou}, \citenamefont {Han},\
  and\ \citenamefont {Landry}}]{Essick:2023fso}%
  \BibitemOpen
  \bibfield  {author} {\bibinfo {author} {\bibfnamefont {R.}~\bibnamefont
  {Essick}}, \bibinfo {author} {\bibfnamefont {I.}~\bibnamefont {Legred}},
  \bibinfo {author} {\bibfnamefont {K.}~\bibnamefont {Chatziioannou}}, \bibinfo
  {author} {\bibfnamefont {S.}~\bibnamefont {Han}},\ and\ \bibinfo {author}
  {\bibfnamefont {P.}~\bibnamefont {Landry}},\ }\bibfield  {title} {\bibinfo
  {title} {{Phase transition phenomenology with nonparametric representations
  of the neutron star equation of state}},\ }\href
  {https://doi.org/10.1103/PhysRevD.108.043013} {\bibfield  {journal} {\bibinfo
   {journal} {Phys. Rev. D}\ }\textbf {\bibinfo {volume} {108}},\ \bibinfo
  {pages} {043013} (\bibinfo {year} {2023})},\ \Eprint
  {https://arxiv.org/abs/2305.07411} {arXiv:2305.07411 [astro-ph.HE]}
  \BibitemShut {NoStop}%
\bibitem [{\citenamefont {Ripley}\ \emph {et~al.}(2024)\citenamefont {Ripley},
  \citenamefont {Hegade K.~R.}, \citenamefont {Chandramouli},\ and\
  \citenamefont {Yunes}}]{Ripley:2023lsq}%
  \BibitemOpen
  \bibfield  {author} {\bibinfo {author} {\bibfnamefont {J.~L.}\ \bibnamefont
  {Ripley}}, \bibinfo {author} {\bibfnamefont {A.}~\bibnamefont {Hegade
  K.~R.}}, \bibinfo {author} {\bibfnamefont {R.~S.}\ \bibnamefont
  {Chandramouli}},\ and\ \bibinfo {author} {\bibfnamefont {N.}~\bibnamefont
  {Yunes}},\ }\bibfield  {title} {\bibinfo {title} {{A constraint on the
  dissipative tidal deformability of neutron stars}},\ }\href
  {https://doi.org/10.1038/s41550-024-02323-7} {\bibfield  {journal} {\bibinfo
  {journal} {Nature Astron.}\ }\textbf {\bibinfo {volume} {8}},\ \bibinfo
  {pages} {1277} (\bibinfo {year} {2024})},\ \Eprint
  {https://arxiv.org/abs/2312.11659} {arXiv:2312.11659 [gr-qc]} \BibitemShut
  {NoStop}%
\bibitem [{\citenamefont {Counsell}\ \emph {et~al.}(2025)\citenamefont
  {Counsell}, \citenamefont {Gittins}, \citenamefont {Andersson},\ and\
  \citenamefont {Tews}}]{Counsell:2025hcv}%
  \BibitemOpen
  \bibfield  {author} {\bibinfo {author} {\bibfnamefont {A.~R.}\ \bibnamefont
  {Counsell}}, \bibinfo {author} {\bibfnamefont {F.}~\bibnamefont {Gittins}},
  \bibinfo {author} {\bibfnamefont {N.}~\bibnamefont {Andersson}},\ and\
  \bibinfo {author} {\bibfnamefont {I.}~\bibnamefont {Tews}},\ }\bibfield
  {title} {\bibinfo {title} {{Interface Modes in Inspiralling Neutron Stars: A
  Gravitational-Wave Probe of First-Order Phase Transitions}},\ }\href
  {https://doi.org/10.1103/8hvq-6dy7} {\bibfield  {journal} {\bibinfo
  {journal} {Phys. Rev. Lett.}\ }\textbf {\bibinfo {volume} {135}},\ \bibinfo
  {pages} {081402} (\bibinfo {year} {2025})},\ \Eprint
  {https://arxiv.org/abs/2504.06181} {arXiv:2504.06181 [gr-qc]} \BibitemShut
  {NoStop}%
\bibitem [{\citenamefont {Pnigouras}\ \emph {et~al.}(2025)\citenamefont
  {Pnigouras}, \citenamefont {Andersson}, \citenamefont {Gittins},\ and\
  \citenamefont {Counsell}}]{Pnigouras:2025muo}%
  \BibitemOpen
  \bibfield  {author} {\bibinfo {author} {\bibfnamefont {P.}~\bibnamefont
  {Pnigouras}}, \bibinfo {author} {\bibfnamefont {N.}~\bibnamefont
  {Andersson}}, \bibinfo {author} {\bibfnamefont {F.}~\bibnamefont {Gittins}},\
  and\ \bibinfo {author} {\bibfnamefont {A.~R.}\ \bibnamefont {Counsell}},\
  }\bibfield  {title} {\bibinfo {title} {{Dynamical neutron-star tides: The
  signature of a mode resonance}}\ }\href
  {https://doi.org/10.1093/mnras/staf1285} {10.1093/mnras/staf1285} (\bibinfo
  {year} {2025}),\ \Eprint {https://arxiv.org/abs/2508.06416} {arXiv:2508.06416
  [gr-qc]} \BibitemShut {NoStop}%
\bibitem [{\citenamefont {Pratten}\ \emph {et~al.}(2022)\citenamefont
  {Pratten}, \citenamefont {Schmidt},\ and\ \citenamefont
  {Williams}}]{Pratten:2021pro}%
  \BibitemOpen
  \bibfield  {author} {\bibinfo {author} {\bibfnamefont {G.}~\bibnamefont
  {Pratten}}, \bibinfo {author} {\bibfnamefont {P.}~\bibnamefont {Schmidt}},\
  and\ \bibinfo {author} {\bibfnamefont {N.}~\bibnamefont {Williams}},\
  }\bibfield  {title} {\bibinfo {title} {{Impact of Dynamical Tides on the
  Reconstruction of the Neutron Star Equation of State}},\ }\href
  {https://doi.org/10.1103/PhysRevLett.129.081102} {\bibfield  {journal}
  {\bibinfo  {journal} {Phys. Rev. Lett.}\ }\textbf {\bibinfo {volume} {129}},\
  \bibinfo {pages} {081102} (\bibinfo {year} {2022})},\ \Eprint
  {https://arxiv.org/abs/2109.07566} {arXiv:2109.07566 [astro-ph.HE]}
  \BibitemShut {NoStop}%
\bibitem [{\citenamefont {Dietrich}\ \emph
  {et~al.}(2019{\natexlab{a}})\citenamefont {Dietrich}, \citenamefont
  {Samajdar}, \citenamefont {Khan}, \citenamefont {Johnson-McDaniel},
  \citenamefont {Dudi},\ and\ \citenamefont {Tichy}}]{Dietrich:2019kaq}%
  \BibitemOpen
  \bibfield  {author} {\bibinfo {author} {\bibfnamefont {T.}~\bibnamefont
  {Dietrich}}, \bibinfo {author} {\bibfnamefont {A.}~\bibnamefont {Samajdar}},
  \bibinfo {author} {\bibfnamefont {S.}~\bibnamefont {Khan}}, \bibinfo {author}
  {\bibfnamefont {N.~K.}\ \bibnamefont {Johnson-McDaniel}}, \bibinfo {author}
  {\bibfnamefont {R.}~\bibnamefont {Dudi}},\ and\ \bibinfo {author}
  {\bibfnamefont {W.}~\bibnamefont {Tichy}},\ }\bibfield  {title} {\bibinfo
  {title} {{Improving the NRTidal model for binary neutron star systems}},\
  }\href {https://doi.org/10.1103/PhysRevD.100.044003} {\bibfield  {journal}
  {\bibinfo  {journal} {Phys. Rev. D}\ }\textbf {\bibinfo {volume} {100}},\
  \bibinfo {pages} {044003} (\bibinfo {year} {2019}{\natexlab{a}})},\ \Eprint
  {https://arxiv.org/abs/1905.06011} {arXiv:1905.06011 [gr-qc]} \BibitemShut
  {NoStop}%
\bibitem [{\citenamefont {Gamba}\ \emph {et~al.}(2021)\citenamefont {Gamba},
  \citenamefont {Breschi}, \citenamefont {Bernuzzi}, \citenamefont {Agathos},\
  and\ \citenamefont {Nagar}}]{Gamba:2020wgg}%
  \BibitemOpen
  \bibfield  {author} {\bibinfo {author} {\bibfnamefont {R.}~\bibnamefont
  {Gamba}}, \bibinfo {author} {\bibfnamefont {M.}~\bibnamefont {Breschi}},
  \bibinfo {author} {\bibfnamefont {S.}~\bibnamefont {Bernuzzi}}, \bibinfo
  {author} {\bibfnamefont {M.}~\bibnamefont {Agathos}},\ and\ \bibinfo {author}
  {\bibfnamefont {A.}~\bibnamefont {Nagar}},\ }\bibfield  {title} {\bibinfo
  {title} {{Waveform systematics in the gravitational-wave inference of tidal
  parameters and equation of state from binary neutron star signals}},\ }\href
  {https://doi.org/10.1103/PhysRevD.103.124015} {\bibfield  {journal} {\bibinfo
   {journal} {Phys. Rev. D}\ }\textbf {\bibinfo {volume} {103}},\ \bibinfo
  {pages} {124015} (\bibinfo {year} {2021})},\ \Eprint
  {https://arxiv.org/abs/2009.08467} {arXiv:2009.08467 [gr-qc]} \BibitemShut
  {NoStop}%
\bibitem [{\citenamefont {Dietrich}\ \emph {et~al.}(2017)\citenamefont
  {Dietrich}, \citenamefont {Bernuzzi},\ and\ \citenamefont
  {Tichy}}]{Dietrich:2017aum}%
  \BibitemOpen
  \bibfield  {author} {\bibinfo {author} {\bibfnamefont {T.}~\bibnamefont
  {Dietrich}}, \bibinfo {author} {\bibfnamefont {S.}~\bibnamefont {Bernuzzi}},\
  and\ \bibinfo {author} {\bibfnamefont {W.}~\bibnamefont {Tichy}},\ }\bibfield
   {title} {\bibinfo {title} {{Closed-form tidal approximants for binary
  neutron star gravitational waveforms constructed from high-resolution
  numerical relativity simulations}},\ }\href
  {https://doi.org/10.1103/PhysRevD.96.121501} {\bibfield  {journal} {\bibinfo
  {journal} {Phys. Rev. D}\ }\textbf {\bibinfo {volume} {96}},\ \bibinfo
  {pages} {121501} (\bibinfo {year} {2017})},\ \Eprint
  {https://arxiv.org/abs/1706.02969} {arXiv:1706.02969 [gr-qc]} \BibitemShut
  {NoStop}%
\bibitem [{\citenamefont {Dietrich}\ \emph
  {et~al.}(2019{\natexlab{b}})\citenamefont {Dietrich} \emph
  {et~al.}}]{Dietrich:2018uni}%
  \BibitemOpen
  \bibfield  {author} {\bibinfo {author} {\bibfnamefont {T.}~\bibnamefont
  {Dietrich}} \emph {et~al.},\ }\bibfield  {title} {\bibinfo {title} {{Matter
  imprints in waveform models for neutron star binaries: Tidal and self-spin
  effects}},\ }\href {https://doi.org/10.1103/PhysRevD.99.024029} {\bibfield
  {journal} {\bibinfo  {journal} {Phys. Rev. D}\ }\textbf {\bibinfo {volume}
  {99}},\ \bibinfo {pages} {024029} (\bibinfo {year} {2019}{\natexlab{b}})},\
  \Eprint {https://arxiv.org/abs/1804.02235} {arXiv:1804.02235 [gr-qc]}
  \BibitemShut {NoStop}%
\bibitem [{\citenamefont {Dietrich}\ \emph {et~al.}(2018)\citenamefont
  {Dietrich}, \citenamefont {Radice}, \citenamefont {Bernuzzi}, \citenamefont
  {Zappa}, \citenamefont {Perego}, \citenamefont {Br\"ugmann}, \citenamefont
  {Chaurasia}, \citenamefont {Dudi}, \citenamefont {Tichy},\ and\ \citenamefont
  {Ujevic}}]{Dietrich:2018phi}%
  \BibitemOpen
  \bibfield  {author} {\bibinfo {author} {\bibfnamefont {T.}~\bibnamefont
  {Dietrich}}, \bibinfo {author} {\bibfnamefont {D.}~\bibnamefont {Radice}},
  \bibinfo {author} {\bibfnamefont {S.}~\bibnamefont {Bernuzzi}}, \bibinfo
  {author} {\bibfnamefont {F.}~\bibnamefont {Zappa}}, \bibinfo {author}
  {\bibfnamefont {A.}~\bibnamefont {Perego}}, \bibinfo {author} {\bibfnamefont
  {B.}~\bibnamefont {Br\"ugmann}}, \bibinfo {author} {\bibfnamefont {S.~V.}\
  \bibnamefont {Chaurasia}}, \bibinfo {author} {\bibfnamefont {R.}~\bibnamefont
  {Dudi}}, \bibinfo {author} {\bibfnamefont {W.}~\bibnamefont {Tichy}},\ and\
  \bibinfo {author} {\bibfnamefont {M.}~\bibnamefont {Ujevic}},\ }\bibfield
  {title} {\bibinfo {title} {{CoRe database of binary neutron star merger
  waveforms}},\ }\href {https://doi.org/10.1088/1361-6382/aaebc0} {\bibfield
  {journal} {\bibinfo  {journal} {Class. Quant. Grav.}\ }\textbf {\bibinfo
  {volume} {35}},\ \bibinfo {pages} {24LT01} (\bibinfo {year} {2018})},\
  \Eprint {https://arxiv.org/abs/1806.01625} {arXiv:1806.01625 [gr-qc]}
  \BibitemShut {NoStop}%
\bibitem [{\citenamefont {Gonzalez}\ \emph {et~al.}(2023)\citenamefont
  {Gonzalez} \emph {et~al.}}]{Gonzalez:2022mgo}%
  \BibitemOpen
  \bibfield  {author} {\bibinfo {author} {\bibfnamefont {A.}~\bibnamefont
  {Gonzalez}} \emph {et~al.},\ }\bibfield  {title} {\bibinfo {title} {{Second
  release of the CoRe database of binary neutron star merger waveforms}},\
  }\href {https://doi.org/10.1088/1361-6382/acc231} {\bibfield  {journal}
  {\bibinfo  {journal} {Class. Quant. Grav.}\ }\textbf {\bibinfo {volume}
  {40}},\ \bibinfo {pages} {085011} (\bibinfo {year} {2023})},\ \Eprint
  {https://arxiv.org/abs/2210.16366} {arXiv:2210.16366 [gr-qc]} \BibitemShut
  {NoStop}%
\bibitem [{\citenamefont {Kiuchi}\ \emph {et~al.}(2017)\citenamefont {Kiuchi},
  \citenamefont {Kawaguchi}, \citenamefont {Kyutoku}, \citenamefont
  {Sekiguchi}, \citenamefont {Shibata},\ and\ \citenamefont
  {Taniguchi}}]{Kiuchi:2017pte}%
  \BibitemOpen
  \bibfield  {author} {\bibinfo {author} {\bibfnamefont {K.}~\bibnamefont
  {Kiuchi}}, \bibinfo {author} {\bibfnamefont {K.}~\bibnamefont {Kawaguchi}},
  \bibinfo {author} {\bibfnamefont {K.}~\bibnamefont {Kyutoku}}, \bibinfo
  {author} {\bibfnamefont {Y.}~\bibnamefont {Sekiguchi}}, \bibinfo {author}
  {\bibfnamefont {M.}~\bibnamefont {Shibata}},\ and\ \bibinfo {author}
  {\bibfnamefont {K.}~\bibnamefont {Taniguchi}},\ }\bibfield  {title} {\bibinfo
  {title} {{Sub-radian-accuracy gravitational waveforms of coalescing binary
  neutron stars in numerical relativity}},\ }\href
  {https://doi.org/10.1103/PhysRevD.96.084060} {\bibfield  {journal} {\bibinfo
  {journal} {Phys. Rev. D}\ }\textbf {\bibinfo {volume} {96}},\ \bibinfo
  {pages} {084060} (\bibinfo {year} {2017})},\ \Eprint
  {https://arxiv.org/abs/1708.08926} {arXiv:1708.08926 [astro-ph.HE]}
  \BibitemShut {NoStop}%
\bibitem [{\citenamefont {Kiuchi}\ \emph {et~al.}(2020)\citenamefont {Kiuchi},
  \citenamefont {Kawaguchi}, \citenamefont {Kyutoku}, \citenamefont
  {Sekiguchi},\ and\ \citenamefont {Shibata}}]{Kiuchi:2019kzt}%
  \BibitemOpen
  \bibfield  {author} {\bibinfo {author} {\bibfnamefont {K.}~\bibnamefont
  {Kiuchi}}, \bibinfo {author} {\bibfnamefont {K.}~\bibnamefont {Kawaguchi}},
  \bibinfo {author} {\bibfnamefont {K.}~\bibnamefont {Kyutoku}}, \bibinfo
  {author} {\bibfnamefont {Y.}~\bibnamefont {Sekiguchi}},\ and\ \bibinfo
  {author} {\bibfnamefont {M.}~\bibnamefont {Shibata}},\ }\bibfield  {title}
  {\bibinfo {title} {{Sub-radian-accuracy gravitational waves from coalescing
  binary neutron stars in numerical relativity. II. Systematic study on the
  equation of state, binary mass, and mass ratio}},\ }\href
  {https://doi.org/10.1103/PhysRevD.101.084006} {\bibfield  {journal} {\bibinfo
   {journal} {Phys. Rev. D}\ }\textbf {\bibinfo {volume} {101}},\ \bibinfo
  {pages} {084006} (\bibinfo {year} {2020})},\ \Eprint
  {https://arxiv.org/abs/1907.03790} {arXiv:1907.03790 [astro-ph.HE]}
  \BibitemShut {NoStop}%
\bibitem [{\citenamefont {Chatziioannou}\ \emph
  {et~al.}(2017{\natexlab{a}})\citenamefont {Chatziioannou}, \citenamefont
  {Clark}, \citenamefont {Bauswein}, \citenamefont {Millhouse}, \citenamefont
  {Littenberg},\ and\ \citenamefont {Cornish}}]{Chatziioannou:2017ixj}%
  \BibitemOpen
  \bibfield  {author} {\bibinfo {author} {\bibfnamefont {K.}~\bibnamefont
  {Chatziioannou}}, \bibinfo {author} {\bibfnamefont {J.~A.}\ \bibnamefont
  {Clark}}, \bibinfo {author} {\bibfnamefont {A.}~\bibnamefont {Bauswein}},
  \bibinfo {author} {\bibfnamefont {M.}~\bibnamefont {Millhouse}}, \bibinfo
  {author} {\bibfnamefont {T.~B.}\ \bibnamefont {Littenberg}},\ and\ \bibinfo
  {author} {\bibfnamefont {N.}~\bibnamefont {Cornish}},\ }\bibfield  {title}
  {\bibinfo {title} {{Inferring the post-merger gravitational wave emission
  from binary neutron star coalescences}},\ }\href
  {https://doi.org/10.1103/PhysRevD.96.124035} {\bibfield  {journal} {\bibinfo
  {journal} {Phys. Rev. D}\ }\textbf {\bibinfo {volume} {96}},\ \bibinfo
  {pages} {124035} (\bibinfo {year} {2017}{\natexlab{a}})},\ \Eprint
  {https://arxiv.org/abs/1711.00040} {arXiv:1711.00040 [gr-qc]} \BibitemShut
  {NoStop}%
\bibitem [{\citenamefont {Wijngaarden}\ \emph {et~al.}(2022)\citenamefont
  {Wijngaarden}, \citenamefont {Chatziioannou}, \citenamefont {Bauswein},
  \citenamefont {Clark},\ and\ \citenamefont {Cornish}}]{Wijngaarden:2022sah}%
  \BibitemOpen
  \bibfield  {author} {\bibinfo {author} {\bibfnamefont {M.}~\bibnamefont
  {Wijngaarden}}, \bibinfo {author} {\bibfnamefont {K.}~\bibnamefont
  {Chatziioannou}}, \bibinfo {author} {\bibfnamefont {A.}~\bibnamefont
  {Bauswein}}, \bibinfo {author} {\bibfnamefont {J.~A.}\ \bibnamefont
  {Clark}},\ and\ \bibinfo {author} {\bibfnamefont {N.~J.}\ \bibnamefont
  {Cornish}},\ }\bibfield  {title} {\bibinfo {title} {{Probing neutron stars
  with the full premerger and postmerger gravitational wave signal from binary
  coalescences}},\ }\href {https://doi.org/10.1103/PhysRevD.105.104019}
  {\bibfield  {journal} {\bibinfo  {journal} {Phys. Rev. D}\ }\textbf {\bibinfo
  {volume} {105}},\ \bibinfo {pages} {104019} (\bibinfo {year} {2022})},\
  \Eprint {https://arxiv.org/abs/2202.09382} {arXiv:2202.09382 [gr-qc]}
  \BibitemShut {NoStop}%
\bibitem [{\citenamefont {Criswell}\ \emph {et~al.}(2023)\citenamefont
  {Criswell}, \citenamefont {Miller}, \citenamefont {Woldemariam},
  \citenamefont {Soultanis}, \citenamefont {Bauswein}, \citenamefont
  {Chatziioannou}, \citenamefont {Coughlin}, \citenamefont {Jones},\ and\
  \citenamefont {Mandic}}]{Criswell:2022ewn}%
  \BibitemOpen
  \bibfield  {author} {\bibinfo {author} {\bibfnamefont {A.~W.}\ \bibnamefont
  {Criswell}}, \bibinfo {author} {\bibfnamefont {J.}~\bibnamefont {Miller}},
  \bibinfo {author} {\bibfnamefont {N.}~\bibnamefont {Woldemariam}}, \bibinfo
  {author} {\bibfnamefont {T.}~\bibnamefont {Soultanis}}, \bibinfo {author}
  {\bibfnamefont {A.}~\bibnamefont {Bauswein}}, \bibinfo {author}
  {\bibfnamefont {K.}~\bibnamefont {Chatziioannou}}, \bibinfo {author}
  {\bibfnamefont {M.~W.}\ \bibnamefont {Coughlin}}, \bibinfo {author}
  {\bibfnamefont {G.}~\bibnamefont {Jones}},\ and\ \bibinfo {author}
  {\bibfnamefont {V.}~\bibnamefont {Mandic}},\ }\bibfield  {title} {\bibinfo
  {title} {{Hierarchical Bayesian method for constraining the neutron star
  equation of state with an ensemble of binary neutron star postmerger
  remnants}},\ }\href {https://doi.org/10.1103/PhysRevD.107.043021} {\bibfield
  {journal} {\bibinfo  {journal} {Phys. Rev. D}\ }\textbf {\bibinfo {volume}
  {107}},\ \bibinfo {pages} {043021} (\bibinfo {year} {2023})},\ \Eprint
  {https://arxiv.org/abs/2211.05250} {arXiv:2211.05250 [astro-ph.HE]}
  \BibitemShut {NoStop}%
\bibitem [{\citenamefont {Boyle}\ \emph {et~al.}(2019)\citenamefont {Boyle}
  \emph {et~al.}}]{Boyle:2019kee}%
  \BibitemOpen
  \bibfield  {author} {\bibinfo {author} {\bibfnamefont {M.}~\bibnamefont
  {Boyle}} \emph {et~al.},\ }\bibfield  {title} {\bibinfo {title} {{The SXS
  Collaboration catalog of binary black hole simulations}},\ }\href
  {https://doi.org/10.1088/1361-6382/ab34e2} {\bibfield  {journal} {\bibinfo
  {journal} {Class. Quant. Grav.}\ }\textbf {\bibinfo {volume} {36}},\ \bibinfo
  {pages} {195006} (\bibinfo {year} {2019})},\ \Eprint
  {https://arxiv.org/abs/1904.04831} {arXiv:1904.04831 [gr-qc]} \BibitemShut
  {NoStop}%
\bibitem [{\citenamefont {Scheel}\ \emph {et~al.}(2025)\citenamefont {Scheel}
  \emph {et~al.}}]{Scheel:2025jct}%
  \BibitemOpen
  \bibfield  {author} {\bibinfo {author} {\bibfnamefont {M.~A.}\ \bibnamefont
  {Scheel}} \emph {et~al.},\ }\bibfield  {title} {\bibinfo {title} {{The SXS
  Collaboration's third catalog of binary black hole simulations}},\
  }\href@noop {} {\  (\bibinfo {year} {2025})},\ \Eprint
  {https://arxiv.org/abs/2505.13378} {arXiv:2505.13378 [gr-qc]} \BibitemShut
  {NoStop}%
\bibitem [{\citenamefont {Radice}\ \emph
  {et~al.}(2014{\natexlab{a}})\citenamefont {Radice}, \citenamefont
  {Rezzolla},\ and\ \citenamefont {Galeazzi}}]{Radice:2013hxh}%
  \BibitemOpen
  \bibfield  {author} {\bibinfo {author} {\bibfnamefont {D.}~\bibnamefont
  {Radice}}, \bibinfo {author} {\bibfnamefont {L.}~\bibnamefont {Rezzolla}},\
  and\ \bibinfo {author} {\bibfnamefont {F.}~\bibnamefont {Galeazzi}},\
  }\bibfield  {title} {\bibinfo {title} {{Beyond second-order convergence in
  simulations of binary neutron stars in full general-relativity}},\ }\href
  {https://doi.org/10.1093/mnrasl/slt137} {\bibfield  {journal} {\bibinfo
  {journal} {Mon. Not. Roy. Astron. Soc.}\ }\textbf {\bibinfo {volume} {437}},\
  \bibinfo {pages} {L46} (\bibinfo {year} {2014}{\natexlab{a}})},\ \Eprint
  {https://arxiv.org/abs/1306.6052} {arXiv:1306.6052 [gr-qc]} \BibitemShut
  {NoStop}%
\bibitem [{\citenamefont {Bernuzzi}\ and\ \citenamefont
  {Dietrich}(2016)}]{Bernuzzi:2016pie}%
  \BibitemOpen
  \bibfield  {author} {\bibinfo {author} {\bibfnamefont {S.}~\bibnamefont
  {Bernuzzi}}\ and\ \bibinfo {author} {\bibfnamefont {T.}~\bibnamefont
  {Dietrich}},\ }\bibfield  {title} {\bibinfo {title} {{Gravitational waveforms
  from binary neutron star mergers with high-order
  weighted-essentially-nonoscillatory schemes in numerical relativity}},\
  }\href {https://doi.org/10.1103/PhysRevD.94.064062} {\bibfield  {journal}
  {\bibinfo  {journal} {Phys. Rev. D}\ }\textbf {\bibinfo {volume} {94}},\
  \bibinfo {pages} {064062} (\bibinfo {year} {2016})},\ \Eprint
  {https://arxiv.org/abs/1604.07999} {arXiv:1604.07999 [gr-qc]} \BibitemShut
  {NoStop}%
\bibitem [{\citenamefont {Most}\ \emph {et~al.}(2019)\citenamefont {Most},
  \citenamefont {Papenfort},\ and\ \citenamefont {Rezzolla}}]{Most:2019kfe}%
  \BibitemOpen
  \bibfield  {author} {\bibinfo {author} {\bibfnamefont {E.~R.}\ \bibnamefont
  {Most}}, \bibinfo {author} {\bibfnamefont {L.~J.}\ \bibnamefont
  {Papenfort}},\ and\ \bibinfo {author} {\bibfnamefont {L.}~\bibnamefont
  {Rezzolla}},\ }\bibfield  {title} {\bibinfo {title} {{Beyond second-order
  convergence in simulations of magnetized binary neutron stars with realistic
  microphysics}},\ }\href {https://doi.org/10.1093/mnras/stz2809} {\bibfield
  {journal} {\bibinfo  {journal} {Mon. Not. Roy. Astron. Soc.}\ }\textbf
  {\bibinfo {volume} {490}},\ \bibinfo {pages} {3588} (\bibinfo {year}
  {2019})},\ \Eprint {https://arxiv.org/abs/1907.10328} {arXiv:1907.10328
  [astro-ph.HE]} \BibitemShut {NoStop}%
\bibitem [{\citenamefont {Liebling}\ \emph {et~al.}(2020)\citenamefont
  {Liebling}, \citenamefont {Palenzuela},\ and\ \citenamefont
  {Lehner}}]{Liebling:2020jlq}%
  \BibitemOpen
  \bibfield  {author} {\bibinfo {author} {\bibfnamefont {S.~L.}\ \bibnamefont
  {Liebling}}, \bibinfo {author} {\bibfnamefont {C.}~\bibnamefont
  {Palenzuela}},\ and\ \bibinfo {author} {\bibfnamefont {L.}~\bibnamefont
  {Lehner}},\ }\bibfield  {title} {\bibinfo {title} {{Toward fidelity and
  scalability in non-vacuum mergers}},\ }\href
  {https://doi.org/10.1088/1361-6382/ab8fcd} {\bibfield  {journal} {\bibinfo
  {journal} {Class. Quant. Grav.}\ }\textbf {\bibinfo {volume} {37}},\ \bibinfo
  {pages} {135006} (\bibinfo {year} {2020})},\ \Eprint
  {https://arxiv.org/abs/2002.07554} {arXiv:2002.07554 [gr-qc]} \BibitemShut
  {NoStop}%
\bibitem [{\citenamefont {Kiuchi}(2025)}]{Kiuchi:2025ksk}%
  \BibitemOpen
  \bibfield  {author} {\bibinfo {author} {\bibfnamefont {K.}~\bibnamefont
  {Kiuchi}},\ }\bibfield  {title} {\bibinfo {title} {{Towards high-precision
  inspiral gravitational waveforms from binary neutron star mergers in
  numerical relativity}},\ }\href@noop {} {\  (\bibinfo {year} {2025})},\
  \Eprint {https://arxiv.org/abs/2508.10981} {arXiv:2508.10981 [astro-ph.HE]}
  \BibitemShut {NoStop}%
\bibitem [{\citenamefont {Doulis}\ \emph {et~al.}(2022)\citenamefont {Doulis},
  \citenamefont {Atteneder}, \citenamefont {Bernuzzi},\ and\ \citenamefont
  {Br\"ugmann}}]{Doulis:2022vkx}%
  \BibitemOpen
  \bibfield  {author} {\bibinfo {author} {\bibfnamefont {G.}~\bibnamefont
  {Doulis}}, \bibinfo {author} {\bibfnamefont {F.}~\bibnamefont {Atteneder}},
  \bibinfo {author} {\bibfnamefont {S.}~\bibnamefont {Bernuzzi}},\ and\
  \bibinfo {author} {\bibfnamefont {B.}~\bibnamefont {Br\"ugmann}},\ }\bibfield
   {title} {\bibinfo {title} {{Entropy-limited higher-order central scheme for
  neutron star merger simulations}},\ }\href
  {https://doi.org/10.1103/PhysRevD.106.024001} {\bibfield  {journal} {\bibinfo
   {journal} {Phys. Rev. D}\ }\textbf {\bibinfo {volume} {106}},\ \bibinfo
  {pages} {024001} (\bibinfo {year} {2022})},\ \Eprint
  {https://arxiv.org/abs/2202.08839} {arXiv:2202.08839 [gr-qc]} \BibitemShut
  {NoStop}%
\bibitem [{\citenamefont {Doulis}\ \emph {et~al.}(2024)\citenamefont {Doulis},
  \citenamefont {Bernuzzi},\ and\ \citenamefont {Tichy}}]{Doulis:2024aew}%
  \BibitemOpen
  \bibfield  {author} {\bibinfo {author} {\bibfnamefont {G.}~\bibnamefont
  {Doulis}}, \bibinfo {author} {\bibfnamefont {S.}~\bibnamefont {Bernuzzi}},\
  and\ \bibinfo {author} {\bibfnamefont {W.}~\bibnamefont {Tichy}},\ }\bibfield
   {title} {\bibinfo {title} {{Entropy based flux limiting scheme for
  conservation laws}},\ }\href@noop {} {\  (\bibinfo {year} {2024})},\ \Eprint
  {https://arxiv.org/abs/2401.04770} {arXiv:2401.04770 [gr-qc]} \BibitemShut
  {NoStop}%
\bibitem [{\citenamefont {Duez}\ \emph {et~al.}(2008)\citenamefont {Duez},
  \citenamefont {Foucart}, \citenamefont {Kidder}, \citenamefont {Pfeiffer},
  \citenamefont {Scheel},\ and\ \citenamefont {Teukolsky}}]{Duez:2008rb}%
  \BibitemOpen
  \bibfield  {author} {\bibinfo {author} {\bibfnamefont {M.~D.}\ \bibnamefont
  {Duez}}, \bibinfo {author} {\bibfnamefont {F.}~\bibnamefont {Foucart}},
  \bibinfo {author} {\bibfnamefont {L.~E.}\ \bibnamefont {Kidder}}, \bibinfo
  {author} {\bibfnamefont {H.~P.}\ \bibnamefont {Pfeiffer}}, \bibinfo {author}
  {\bibfnamefont {M.~A.}\ \bibnamefont {Scheel}},\ and\ \bibinfo {author}
  {\bibfnamefont {S.~A.}\ \bibnamefont {Teukolsky}},\ }\bibfield  {title}
  {\bibinfo {title} {{Evolving black hole-neutron star binaries in general
  relativity using pseudospectral and finite difference methods}},\ }\href
  {https://doi.org/10.1103/PhysRevD.78.104015} {\bibfield  {journal} {\bibinfo
  {journal} {Phys. Rev. D}\ }\textbf {\bibinfo {volume} {78}},\ \bibinfo
  {pages} {104015} (\bibinfo {year} {2008})},\ \Eprint
  {https://arxiv.org/abs/0809.0002} {arXiv:0809.0002 [gr-qc]} \BibitemShut
  {NoStop}%
\bibitem [{\citenamefont {Deppe}\ \emph {et~al.}(2024)\citenamefont {Deppe}
  \emph {et~al.}}]{Deppe:2024ckt}%
  \BibitemOpen
  \bibfield  {author} {\bibinfo {author} {\bibfnamefont {N.}~\bibnamefont
  {Deppe}} \emph {et~al.},\ }\bibfield  {title} {\bibinfo {title} {{Binary
  neutron star mergers using a discontinuous Galerkin-finite difference hybrid
  method}},\ }\href {https://doi.org/10.1088/1361-6382/ad88cf} {\bibfield
  {journal} {\bibinfo  {journal} {Class. Quant. Grav.}\ }\textbf {\bibinfo
  {volume} {41}},\ \bibinfo {pages} {245002} (\bibinfo {year} {2024})},\
  \Eprint {https://arxiv.org/abs/2406.19038} {arXiv:2406.19038 [gr-qc]}
  \BibitemShut {NoStop}%
\bibitem [{\citenamefont {Adhikari}\ \emph {et~al.}(2025)\citenamefont
  {Adhikari}, \citenamefont {Tichy}, \citenamefont {Ji},\ and\ \citenamefont
  {Poudel}}]{Adhikari:2025nio}%
  \BibitemOpen
  \bibfield  {author} {\bibinfo {author} {\bibfnamefont {A.}~\bibnamefont
  {Adhikari}}, \bibinfo {author} {\bibfnamefont {W.}~\bibnamefont {Tichy}},
  \bibinfo {author} {\bibfnamefont {L.}~\bibnamefont {Ji}},\ and\ \bibinfo
  {author} {\bibfnamefont {A.}~\bibnamefont {Poudel}},\ }\bibfield  {title}
  {\bibinfo {title} {{Neutron star evolution by combining discontinuous
  Galerkin and finite volume methods}},\ }\href@noop {} {\  (\bibinfo {year}
  {2025})},\ \Eprint {https://arxiv.org/abs/2502.07204} {arXiv:2502.07204
  [gr-qc]} \BibitemShut {NoStop}%
\bibitem [{\citenamefont {Shibata}\ and\ \citenamefont
  {Nakamura}(1995)}]{Shibata:1995we}%
  \BibitemOpen
  \bibfield  {author} {\bibinfo {author} {\bibfnamefont {M.}~\bibnamefont
  {Shibata}}\ and\ \bibinfo {author} {\bibfnamefont {T.}~\bibnamefont
  {Nakamura}},\ }\bibfield  {title} {\bibinfo {title} {{Evolution of
  three-dimensional gravitational waves: Harmonic slicing case}},\ }\href
  {https://doi.org/10.1103/PhysRevD.52.5428} {\bibfield  {journal} {\bibinfo
  {journal} {Phys. Rev. D}\ }\textbf {\bibinfo {volume} {52}},\ \bibinfo
  {pages} {5428} (\bibinfo {year} {1995})}\BibitemShut {NoStop}%
\bibitem [{\citenamefont {Baumgarte}\ and\ \citenamefont
  {Shapiro}(1998)}]{Baumgarte:1998te}%
  \BibitemOpen
  \bibfield  {author} {\bibinfo {author} {\bibfnamefont {T.~W.}\ \bibnamefont
  {Baumgarte}}\ and\ \bibinfo {author} {\bibfnamefont {S.~L.}\ \bibnamefont
  {Shapiro}},\ }\bibfield  {title} {\bibinfo {title} {{On the numerical
  integration of Einstein's field equations}},\ }\href
  {https://doi.org/10.1103/PhysRevD.59.024007} {\bibfield  {journal} {\bibinfo
  {journal} {Phys. Rev. D}\ }\textbf {\bibinfo {volume} {59}},\ \bibinfo
  {pages} {024007} (\bibinfo {year} {1998})},\ \Eprint
  {https://arxiv.org/abs/gr-qc/9810065} {arXiv:gr-qc/9810065} \BibitemShut
  {NoStop}%
\bibitem [{\citenamefont {Bona}\ \emph {et~al.}(2003)\citenamefont {Bona},
  \citenamefont {Ledvinka}, \citenamefont {Palenzuela},\ and\ \citenamefont
  {Zacek}}]{Bona:2003fj}%
  \BibitemOpen
  \bibfield  {author} {\bibinfo {author} {\bibfnamefont {C.}~\bibnamefont
  {Bona}}, \bibinfo {author} {\bibfnamefont {T.}~\bibnamefont {Ledvinka}},
  \bibinfo {author} {\bibfnamefont {C.}~\bibnamefont {Palenzuela}},\ and\
  \bibinfo {author} {\bibfnamefont {M.}~\bibnamefont {Zacek}},\ }\bibfield
  {title} {\bibinfo {title} {{General covariant evolution formalism for
  numerical relativity}},\ }\href {https://doi.org/10.1103/PhysRevD.67.104005}
  {\bibfield  {journal} {\bibinfo  {journal} {Phys. Rev. D}\ }\textbf {\bibinfo
  {volume} {67}},\ \bibinfo {pages} {104005} (\bibinfo {year} {2003})},\
  \Eprint {https://arxiv.org/abs/gr-qc/0302083} {arXiv:gr-qc/0302083}
  \BibitemShut {NoStop}%
\bibitem [{\citenamefont {Bernuzzi}\ and\ \citenamefont
  {Hilditch}(2010)}]{Bernuzzi:2009ex}%
  \BibitemOpen
  \bibfield  {author} {\bibinfo {author} {\bibfnamefont {S.}~\bibnamefont
  {Bernuzzi}}\ and\ \bibinfo {author} {\bibfnamefont {D.}~\bibnamefont
  {Hilditch}},\ }\bibfield  {title} {\bibinfo {title} {{Constraint violation in
  free evolution schemes: Comparing BSSNOK with a conformal decomposition of
  Z4}},\ }\href {https://doi.org/10.1103/PhysRevD.81.084003} {\bibfield
  {journal} {\bibinfo  {journal} {Phys. Rev. D}\ }\textbf {\bibinfo {volume}
  {81}},\ \bibinfo {pages} {084003} (\bibinfo {year} {2010})},\ \Eprint
  {https://arxiv.org/abs/0912.2920} {arXiv:0912.2920 [gr-qc]} \BibitemShut
  {NoStop}%
\bibitem [{\citenamefont {Hilditch}\ \emph {et~al.}(2013)\citenamefont
  {Hilditch}, \citenamefont {Bernuzzi}, \citenamefont {Thierfelder},
  \citenamefont {Cao}, \citenamefont {Tichy},\ and\ \citenamefont
  {Bruegmann}}]{Hilditch:2012fp}%
  \BibitemOpen
  \bibfield  {author} {\bibinfo {author} {\bibfnamefont {D.}~\bibnamefont
  {Hilditch}}, \bibinfo {author} {\bibfnamefont {S.}~\bibnamefont {Bernuzzi}},
  \bibinfo {author} {\bibfnamefont {M.}~\bibnamefont {Thierfelder}}, \bibinfo
  {author} {\bibfnamefont {Z.}~\bibnamefont {Cao}}, \bibinfo {author}
  {\bibfnamefont {W.}~\bibnamefont {Tichy}},\ and\ \bibinfo {author}
  {\bibfnamefont {B.}~\bibnamefont {Bruegmann}},\ }\bibfield  {title} {\bibinfo
  {title} {{Compact binary evolutions with the Z4c formulation}},\ }\href
  {https://doi.org/10.1103/PhysRevD.88.084057} {\bibfield  {journal} {\bibinfo
  {journal} {Phys. Rev. D}\ }\textbf {\bibinfo {volume} {88}},\ \bibinfo
  {pages} {084057} (\bibinfo {year} {2013})},\ \Eprint
  {https://arxiv.org/abs/1212.2901} {arXiv:1212.2901 [gr-qc]} \BibitemShut
  {NoStop}%
\bibitem [{\citenamefont {Alic}\ \emph {et~al.}(2012)\citenamefont {Alic},
  \citenamefont {Bona-Casas}, \citenamefont {Bona}, \citenamefont {Rezzolla},\
  and\ \citenamefont {Palenzuela}}]{Alic:2011gg}%
  \BibitemOpen
  \bibfield  {author} {\bibinfo {author} {\bibfnamefont {D.}~\bibnamefont
  {Alic}}, \bibinfo {author} {\bibfnamefont {C.}~\bibnamefont {Bona-Casas}},
  \bibinfo {author} {\bibfnamefont {C.}~\bibnamefont {Bona}}, \bibinfo {author}
  {\bibfnamefont {L.}~\bibnamefont {Rezzolla}},\ and\ \bibinfo {author}
  {\bibfnamefont {C.}~\bibnamefont {Palenzuela}},\ }\bibfield  {title}
  {\bibinfo {title} {{Conformal and covariant formulation of the Z4 system with
  constraint-violation damping}},\ }\href
  {https://doi.org/10.1103/PhysRevD.85.064040} {\bibfield  {journal} {\bibinfo
  {journal} {Phys. Rev. D}\ }\textbf {\bibinfo {volume} {85}},\ \bibinfo
  {pages} {064040} (\bibinfo {year} {2012})},\ \Eprint
  {https://arxiv.org/abs/1106.2254} {arXiv:1106.2254 [gr-qc]} \BibitemShut
  {NoStop}%
\bibitem [{\citenamefont {Pretorius}(2005)}]{Pretorius:2004jg}%
  \BibitemOpen
  \bibfield  {author} {\bibinfo {author} {\bibfnamefont {F.}~\bibnamefont
  {Pretorius}},\ }\bibfield  {title} {\bibinfo {title} {{Numerical relativity
  using a generalized harmonic decomposition}},\ }\href
  {https://doi.org/10.1088/0264-9381/22/2/014} {\bibfield  {journal} {\bibinfo
  {journal} {Class. Quant. Grav.}\ }\textbf {\bibinfo {volume} {22}},\ \bibinfo
  {pages} {425} (\bibinfo {year} {2005})},\ \Eprint
  {https://arxiv.org/abs/gr-qc/0407110} {arXiv:gr-qc/0407110} \BibitemShut
  {NoStop}%
\bibitem [{\citenamefont {Lindblom}\ \emph {et~al.}(2006)\citenamefont
  {Lindblom}, \citenamefont {Scheel}, \citenamefont {Kidder}, \citenamefont
  {Owen},\ and\ \citenamefont {Rinne}}]{Lindblom:2005qh}%
  \BibitemOpen
  \bibfield  {author} {\bibinfo {author} {\bibfnamefont {L.}~\bibnamefont
  {Lindblom}}, \bibinfo {author} {\bibfnamefont {M.~A.}\ \bibnamefont
  {Scheel}}, \bibinfo {author} {\bibfnamefont {L.~E.}\ \bibnamefont {Kidder}},
  \bibinfo {author} {\bibfnamefont {R.}~\bibnamefont {Owen}},\ and\ \bibinfo
  {author} {\bibfnamefont {O.}~\bibnamefont {Rinne}},\ }\bibfield  {title}
  {\bibinfo {title} {{A New generalized harmonic evolution system}},\ }\href
  {https://doi.org/10.1088/0264-9381/23/16/S09} {\bibfield  {journal} {\bibinfo
   {journal} {Class. Quant. Grav.}\ }\textbf {\bibinfo {volume} {23}},\
  \bibinfo {pages} {S447} (\bibinfo {year} {2006})},\ \Eprint
  {https://arxiv.org/abs/gr-qc/0512093} {arXiv:gr-qc/0512093} \BibitemShut
  {NoStop}%
\bibitem [{\citenamefont {Pfeiffer}\ and\ \citenamefont
  {York}(2003)}]{Pfeiffer:2002iy}%
  \BibitemOpen
  \bibfield  {author} {\bibinfo {author} {\bibfnamefont {H.~P.}\ \bibnamefont
  {Pfeiffer}}\ and\ \bibinfo {author} {\bibfnamefont {J.~W.}\ \bibnamefont
  {York}, \bibfnamefont {Jr.}},\ }\bibfield  {title} {\bibinfo {title}
  {{Extrinsic curvature and the Einstein constraints}},\ }\href
  {https://doi.org/10.1103/PhysRevD.67.044022} {\bibfield  {journal} {\bibinfo
  {journal} {Phys. Rev. D}\ }\textbf {\bibinfo {volume} {67}},\ \bibinfo
  {pages} {044022} (\bibinfo {year} {2003})},\ \Eprint
  {https://arxiv.org/abs/gr-qc/0207095} {arXiv:gr-qc/0207095} \BibitemShut
  {NoStop}%
\bibitem [{\citenamefont {Gourgoulhon}\ \emph {et~al.}(2001)\citenamefont
  {Gourgoulhon}, \citenamefont {Grandclement}, \citenamefont {Taniguchi},
  \citenamefont {Marck},\ and\ \citenamefont {Bonazzola}}]{Gourgoulhon:2000nn}%
  \BibitemOpen
  \bibfield  {author} {\bibinfo {author} {\bibfnamefont {E.}~\bibnamefont
  {Gourgoulhon}}, \bibinfo {author} {\bibfnamefont {P.}~\bibnamefont
  {Grandclement}}, \bibinfo {author} {\bibfnamefont {K.}~\bibnamefont
  {Taniguchi}}, \bibinfo {author} {\bibfnamefont {J.-A.}\ \bibnamefont
  {Marck}},\ and\ \bibinfo {author} {\bibfnamefont {S.}~\bibnamefont
  {Bonazzola}},\ }\bibfield  {title} {\bibinfo {title} {{Quasiequilibrium
  sequences of synchronized and irrotational binary neutron stars in general
  relativity: 1. Method and tests}},\ }\href
  {https://doi.org/10.1103/PhysRevD.63.064029} {\bibfield  {journal} {\bibinfo
  {journal} {Phys. Rev. D}\ }\textbf {\bibinfo {volume} {63}},\ \bibinfo
  {pages} {064029} (\bibinfo {year} {2001})},\ \Eprint
  {https://arxiv.org/abs/gr-qc/0007028} {arXiv:gr-qc/0007028} \BibitemShut
  {NoStop}%
\bibitem [{\citenamefont {Tichy}(2009)}]{Tichy:2009yr}%
  \BibitemOpen
  \bibfield  {author} {\bibinfo {author} {\bibfnamefont {W.}~\bibnamefont
  {Tichy}},\ }\bibfield  {title} {\bibinfo {title} {{A New numerical method to
  construct binary neutron star initial data}},\ }\href
  {https://doi.org/10.1088/0264-9381/26/17/175018} {\bibfield  {journal}
  {\bibinfo  {journal} {Class. Quant. Grav.}\ }\textbf {\bibinfo {volume}
  {26}},\ \bibinfo {pages} {175018} (\bibinfo {year} {2009})},\ \Eprint
  {https://arxiv.org/abs/0908.0620} {arXiv:0908.0620 [gr-qc]} \BibitemShut
  {NoStop}%
\bibitem [{\citenamefont {Tacik}\ \emph {et~al.}(2015)\citenamefont {Tacik}
  \emph {et~al.}}]{Tacik:2015tja}%
  \BibitemOpen
  \bibfield  {author} {\bibinfo {author} {\bibfnamefont {N.}~\bibnamefont
  {Tacik}} \emph {et~al.},\ }\bibfield  {title} {\bibinfo {title} {{Binary
  Neutron Stars with Arbitrary Spins in Numerical Relativity}},\ }\href
  {https://doi.org/10.1103/PhysRevD.92.124012} {\bibfield  {journal} {\bibinfo
  {journal} {Phys. Rev. D}\ }\textbf {\bibinfo {volume} {92}},\ \bibinfo
  {pages} {124012} (\bibinfo {year} {2015})},\ \bibinfo {note} {[Erratum:
  Phys.Rev.D 94, 049903 (2016)]},\ \Eprint {https://arxiv.org/abs/1508.06986}
  {arXiv:1508.06986 [gr-qc]} \BibitemShut {NoStop}%
\bibitem [{\citenamefont {Papenfort}\ \emph {et~al.}(2021)\citenamefont
  {Papenfort}, \citenamefont {Tootle}, \citenamefont {Grandcl\'ement},
  \citenamefont {Most},\ and\ \citenamefont {Rezzolla}}]{Papenfort:2021hod}%
  \BibitemOpen
  \bibfield  {author} {\bibinfo {author} {\bibfnamefont {L.~J.}\ \bibnamefont
  {Papenfort}}, \bibinfo {author} {\bibfnamefont {S.~D.}\ \bibnamefont
  {Tootle}}, \bibinfo {author} {\bibfnamefont {P.}~\bibnamefont
  {Grandcl\'ement}}, \bibinfo {author} {\bibfnamefont {E.~R.}\ \bibnamefont
  {Most}},\ and\ \bibinfo {author} {\bibfnamefont {L.}~\bibnamefont
  {Rezzolla}},\ }\bibfield  {title} {\bibinfo {title} {{New public code for
  initial data of unequal-mass, spinning compact-object binaries}},\ }\href
  {https://doi.org/10.1103/PhysRevD.104.024057} {\bibfield  {journal} {\bibinfo
   {journal} {Phys. Rev. D}\ }\textbf {\bibinfo {volume} {104}},\ \bibinfo
  {pages} {024057} (\bibinfo {year} {2021})},\ \Eprint
  {https://arxiv.org/abs/2103.09911} {arXiv:2103.09911 [gr-qc]} \BibitemShut
  {NoStop}%
\bibitem [{\citenamefont {East}\ \emph {et~al.}(2012)\citenamefont {East},
  \citenamefont {Ramazanoglu},\ and\ \citenamefont {Pretorius}}]{East:2012zn}%
  \BibitemOpen
  \bibfield  {author} {\bibinfo {author} {\bibfnamefont {W.~E.}\ \bibnamefont
  {East}}, \bibinfo {author} {\bibfnamefont {F.~M.}\ \bibnamefont
  {Ramazanoglu}},\ and\ \bibinfo {author} {\bibfnamefont {F.}~\bibnamefont
  {Pretorius}},\ }\bibfield  {title} {\bibinfo {title} {{Conformal
  Thin-Sandwich Solver for Generic Initial Data}},\ }\href
  {https://doi.org/10.1103/PhysRevD.86.104053} {\bibfield  {journal} {\bibinfo
  {journal} {Phys. Rev. D}\ }\textbf {\bibinfo {volume} {86}},\ \bibinfo
  {pages} {104053} (\bibinfo {year} {2012})},\ \Eprint
  {https://arxiv.org/abs/1208.3473} {arXiv:1208.3473 [gr-qc]} \BibitemShut
  {NoStop}%
\bibitem [{\citenamefont {Uryu}\ and\ \citenamefont
  {Tsokaros}(2012)}]{Uryu:2011ky}%
  \BibitemOpen
  \bibfield  {author} {\bibinfo {author} {\bibfnamefont {K.}~\bibnamefont
  {Uryu}}\ and\ \bibinfo {author} {\bibfnamefont {A.}~\bibnamefont
  {Tsokaros}},\ }\bibfield  {title} {\bibinfo {title} {{A new code for
  equilibriums and quasiequilibrium initial data of compact objects}},\ }\href
  {https://doi.org/10.1103/PhysRevD.85.064014} {\bibfield  {journal} {\bibinfo
  {journal} {Phys. Rev. D}\ }\textbf {\bibinfo {volume} {85}},\ \bibinfo
  {pages} {064014} (\bibinfo {year} {2012})},\ \Eprint
  {https://arxiv.org/abs/1108.3065} {arXiv:1108.3065 [gr-qc]} \BibitemShut
  {NoStop}%
\bibitem [{\citenamefont {Espino}\ \emph {et~al.}(2023)\citenamefont {Espino},
  \citenamefont {Bozzola},\ and\ \citenamefont {Paschalidis}}]{Espino:2022mtb}%
  \BibitemOpen
  \bibfield  {author} {\bibinfo {author} {\bibfnamefont {P.~L.}\ \bibnamefont
  {Espino}}, \bibinfo {author} {\bibfnamefont {G.}~\bibnamefont {Bozzola}},\
  and\ \bibinfo {author} {\bibfnamefont {V.}~\bibnamefont {Paschalidis}},\
  }\bibfield  {title} {\bibinfo {title} {{Quantifying uncertainties in general
  relativistic magnetohydrodynamic codes}},\ }\href
  {https://doi.org/10.1103/PhysRevD.107.104059} {\bibfield  {journal} {\bibinfo
   {journal} {Phys. Rev. D}\ }\textbf {\bibinfo {volume} {107}},\ \bibinfo
  {pages} {104059} (\bibinfo {year} {2023})},\ \Eprint
  {https://arxiv.org/abs/2210.13481} {arXiv:2210.13481 [gr-qc]} \BibitemShut
  {NoStop}%
\bibitem [{\citenamefont {Hamilton}\ and\ \citenamefont
  {Messman}(2024)}]{Hamilton:2024ziw}%
  \BibitemOpen
  \bibfield  {author} {\bibinfo {author} {\bibfnamefont {M.~C.~B.}\
  \bibnamefont {Hamilton}}\ and\ \bibinfo {author} {\bibfnamefont {W.~A.}\
  \bibnamefont {Messman}},\ }\bibfield  {title} {\bibinfo {title} {{Insights
  into Binary Neutron Star Merger Simulations: A Multi-Code Comparison}},\
  }\href@noop {} {\  (\bibinfo {year} {2024})},\ \Eprint
  {https://arxiv.org/abs/2411.10552} {arXiv:2411.10552 [gr-qc]} \BibitemShut
  {NoStop}%
\bibitem [{\citenamefont {Neuweiler}\ \emph {et~al.}(2024)\citenamefont
  {Neuweiler}, \citenamefont {Dietrich}, \citenamefont {Br\"ugmann},
  \citenamefont {Giangrandi}, \citenamefont {Kiuchi}, \citenamefont
  {Schianchi}, \citenamefont {M\"osta}, \citenamefont {Shankar}, \citenamefont
  {Giacomazzo},\ and\ \citenamefont {Shibata}}]{Neuweiler:2024jae}%
  \BibitemOpen
  \bibfield  {author} {\bibinfo {author} {\bibfnamefont {A.}~\bibnamefont
  {Neuweiler}}, \bibinfo {author} {\bibfnamefont {T.}~\bibnamefont {Dietrich}},
  \bibinfo {author} {\bibfnamefont {B.}~\bibnamefont {Br\"ugmann}}, \bibinfo
  {author} {\bibfnamefont {E.}~\bibnamefont {Giangrandi}}, \bibinfo {author}
  {\bibfnamefont {K.}~\bibnamefont {Kiuchi}}, \bibinfo {author} {\bibfnamefont
  {F.}~\bibnamefont {Schianchi}}, \bibinfo {author} {\bibfnamefont
  {P.}~\bibnamefont {M\"osta}}, \bibinfo {author} {\bibfnamefont
  {S.}~\bibnamefont {Shankar}}, \bibinfo {author} {\bibfnamefont
  {B.}~\bibnamefont {Giacomazzo}},\ and\ \bibinfo {author} {\bibfnamefont
  {M.}~\bibnamefont {Shibata}},\ }\bibfield  {title} {\bibinfo {title}
  {{General relativistic magnetohydrodynamic simulations with bam:
  Implementation and code comparison}},\ }\href
  {https://doi.org/10.1103/PhysRevD.110.084046} {\bibfield  {journal} {\bibinfo
   {journal} {Phys. Rev. D}\ }\textbf {\bibinfo {volume} {110}},\ \bibinfo
  {pages} {084046} (\bibinfo {year} {2024})},\ \Eprint
  {https://arxiv.org/abs/2407.20946} {arXiv:2407.20946 [gr-qc]} \BibitemShut
  {NoStop}%
\bibitem [{\citenamefont {Kuan}\ \emph {et~al.}(2025)\citenamefont {Kuan},
  \citenamefont {Markin}, \citenamefont {Ujevic}, \citenamefont {Dietrich},
  \citenamefont {Kiuchi}, \citenamefont {Shibata},\ and\ \citenamefont
  {Tichy}}]{Kuan:2025bzu}%
  \BibitemOpen
  \bibfield  {author} {\bibinfo {author} {\bibfnamefont {H.-J.}\ \bibnamefont
  {Kuan}}, \bibinfo {author} {\bibfnamefont {I.}~\bibnamefont {Markin}},
  \bibinfo {author} {\bibfnamefont {M.}~\bibnamefont {Ujevic}}, \bibinfo
  {author} {\bibfnamefont {T.}~\bibnamefont {Dietrich}}, \bibinfo {author}
  {\bibfnamefont {K.}~\bibnamefont {Kiuchi}}, \bibinfo {author} {\bibfnamefont
  {M.}~\bibnamefont {Shibata}},\ and\ \bibinfo {author} {\bibfnamefont
  {W.}~\bibnamefont {Tichy}},\ }\bibfield  {title} {\bibinfo {title} {{The
  error budget of binary neutron star merger simulations for configurations
  with high spin}},\ }\href@noop {} {\  (\bibinfo {year} {2025})},\ \Eprint
  {https://arxiv.org/abs/2506.02115} {arXiv:2506.02115 [gr-qc]} \BibitemShut
  {NoStop}%
\bibitem [{\citenamefont {Ajith}\ \emph {et~al.}(2012)\citenamefont {Ajith}
  \emph {et~al.}}]{Ajith:2012az}%
  \BibitemOpen
  \bibfield  {author} {\bibinfo {author} {\bibfnamefont {P.}~\bibnamefont
  {Ajith}} \emph {et~al.},\ }\bibfield  {title} {\bibinfo {title} {{The NINJA-2
  catalog of hybrid post-Newtonian/numerical-relativity waveforms for
  non-precessing black-hole binaries}},\ }\href
  {https://doi.org/10.1088/0264-9381/29/12/124001} {\bibfield  {journal}
  {\bibinfo  {journal} {Class. Quant. Grav.}\ }\textbf {\bibinfo {volume}
  {29}},\ \bibinfo {pages} {124001} (\bibinfo {year} {2012})},\ \bibinfo {note}
  {[Erratum: Class.Quant.Grav. 30, 199401 (2013)]},\ \Eprint
  {https://arxiv.org/abs/1201.5319} {arXiv:1201.5319 [gr-qc]} \BibitemShut
  {NoStop}%
\bibitem [{\citenamefont {Etienne}\ \emph {et~al.}(2015)\citenamefont
  {Etienne}, \citenamefont {Paschalidis}, \citenamefont {Haas}, \citenamefont
  {M\"osta},\ and\ \citenamefont {Shapiro}}]{Etienne:2015cea}%
  \BibitemOpen
  \bibfield  {author} {\bibinfo {author} {\bibfnamefont {Z.~B.}\ \bibnamefont
  {Etienne}}, \bibinfo {author} {\bibfnamefont {V.}~\bibnamefont
  {Paschalidis}}, \bibinfo {author} {\bibfnamefont {R.}~\bibnamefont {Haas}},
  \bibinfo {author} {\bibfnamefont {P.}~\bibnamefont {M\"osta}},\ and\ \bibinfo
  {author} {\bibfnamefont {S.~L.}\ \bibnamefont {Shapiro}},\ }\bibfield
  {title} {\bibinfo {title} {{IllinoisGRMHD: An Open-Source, User-Friendly
  GRMHD Code for Dynamical Spacetimes}},\ }\href
  {https://doi.org/10.1088/0264-9381/32/17/175009} {\bibfield  {journal}
  {\bibinfo  {journal} {Class. Quant. Grav.}\ }\textbf {\bibinfo {volume}
  {32}},\ \bibinfo {pages} {175009} (\bibinfo {year} {2015})},\ \Eprint
  {https://arxiv.org/abs/1501.07276} {arXiv:1501.07276 [astro-ph.HE]}
  \BibitemShut {NoStop}%
\bibitem [{\citenamefont {Scheel}\ \emph {et~al.}(2006)\citenamefont {Scheel},
  \citenamefont {Pfeiffer}, \citenamefont {Lindblom}, \citenamefont {Kidder},
  \citenamefont {Rinne},\ and\ \citenamefont {Teukolsky}}]{Scheel:2006gg}%
  \BibitemOpen
  \bibfield  {author} {\bibinfo {author} {\bibfnamefont {M.~A.}\ \bibnamefont
  {Scheel}}, \bibinfo {author} {\bibfnamefont {H.~P.}\ \bibnamefont
  {Pfeiffer}}, \bibinfo {author} {\bibfnamefont {L.}~\bibnamefont {Lindblom}},
  \bibinfo {author} {\bibfnamefont {L.~E.}\ \bibnamefont {Kidder}}, \bibinfo
  {author} {\bibfnamefont {O.}~\bibnamefont {Rinne}},\ and\ \bibinfo {author}
  {\bibfnamefont {S.~A.}\ \bibnamefont {Teukolsky}},\ }\bibfield  {title}
  {\bibinfo {title} {{Solving Einstein's equations with dual coordinate
  frames}},\ }\href {https://doi.org/10.1103/PhysRevD.74.104006} {\bibfield
  {journal} {\bibinfo  {journal} {Phys. Rev. D}\ }\textbf {\bibinfo {volume}
  {74}},\ \bibinfo {pages} {104006} (\bibinfo {year} {2006})},\ \Eprint
  {https://arxiv.org/abs/gr-qc/0607056} {arXiv:gr-qc/0607056} \BibitemShut
  {NoStop}%
\bibitem [{\citenamefont {Scheel}\ \emph {et~al.}(2009)\citenamefont {Scheel},
  \citenamefont {Boyle}, \citenamefont {Chu}, \citenamefont {Kidder},
  \citenamefont {Matthews},\ and\ \citenamefont {Pfeiffer}}]{Scheel:2008rj}%
  \BibitemOpen
  \bibfield  {author} {\bibinfo {author} {\bibfnamefont {M.~A.}\ \bibnamefont
  {Scheel}}, \bibinfo {author} {\bibfnamefont {M.}~\bibnamefont {Boyle}},
  \bibinfo {author} {\bibfnamefont {T.}~\bibnamefont {Chu}}, \bibinfo {author}
  {\bibfnamefont {L.~E.}\ \bibnamefont {Kidder}}, \bibinfo {author}
  {\bibfnamefont {K.~D.}\ \bibnamefont {Matthews}},\ and\ \bibinfo {author}
  {\bibfnamefont {H.~P.}\ \bibnamefont {Pfeiffer}},\ }\bibfield  {title}
  {\bibinfo {title} {{High-accuracy waveforms for binary black hole inspiral,
  merger, and ringdown}},\ }\href {https://doi.org/10.1103/PhysRevD.79.024003}
  {\bibfield  {journal} {\bibinfo  {journal} {Phys. Rev. D}\ }\textbf {\bibinfo
  {volume} {79}},\ \bibinfo {pages} {024003} (\bibinfo {year} {2009})},\
  \Eprint {https://arxiv.org/abs/0810.1767} {arXiv:0810.1767 [gr-qc]}
  \BibitemShut {NoStop}%
\bibitem [{\citenamefont {Szilagyi}\ \emph {et~al.}(2009)\citenamefont
  {Szilagyi}, \citenamefont {Lindblom},\ and\ \citenamefont
  {Scheel}}]{Szilagyi:2009qz}%
  \BibitemOpen
  \bibfield  {author} {\bibinfo {author} {\bibfnamefont {B.}~\bibnamefont
  {Szilagyi}}, \bibinfo {author} {\bibfnamefont {L.}~\bibnamefont {Lindblom}},\
  and\ \bibinfo {author} {\bibfnamefont {M.~A.}\ \bibnamefont {Scheel}},\
  }\bibfield  {title} {\bibinfo {title} {{Simulations of Binary Black Hole
  Mergers Using Spectral Methods}},\ }\href
  {https://doi.org/10.1103/PhysRevD.80.124010} {\bibfield  {journal} {\bibinfo
  {journal} {Phys. Rev. D}\ }\textbf {\bibinfo {volume} {80}},\ \bibinfo
  {pages} {124010} (\bibinfo {year} {2009})},\ \Eprint
  {https://arxiv.org/abs/0909.3557} {arXiv:0909.3557 [gr-qc]} \BibitemShut
  {NoStop}%
\bibitem [{\citenamefont {Szil\'agyi}(2014)}]{Szilagyi:2014fna}%
  \BibitemOpen
  \bibfield  {author} {\bibinfo {author} {\bibfnamefont {B.}~\bibnamefont
  {Szil\'agyi}},\ }\bibfield  {title} {\bibinfo {title} {{Key Elements of
  Robustness in Binary Black Hole Evolutions using Spectral Methods}},\ }\href
  {https://doi.org/10.1142/S0218271814300146} {\bibfield  {journal} {\bibinfo
  {journal} {Int. J. Mod. Phys. D}\ }\textbf {\bibinfo {volume} {23}},\
  \bibinfo {pages} {1430014} (\bibinfo {year} {2014})},\ \Eprint
  {https://arxiv.org/abs/1405.3693} {arXiv:1405.3693 [gr-qc]} \BibitemShut
  {NoStop}%
\bibitem [{\citenamefont {Grandclement}(2010)}]{Grandclement:2009ju}%
  \BibitemOpen
  \bibfield  {author} {\bibinfo {author} {\bibfnamefont {P.}~\bibnamefont
  {Grandclement}},\ }\bibfield  {title} {\bibinfo {title} {{Kadath: A Spectral
  solver for theoretical physics}},\ }\href
  {https://doi.org/10.1016/j.jcp.2010.01.005} {\bibfield  {journal} {\bibinfo
  {journal} {J. Comput. Phys.}\ }\textbf {\bibinfo {volume} {229}},\ \bibinfo
  {pages} {3334} (\bibinfo {year} {2010})},\ \Eprint
  {https://arxiv.org/abs/0909.1228} {arXiv:0909.1228 [gr-qc]} \BibitemShut
  {NoStop}%
\bibitem [{\citenamefont {Buonanno}\ \emph {et~al.}(2011)\citenamefont
  {Buonanno}, \citenamefont {Kidder}, \citenamefont {Mroue}, \citenamefont
  {Pfeiffer},\ and\ \citenamefont {Taracchini}}]{Buonanno:2010yk}%
  \BibitemOpen
  \bibfield  {author} {\bibinfo {author} {\bibfnamefont {A.}~\bibnamefont
  {Buonanno}}, \bibinfo {author} {\bibfnamefont {L.~E.}\ \bibnamefont
  {Kidder}}, \bibinfo {author} {\bibfnamefont {A.~H.}\ \bibnamefont {Mroue}},
  \bibinfo {author} {\bibfnamefont {H.~P.}\ \bibnamefont {Pfeiffer}},\ and\
  \bibinfo {author} {\bibfnamefont {A.}~\bibnamefont {Taracchini}},\ }\bibfield
   {title} {\bibinfo {title} {{Reducing orbital eccentricity of precessing
  black-hole binaries}},\ }\href {https://doi.org/10.1103/PhysRevD.83.104034}
  {\bibfield  {journal} {\bibinfo  {journal} {Phys. Rev. D}\ }\textbf {\bibinfo
  {volume} {83}},\ \bibinfo {pages} {104034} (\bibinfo {year} {2011})},\
  \Eprint {https://arxiv.org/abs/1012.1549} {arXiv:1012.1549 [gr-qc]}
  \BibitemShut {NoStop}%
\bibitem [{\citenamefont {Douchin}\ and\ \citenamefont
  {Haensel}(2001)}]{Douchin:2001sv}%
  \BibitemOpen
  \bibfield  {author} {\bibinfo {author} {\bibfnamefont {F.}~\bibnamefont
  {Douchin}}\ and\ \bibinfo {author} {\bibfnamefont {P.}~\bibnamefont
  {Haensel}},\ }\bibfield  {title} {\bibinfo {title} {{A unified equation of
  state of dense matter and neutron star structure}},\ }\href
  {https://doi.org/10.1051/0004-6361:20011402} {\bibfield  {journal} {\bibinfo
  {journal} {Astron. Astrophys.}\ }\textbf {\bibinfo {volume} {380}},\ \bibinfo
  {pages} {151} (\bibinfo {year} {2001})},\ \Eprint
  {https://arxiv.org/abs/astro-ph/0111092} {arXiv:astro-ph/0111092}
  \BibitemShut {NoStop}%
\bibitem [{\citenamefont {Foucart}\ \emph {et~al.}(2019)\citenamefont
  {Foucart}, \citenamefont {Duez}, \citenamefont {Gudinas}, \citenamefont
  {Hebert}, \citenamefont {Kidder}, \citenamefont {Pfeiffer},\ and\
  \citenamefont {Scheel}}]{Foucart:2019yzo}%
  \BibitemOpen
  \bibfield  {author} {\bibinfo {author} {\bibfnamefont {F.}~\bibnamefont
  {Foucart}}, \bibinfo {author} {\bibfnamefont {M.~D.}\ \bibnamefont {Duez}},
  \bibinfo {author} {\bibfnamefont {A.}~\bibnamefont {Gudinas}}, \bibinfo
  {author} {\bibfnamefont {F.}~\bibnamefont {Hebert}}, \bibinfo {author}
  {\bibfnamefont {L.~E.}\ \bibnamefont {Kidder}}, \bibinfo {author}
  {\bibfnamefont {H.~P.}\ \bibnamefont {Pfeiffer}},\ and\ \bibinfo {author}
  {\bibfnamefont {M.~A.}\ \bibnamefont {Scheel}},\ }\bibfield  {title}
  {\bibinfo {title} {{Smooth Equations of State for High-Accuracy Simulations
  of Neutron Star Binaries}},\ }\href
  {https://doi.org/10.1103/PhysRevD.100.104048} {\bibfield  {journal} {\bibinfo
   {journal} {Phys. Rev. D}\ }\textbf {\bibinfo {volume} {100}},\ \bibinfo
  {pages} {104048} (\bibinfo {year} {2019})},\ \Eprint
  {https://arxiv.org/abs/1908.05277} {arXiv:1908.05277 [gr-qc]} \BibitemShut
  {NoStop}%
\bibitem [{\citenamefont {Abbott}\ \emph {et~al.}(2019)\citenamefont {Abbott}
  \emph {et~al.}}]{LIGOScientific:2018hze}%
  \BibitemOpen
  \bibfield  {author} {\bibinfo {author} {\bibfnamefont {B.~P.}\ \bibnamefont
  {Abbott}} \emph {et~al.} (\bibinfo {collaboration} {LIGO Scientific,
  Virgo}),\ }\bibfield  {title} {\bibinfo {title} {{Properties of the binary
  neutron star merger GW170817}},\ }\href
  {https://doi.org/10.1103/PhysRevX.9.011001} {\bibfield  {journal} {\bibinfo
  {journal} {Phys. Rev. X}\ }\textbf {\bibinfo {volume} {9}},\ \bibinfo {pages}
  {011001} (\bibinfo {year} {2019})},\ \Eprint
  {https://arxiv.org/abs/1805.11579} {arXiv:1805.11579 [gr-qc]} \BibitemShut
  {NoStop}%
\bibitem [{\citenamefont {Lindblom}(2010)}]{Lindblom:2010bb}%
  \BibitemOpen
  \bibfield  {author} {\bibinfo {author} {\bibfnamefont {L.}~\bibnamefont
  {Lindblom}},\ }\bibfield  {title} {\bibinfo {title} {{Spectral
  Representations of Neutron-Star Equations of State}},\ }\href
  {https://doi.org/10.1103/PhysRevD.82.103011} {\bibfield  {journal} {\bibinfo
  {journal} {Phys. Rev. D}\ }\textbf {\bibinfo {volume} {82}},\ \bibinfo
  {pages} {103011} (\bibinfo {year} {2010})},\ \Eprint
  {https://arxiv.org/abs/1009.0738} {arXiv:1009.0738 [astro-ph.HE]}
  \BibitemShut {NoStop}%
\bibitem [{\citenamefont {Raithel}\ and\ \citenamefont
  {Paschalidis}(2022)}]{Raithel:2022san}%
  \BibitemOpen
  \bibfield  {author} {\bibinfo {author} {\bibfnamefont {C.~A.}\ \bibnamefont
  {Raithel}}\ and\ \bibinfo {author} {\bibfnamefont {V.}~\bibnamefont
  {Paschalidis}},\ }\bibfield  {title} {\bibinfo {title} {{Improving the
  convergence order of binary neutron star merger simulations in the Baumgarte-
  Shapiro-Shibata-Nakamura formulation}},\ }\href
  {https://doi.org/10.1103/PhysRevD.106.023015} {\bibfield  {journal} {\bibinfo
   {journal} {Phys. Rev. D}\ }\textbf {\bibinfo {volume} {106}},\ \bibinfo
  {pages} {023015} (\bibinfo {year} {2022})},\ \Eprint
  {https://arxiv.org/abs/2204.00698} {arXiv:2204.00698 [gr-qc]} \BibitemShut
  {NoStop}%
\bibitem [{\citenamefont {Tsokaros}\ \emph {et~al.}(2016)\citenamefont
  {Tsokaros}, \citenamefont {Mundim}, \citenamefont {Galeazzi}, \citenamefont
  {Rezzolla},\ and\ \citenamefont {Ury{\={u}}}}]{Tsokaros:2016eik}%
  \BibitemOpen
  \bibfield  {author} {\bibinfo {author} {\bibfnamefont {A.}~\bibnamefont
  {Tsokaros}}, \bibinfo {author} {\bibfnamefont {B.~C.}\ \bibnamefont
  {Mundim}}, \bibinfo {author} {\bibfnamefont {F.}~\bibnamefont {Galeazzi}},
  \bibinfo {author} {\bibfnamefont {L.}~\bibnamefont {Rezzolla}},\ and\
  \bibinfo {author} {\bibfnamefont {K.}~\bibnamefont {Ury{\={u}}}},\ }\bibfield
   {title} {\bibinfo {title} {{Initial-data contribution to the error budget of
  gravitational waves from neutron-star binaries}},\ }\href
  {https://doi.org/10.1103/PhysRevD.94.044049} {\bibfield  {journal} {\bibinfo
  {journal} {Phys. Rev. D}\ }\textbf {\bibinfo {volume} {94}},\ \bibinfo
  {pages} {044049} (\bibinfo {year} {2016})},\ \Eprint
  {https://arxiv.org/abs/1605.07205} {arXiv:1605.07205 [gr-qc]} \BibitemShut
  {NoStop}%
\bibitem [{\citenamefont {Duez}\ \emph {et~al.}(2005)\citenamefont {Duez},
  \citenamefont {Liu}, \citenamefont {Shapiro},\ and\ \citenamefont
  {Stephens}}]{Duez:2005sf}%
  \BibitemOpen
  \bibfield  {author} {\bibinfo {author} {\bibfnamefont {M.~D.}\ \bibnamefont
  {Duez}}, \bibinfo {author} {\bibfnamefont {Y.~T.}\ \bibnamefont {Liu}},
  \bibinfo {author} {\bibfnamefont {S.~L.}\ \bibnamefont {Shapiro}},\ and\
  \bibinfo {author} {\bibfnamefont {B.~C.}\ \bibnamefont {Stephens}},\
  }\bibfield  {title} {\bibinfo {title} {{Relativistic magnetohydrodynamics in
  dynamical spacetimes: Numerical methods and tests}},\ }\href
  {https://doi.org/10.1103/PhysRevD.72.024028} {\bibfield  {journal} {\bibinfo
  {journal} {Phys. Rev. D}\ }\textbf {\bibinfo {volume} {72}},\ \bibinfo
  {pages} {024028} (\bibinfo {year} {2005})},\ \Eprint
  {https://arxiv.org/abs/astro-ph/0503420} {arXiv:astro-ph/0503420}
  \BibitemShut {NoStop}%
\bibitem [{\citenamefont {Loffler}\ \emph {et~al.}(2012)\citenamefont {Loffler}
  \emph {et~al.}}]{Loffler:2011ay}%
  \BibitemOpen
  \bibfield  {author} {\bibinfo {author} {\bibfnamefont {F.}~\bibnamefont
  {Loffler}} \emph {et~al.},\ }\bibfield  {title} {\bibinfo {title} {{The
  Einstein Toolkit: A Community Computational Infrastructure for Relativistic
  Astrophysics}},\ }\href {https://doi.org/10.1088/0264-9381/29/11/115001}
  {\bibfield  {journal} {\bibinfo  {journal} {Class. Quant. Grav.}\ }\textbf
  {\bibinfo {volume} {29}},\ \bibinfo {pages} {115001} (\bibinfo {year}
  {2012})},\ \Eprint {https://arxiv.org/abs/1111.3344} {arXiv:1111.3344
  [gr-qc]} \BibitemShut {NoStop}%
\bibitem [{\citenamefont {Del~Zanna}\ \emph {et~al.}(2007)\citenamefont
  {Del~Zanna}, \citenamefont {Zanotti}, \citenamefont {Bucciantini},\ and\
  \citenamefont {Londrillo}}]{DelZanna:2007pk}%
  \BibitemOpen
  \bibfield  {author} {\bibinfo {author} {\bibfnamefont {L.}~\bibnamefont
  {Del~Zanna}}, \bibinfo {author} {\bibfnamefont {O.}~\bibnamefont {Zanotti}},
  \bibinfo {author} {\bibfnamefont {N.}~\bibnamefont {Bucciantini}},\ and\
  \bibinfo {author} {\bibfnamefont {P.}~\bibnamefont {Londrillo}},\ }\bibfield
  {title} {\bibinfo {title} {{ECHO: an Eulerian Conservative High Order scheme
  for general relativistic magnetohydrodynamics and magnetodynamics}},\ }\href
  {https://doi.org/10.1051/0004-6361:20077093} {\bibfield  {journal} {\bibinfo
  {journal} {Astron. Astrophys.}\ }\textbf {\bibinfo {volume} {473}},\ \bibinfo
  {pages} {11} (\bibinfo {year} {2007})},\ \Eprint
  {https://arxiv.org/abs/0704.3206} {arXiv:0704.3206 [astro-ph]} \BibitemShut
  {NoStop}%
\bibitem [{\citenamefont {Borges}\ \emph {et~al.}(2008)\citenamefont {Borges},
  \citenamefont {Carmona}, \citenamefont {Costa},\ and\ \citenamefont
  {Don}}]{borges2008improved}%
  \BibitemOpen
  \bibfield  {author} {\bibinfo {author} {\bibfnamefont {R.}~\bibnamefont
  {Borges}}, \bibinfo {author} {\bibfnamefont {M.}~\bibnamefont {Carmona}},
  \bibinfo {author} {\bibfnamefont {B.}~\bibnamefont {Costa}},\ and\ \bibinfo
  {author} {\bibfnamefont {W.~S.}\ \bibnamefont {Don}},\ }\bibfield  {title}
  {\bibinfo {title} {An improved weighted essentially non-oscillatory scheme
  for hyperbolic conservation laws},\ }\href@noop {} {\bibfield  {journal}
  {\bibinfo  {journal} {Journal of computational physics}\ }\textbf {\bibinfo
  {volume} {227}},\ \bibinfo {pages} {3191} (\bibinfo {year}
  {2008})}\BibitemShut {NoStop}%
\bibitem [{\citenamefont {Einfeldt}(1988)}]{einfeldt1988godunov}%
  \BibitemOpen
  \bibfield  {author} {\bibinfo {author} {\bibfnamefont {B.}~\bibnamefont
  {Einfeldt}},\ }\bibfield  {title} {\bibinfo {title} {On godunov-type methods
  for gas dynamics},\ }\href@noop {} {\bibfield  {journal} {\bibinfo  {journal}
  {SIAM Journal on numerical analysis}\ }\textbf {\bibinfo {volume} {25}},\
  \bibinfo {pages} {294} (\bibinfo {year} {1988})}\BibitemShut {NoStop}%
\bibitem [{\citenamefont {Radice}\ \emph
  {et~al.}(2014{\natexlab{b}})\citenamefont {Radice}, \citenamefont
  {Rezzolla},\ and\ \citenamefont {Galeazzi}}]{Radice:2013xpa}%
  \BibitemOpen
  \bibfield  {author} {\bibinfo {author} {\bibfnamefont {D.}~\bibnamefont
  {Radice}}, \bibinfo {author} {\bibfnamefont {L.}~\bibnamefont {Rezzolla}},\
  and\ \bibinfo {author} {\bibfnamefont {F.}~\bibnamefont {Galeazzi}},\
  }\bibfield  {title} {\bibinfo {title} {{High-Order Fully General-Relativistic
  Hydrodynamics: new Approaches and Tests}},\ }\href
  {https://doi.org/10.1088/0264-9381/31/7/075012} {\bibfield  {journal}
  {\bibinfo  {journal} {Class. Quant. Grav.}\ }\textbf {\bibinfo {volume}
  {31}},\ \bibinfo {pages} {075012} (\bibinfo {year} {2014}{\natexlab{b}})},\
  \Eprint {https://arxiv.org/abs/1312.5004} {arXiv:1312.5004 [gr-qc]}
  \BibitemShut {NoStop}%
\bibitem [{\citenamefont {Kastaun}\ \emph {et~al.}(2021)\citenamefont
  {Kastaun}, \citenamefont {Kalinani},\ and\ \citenamefont
  {Ciolfi}}]{Kastaun:2020uxr}%
  \BibitemOpen
  \bibfield  {author} {\bibinfo {author} {\bibfnamefont {W.}~\bibnamefont
  {Kastaun}}, \bibinfo {author} {\bibfnamefont {J.~V.}\ \bibnamefont
  {Kalinani}},\ and\ \bibinfo {author} {\bibfnamefont {R.}~\bibnamefont
  {Ciolfi}},\ }\bibfield  {title} {\bibinfo {title} {{Robust Recovery of
  Primitive Variables in Relativistic Ideal Magnetohydrodynamics}},\ }\href
  {https://doi.org/10.1103/PhysRevD.103.023018} {\bibfield  {journal} {\bibinfo
   {journal} {Phys. Rev. D}\ }\textbf {\bibinfo {volume} {103}},\ \bibinfo
  {pages} {023018} (\bibinfo {year} {2021})},\ \Eprint
  {https://arxiv.org/abs/2005.01821} {arXiv:2005.01821 [gr-qc]} \BibitemShut
  {NoStop}%
\bibitem [{\citenamefont {Galeazzi}\ \emph {et~al.}(2013)\citenamefont
  {Galeazzi}, \citenamefont {Kastaun}, \citenamefont {Rezzolla},\ and\
  \citenamefont {Font}}]{Galeazzi:2013mia}%
  \BibitemOpen
  \bibfield  {author} {\bibinfo {author} {\bibfnamefont {F.}~\bibnamefont
  {Galeazzi}}, \bibinfo {author} {\bibfnamefont {W.}~\bibnamefont {Kastaun}},
  \bibinfo {author} {\bibfnamefont {L.}~\bibnamefont {Rezzolla}},\ and\
  \bibinfo {author} {\bibfnamefont {J.~A.}\ \bibnamefont {Font}},\ }\bibfield
  {title} {\bibinfo {title} {{Implementation of a simplified approach to
  radiative transfer in general relativity}},\ }\href
  {https://doi.org/10.1103/PhysRevD.88.064009} {\bibfield  {journal} {\bibinfo
  {journal} {Phys. Rev. D}\ }\textbf {\bibinfo {volume} {88}},\ \bibinfo
  {pages} {064009} (\bibinfo {year} {2013})},\ \Eprint
  {https://arxiv.org/abs/1306.4953} {arXiv:1306.4953 [gr-qc]} \BibitemShut
  {NoStop}%
\bibitem [{\citenamefont {Alcubierre}\ \emph {et~al.}(2003)\citenamefont
  {Alcubierre}, \citenamefont {Bruegmann}, \citenamefont {Diener},
  \citenamefont {Koppitz}, \citenamefont {Pollney}, \citenamefont {Seidel},\
  and\ \citenamefont {Takahashi}}]{Alcubierre:2002kk}%
  \BibitemOpen
  \bibfield  {author} {\bibinfo {author} {\bibfnamefont {M.}~\bibnamefont
  {Alcubierre}}, \bibinfo {author} {\bibfnamefont {B.}~\bibnamefont
  {Bruegmann}}, \bibinfo {author} {\bibfnamefont {P.}~\bibnamefont {Diener}},
  \bibinfo {author} {\bibfnamefont {M.}~\bibnamefont {Koppitz}}, \bibinfo
  {author} {\bibfnamefont {D.}~\bibnamefont {Pollney}}, \bibinfo {author}
  {\bibfnamefont {E.}~\bibnamefont {Seidel}},\ and\ \bibinfo {author}
  {\bibfnamefont {R.}~\bibnamefont {Takahashi}},\ }\bibfield  {title} {\bibinfo
  {title} {{Gauge conditions for long term numerical black hole evolutions
  without excision}},\ }\href {https://doi.org/10.1103/PhysRevD.67.084023}
  {\bibfield  {journal} {\bibinfo  {journal} {Phys. Rev. D}\ }\textbf {\bibinfo
  {volume} {67}},\ \bibinfo {pages} {084023} (\bibinfo {year} {2003})},\
  \Eprint {https://arxiv.org/abs/gr-qc/0206072} {arXiv:gr-qc/0206072}
  \BibitemShut {NoStop}%
\bibitem [{\citenamefont {Campanelli}\ \emph {et~al.}(2006)\citenamefont
  {Campanelli}, \citenamefont {Lousto}, \citenamefont {Marronetti},\ and\
  \citenamefont {Zlochower}}]{Campanelli:2005dd}%
  \BibitemOpen
  \bibfield  {author} {\bibinfo {author} {\bibfnamefont {M.}~\bibnamefont
  {Campanelli}}, \bibinfo {author} {\bibfnamefont {C.~O.}\ \bibnamefont
  {Lousto}}, \bibinfo {author} {\bibfnamefont {P.}~\bibnamefont {Marronetti}},\
  and\ \bibinfo {author} {\bibfnamefont {Y.}~\bibnamefont {Zlochower}},\
  }\bibfield  {title} {\bibinfo {title} {{Accurate evolutions of orbiting
  black-hole binaries without excision}},\ }\href
  {https://doi.org/10.1103/PhysRevLett.96.111101} {\bibfield  {journal}
  {\bibinfo  {journal} {Phys. Rev. Lett.}\ }\textbf {\bibinfo {volume} {96}},\
  \bibinfo {pages} {111101} (\bibinfo {year} {2006})},\ \Eprint
  {https://arxiv.org/abs/gr-qc/0511048} {arXiv:gr-qc/0511048} \BibitemShut
  {NoStop}%
\bibitem [{\citenamefont {Baker}\ \emph {et~al.}(2006)\citenamefont {Baker},
  \citenamefont {Centrella}, \citenamefont {Choi}, \citenamefont {Koppitz},\
  and\ \citenamefont {van Meter}}]{Baker:2005vv}%
  \BibitemOpen
  \bibfield  {author} {\bibinfo {author} {\bibfnamefont {J.~G.}\ \bibnamefont
  {Baker}}, \bibinfo {author} {\bibfnamefont {J.}~\bibnamefont {Centrella}},
  \bibinfo {author} {\bibfnamefont {D.-I.}\ \bibnamefont {Choi}}, \bibinfo
  {author} {\bibfnamefont {M.}~\bibnamefont {Koppitz}},\ and\ \bibinfo {author}
  {\bibfnamefont {J.}~\bibnamefont {van Meter}},\ }\bibfield  {title} {\bibinfo
  {title} {{Gravitational wave extraction from an inspiraling configuration of
  merging black holes}},\ }\href
  {https://doi.org/10.1103/PhysRevLett.96.111102} {\bibfield  {journal}
  {\bibinfo  {journal} {Phys. Rev. Lett.}\ }\textbf {\bibinfo {volume} {96}},\
  \bibinfo {pages} {111102} (\bibinfo {year} {2006})},\ \Eprint
  {https://arxiv.org/abs/gr-qc/0511103} {arXiv:gr-qc/0511103} \BibitemShut
  {NoStop}%
\bibitem [{\citenamefont {Gottlieb}\ \emph {et~al.}(2001)\citenamefont
  {Gottlieb}, \citenamefont {Shu},\ and\ \citenamefont
  {Tadmor}}]{gottlieb2001strong}%
  \BibitemOpen
  \bibfield  {author} {\bibinfo {author} {\bibfnamefont {S.}~\bibnamefont
  {Gottlieb}}, \bibinfo {author} {\bibfnamefont {C.-W.}\ \bibnamefont {Shu}},\
  and\ \bibinfo {author} {\bibfnamefont {E.}~\bibnamefont {Tadmor}},\
  }\bibfield  {title} {\bibinfo {title} {Strong stability-preserving high-order
  time discretization methods},\ }\href@noop {} {\bibfield  {journal} {\bibinfo
   {journal} {SIAM review}\ }\textbf {\bibinfo {volume} {43}},\ \bibinfo
  {pages} {89} (\bibinfo {year} {2001})}\BibitemShut {NoStop}%
\bibitem [{\citenamefont {Knight}\ \emph {et~al.}(2024)\citenamefont {Knight},
  \citenamefont {Foucart}, \citenamefont {Duez}, \citenamefont {Boyle},
  \citenamefont {Kidder}, \citenamefont {Pfeiffer},\ and\ \citenamefont
  {Scheel}}]{Knight:2023kqw}%
  \BibitemOpen
  \bibfield  {author} {\bibinfo {author} {\bibfnamefont {A.}~\bibnamefont
  {Knight}}, \bibinfo {author} {\bibfnamefont {F.}~\bibnamefont {Foucart}},
  \bibinfo {author} {\bibfnamefont {M.~D.}\ \bibnamefont {Duez}}, \bibinfo
  {author} {\bibfnamefont {M.}~\bibnamefont {Boyle}}, \bibinfo {author}
  {\bibfnamefont {L.~E.}\ \bibnamefont {Kidder}}, \bibinfo {author}
  {\bibfnamefont {H.~P.}\ \bibnamefont {Pfeiffer}},\ and\ \bibinfo {author}
  {\bibfnamefont {M.~A.}\ \bibnamefont {Scheel}},\ }\bibfield  {title}
  {\bibinfo {title} {{Gravitational waves from binary neutron star mergers with
  a spectral equation of state}},\ }\href
  {https://doi.org/10.1103/PhysRevD.110.023034} {\bibfield  {journal} {\bibinfo
   {journal} {Phys. Rev. D}\ }\textbf {\bibinfo {volume} {110}},\ \bibinfo
  {pages} {023034} (\bibinfo {year} {2024})},\ \Eprint
  {https://arxiv.org/abs/2307.03250} {arXiv:2307.03250 [astro-ph.HE]}
  \BibitemShut {NoStop}%
\bibitem [{\citenamefont {Foucart}\ \emph {et~al.}(2013)\citenamefont
  {Foucart}, \citenamefont {Deaton}, \citenamefont {Duez}, \citenamefont
  {Kidder}, \citenamefont {MacDonald}, \citenamefont {Ott}, \citenamefont
  {Pfeiffer}, \citenamefont {Scheel}, \citenamefont {Szilagyi},\ and\
  \citenamefont {Teukolsky}}]{Foucart:2012vn}%
  \BibitemOpen
  \bibfield  {author} {\bibinfo {author} {\bibfnamefont {F.}~\bibnamefont
  {Foucart}}, \bibinfo {author} {\bibfnamefont {M.~B.}\ \bibnamefont {Deaton}},
  \bibinfo {author} {\bibfnamefont {M.~D.}\ \bibnamefont {Duez}}, \bibinfo
  {author} {\bibfnamefont {L.~E.}\ \bibnamefont {Kidder}}, \bibinfo {author}
  {\bibfnamefont {I.}~\bibnamefont {MacDonald}}, \bibinfo {author}
  {\bibfnamefont {C.~D.}\ \bibnamefont {Ott}}, \bibinfo {author} {\bibfnamefont
  {H.~P.}\ \bibnamefont {Pfeiffer}}, \bibinfo {author} {\bibfnamefont {M.~A.}\
  \bibnamefont {Scheel}}, \bibinfo {author} {\bibfnamefont {B.}~\bibnamefont
  {Szilagyi}},\ and\ \bibinfo {author} {\bibfnamefont {S.~A.}\ \bibnamefont
  {Teukolsky}},\ }\bibfield  {title} {\bibinfo {title} {{Black hole-neutron
  star mergers at realistic mass ratios: Equation of state and spin orientation
  effects}},\ }\href {https://doi.org/10.1103/PhysRevD.87.084006} {\bibfield
  {journal} {\bibinfo  {journal} {Phys. Rev. D}\ }\textbf {\bibinfo {volume}
  {87}},\ \bibinfo {pages} {084006} (\bibinfo {year} {2013})},\ \Eprint
  {https://arxiv.org/abs/1212.4810} {arXiv:1212.4810 [gr-qc]} \BibitemShut
  {NoStop}%
\bibitem [{\citenamefont {Hemberger}\ \emph {et~al.}(2013)\citenamefont
  {Hemberger}, \citenamefont {Scheel}, \citenamefont {Kidder}, \citenamefont
  {Szil\'agyi}, \citenamefont {Lovelace}, \citenamefont {Taylor},\ and\
  \citenamefont {Teukolsky}}]{Hemberger:2012jz}%
  \BibitemOpen
  \bibfield  {author} {\bibinfo {author} {\bibfnamefont {D.~A.}\ \bibnamefont
  {Hemberger}}, \bibinfo {author} {\bibfnamefont {M.~A.}\ \bibnamefont
  {Scheel}}, \bibinfo {author} {\bibfnamefont {L.~E.}\ \bibnamefont {Kidder}},
  \bibinfo {author} {\bibfnamefont {B.}~\bibnamefont {Szil\'agyi}}, \bibinfo
  {author} {\bibfnamefont {G.}~\bibnamefont {Lovelace}}, \bibinfo {author}
  {\bibfnamefont {N.~W.}\ \bibnamefont {Taylor}},\ and\ \bibinfo {author}
  {\bibfnamefont {S.~A.}\ \bibnamefont {Teukolsky}},\ }\bibfield  {title}
  {\bibinfo {title} {{Dynamical Excision Boundaries in Spectral Evolutions of
  Binary Black Hole Spacetimes}},\ }\href
  {https://doi.org/10.1088/0264-9381/30/11/115001} {\bibfield  {journal}
  {\bibinfo  {journal} {Class. Quant. Grav.}\ }\textbf {\bibinfo {volume}
  {30}},\ \bibinfo {pages} {115001} (\bibinfo {year} {2013})},\ \Eprint
  {https://arxiv.org/abs/1211.6079} {arXiv:1211.6079 [gr-qc]} \BibitemShut
  {NoStop}%
\bibitem [{\citenamefont {Rinne}\ \emph {et~al.}(2007)\citenamefont {Rinne},
  \citenamefont {Lindblom},\ and\ \citenamefont {Scheel}}]{Rinne:2007ui}%
  \BibitemOpen
  \bibfield  {author} {\bibinfo {author} {\bibfnamefont {O.}~\bibnamefont
  {Rinne}}, \bibinfo {author} {\bibfnamefont {L.}~\bibnamefont {Lindblom}},\
  and\ \bibinfo {author} {\bibfnamefont {M.~A.}\ \bibnamefont {Scheel}},\
  }\bibfield  {title} {\bibinfo {title} {{Testing outer boundary treatments for
  the Einstein equations}},\ }\href
  {https://doi.org/10.1088/0264-9381/24/16/006} {\bibfield  {journal} {\bibinfo
   {journal} {Class. Quant. Grav.}\ }\textbf {\bibinfo {volume} {24}},\
  \bibinfo {pages} {4053} (\bibinfo {year} {2007})},\ \Eprint
  {https://arxiv.org/abs/0704.0782} {arXiv:0704.0782 [gr-qc]} \BibitemShut
  {NoStop}%
\bibitem [{\citenamefont {Lousto}\ \emph {et~al.}(2010)\citenamefont {Lousto},
  \citenamefont {Nakano}, \citenamefont {Zlochower},\ and\ \citenamefont
  {Campanelli}}]{Lousto:2010qx}%
  \BibitemOpen
  \bibfield  {author} {\bibinfo {author} {\bibfnamefont {C.~O.}\ \bibnamefont
  {Lousto}}, \bibinfo {author} {\bibfnamefont {H.}~\bibnamefont {Nakano}},
  \bibinfo {author} {\bibfnamefont {Y.}~\bibnamefont {Zlochower}},\ and\
  \bibinfo {author} {\bibfnamefont {M.}~\bibnamefont {Campanelli}},\ }\bibfield
   {title} {\bibinfo {title} {{Intermediate-mass-ratio black hole binaries:
  Intertwining numerical and perturbative techniques}},\ }\href
  {https://doi.org/10.1103/PhysRevD.82.104057} {\bibfield  {journal} {\bibinfo
  {journal} {Phys. Rev. D}\ }\textbf {\bibinfo {volume} {82}},\ \bibinfo
  {pages} {104057} (\bibinfo {year} {2010})},\ \Eprint
  {https://arxiv.org/abs/1008.4360} {arXiv:1008.4360 [gr-qc]} \BibitemShut
  {NoStop}%
\bibitem [{\citenamefont {Fontbut{\'e}}\ \emph {et~al.}(2025)\citenamefont
  {Fontbut{\'e}}, \citenamefont {Bernuzzi}, \citenamefont {Albanesi},
  \citenamefont {Radice}, \citenamefont {Rashti}, \citenamefont {Cook},
  \citenamefont {Daszuta},\ and\ \citenamefont {Nagar}}]{Fontbute:2025ixd}%
  \BibitemOpen
  \bibfield  {author} {\bibinfo {author} {\bibfnamefont {J.}~\bibnamefont
  {Fontbut{\'e}}}, \bibinfo {author} {\bibfnamefont {S.}~\bibnamefont
  {Bernuzzi}}, \bibinfo {author} {\bibfnamefont {S.}~\bibnamefont {Albanesi}},
  \bibinfo {author} {\bibfnamefont {D.}~\bibnamefont {Radice}}, \bibinfo
  {author} {\bibfnamefont {A.}~\bibnamefont {Rashti}}, \bibinfo {author}
  {\bibfnamefont {W.}~\bibnamefont {Cook}}, \bibinfo {author} {\bibfnamefont
  {B.}~\bibnamefont {Daszuta}},\ and\ \bibinfo {author} {\bibfnamefont
  {A.}~\bibnamefont {Nagar}},\ }\bibfield  {title} {\bibinfo {title}
  {{Covariant and Gauge-invariant Metric-based Gravitational-waves Extraction
  in Numerical Relativity}},\ }\href@noop {} {\  (\bibinfo {year} {2025})},\
  \Eprint {https://arxiv.org/abs/2508.03799} {arXiv:2508.03799 [gr-qc]}
  \BibitemShut {NoStop}%
\bibitem [{\citenamefont {Reisswig}\ and\ \citenamefont
  {Pollney}(2011)}]{Reisswig:2010di}%
  \BibitemOpen
  \bibfield  {author} {\bibinfo {author} {\bibfnamefont {C.}~\bibnamefont
  {Reisswig}}\ and\ \bibinfo {author} {\bibfnamefont {D.}~\bibnamefont
  {Pollney}},\ }\bibfield  {title} {\bibinfo {title} {{Notes on the integration
  of numerical relativity waveforms}},\ }\href
  {https://doi.org/10.1088/0264-9381/28/19/195015} {\bibfield  {journal}
  {\bibinfo  {journal} {Class. Quant. Grav.}\ }\textbf {\bibinfo {volume}
  {28}},\ \bibinfo {pages} {195015} (\bibinfo {year} {2011})},\ \Eprint
  {https://arxiv.org/abs/1006.1632} {arXiv:1006.1632 [gr-qc]} \BibitemShut
  {NoStop}%
\bibitem [{\citenamefont {Boyle}\ and\ \citenamefont
  {Mroue}(2009)}]{Boyle:2009vi}%
  \BibitemOpen
  \bibfield  {author} {\bibinfo {author} {\bibfnamefont {M.}~\bibnamefont
  {Boyle}}\ and\ \bibinfo {author} {\bibfnamefont {A.~H.}\ \bibnamefont
  {Mroue}},\ }\bibfield  {title} {\bibinfo {title} {{Extrapolating
  gravitational-wave data from numerical simulations}},\ }\href
  {https://doi.org/10.1103/PhysRevD.80.124045} {\bibfield  {journal} {\bibinfo
  {journal} {Phys. Rev. D}\ }\textbf {\bibinfo {volume} {80}},\ \bibinfo
  {pages} {124045} (\bibinfo {year} {2009})},\ \Eprint
  {https://arxiv.org/abs/0905.3177} {arXiv:0905.3177 [gr-qc]} \BibitemShut
  {NoStop}%
\bibitem [{\citenamefont {Sarbach}\ and\ \citenamefont
  {Tiglio}(2001)}]{Sarbach:2001qq}%
  \BibitemOpen
  \bibfield  {author} {\bibinfo {author} {\bibfnamefont {O.}~\bibnamefont
  {Sarbach}}\ and\ \bibinfo {author} {\bibfnamefont {M.}~\bibnamefont
  {Tiglio}},\ }\bibfield  {title} {\bibinfo {title} {{Gauge invariant
  perturbations of Schwarzschild black holes in horizon penetrating
  coordinates}},\ }\href {https://doi.org/10.1103/PhysRevD.64.084016}
  {\bibfield  {journal} {\bibinfo  {journal} {Phys. Rev.}\ }\textbf {\bibinfo
  {volume} {D64}},\ \bibinfo {pages} {084016} (\bibinfo {year} {2001})},\
  \Eprint {https://arxiv.org/abs/gr-qc/0104061} {arXiv:gr-qc/0104061 [gr-qc]}
  \BibitemShut {NoStop}%
\bibitem [{\citenamefont {Regge}\ and\ \citenamefont
  {Wheeler}(1957)}]{Regge:1957td}%
  \BibitemOpen
  \bibfield  {author} {\bibinfo {author} {\bibfnamefont {T.}~\bibnamefont
  {Regge}}\ and\ \bibinfo {author} {\bibfnamefont {J.~A.}\ \bibnamefont
  {Wheeler}},\ }\bibfield  {title} {\bibinfo {title} {{Stability of a
  Schwarzschild singularity}},\ }\href
  {https://doi.org/10.1103/PhysRev.108.1063} {\bibfield  {journal} {\bibinfo
  {journal} {Phys. Rev.}\ }\textbf {\bibinfo {volume} {108}},\ \bibinfo {pages}
  {1063} (\bibinfo {year} {1957})}\BibitemShut {NoStop}%
\bibitem [{\citenamefont {Zerilli}(1970)}]{Zerilli:1970se}%
  \BibitemOpen
  \bibfield  {author} {\bibinfo {author} {\bibfnamefont {F.~J.}\ \bibnamefont
  {Zerilli}},\ }\bibfield  {title} {\bibinfo {title} {{Effective potential for
  even parity Regge-Wheeler gravitational perturbation equations}},\ }\href
  {https://doi.org/10.1103/PhysRevLett.24.737} {\bibfield  {journal} {\bibinfo
  {journal} {Phys. Rev. Lett.}\ }\textbf {\bibinfo {volume} {24}},\ \bibinfo
  {pages} {737} (\bibinfo {year} {1970})}\BibitemShut {NoStop}%
\bibitem [{\citenamefont {Rinne}\ \emph {et~al.}(2009)\citenamefont {Rinne},
  \citenamefont {Buchman}, \citenamefont {Scheel},\ and\ \citenamefont
  {Pfeiffer}}]{Rinne:2008vn}%
  \BibitemOpen
  \bibfield  {author} {\bibinfo {author} {\bibfnamefont {O.}~\bibnamefont
  {Rinne}}, \bibinfo {author} {\bibfnamefont {L.~T.}\ \bibnamefont {Buchman}},
  \bibinfo {author} {\bibfnamefont {M.~A.}\ \bibnamefont {Scheel}},\ and\
  \bibinfo {author} {\bibfnamefont {H.~P.}\ \bibnamefont {Pfeiffer}},\
  }\bibfield  {title} {\bibinfo {title} {{Implementation of higher-order
  absorbing boundary conditions for the Einstein equations}},\ }\href
  {https://doi.org/10.1088/0264-9381/26/7/075009} {\bibfield  {journal}
  {\bibinfo  {journal} {Class. Quant. Grav.}\ }\textbf {\bibinfo {volume}
  {26}},\ \bibinfo {pages} {075009} (\bibinfo {year} {2009})},\ \Eprint
  {https://arxiv.org/abs/0811.3593} {arXiv:0811.3593 [gr-qc]} \BibitemShut
  {NoStop}%
\bibitem [{\citenamefont {Boyle}\ \emph {et~al.}(2007)\citenamefont {Boyle},
  \citenamefont {Brown}, \citenamefont {Kidder}, \citenamefont {Mroue},
  \citenamefont {Pfeiffer}, \citenamefont {Scheel}, \citenamefont {Cook},\ and\
  \citenamefont {Teukolsky}}]{Boyle:2007ft}%
  \BibitemOpen
  \bibfield  {author} {\bibinfo {author} {\bibfnamefont {M.}~\bibnamefont
  {Boyle}}, \bibinfo {author} {\bibfnamefont {D.~A.}\ \bibnamefont {Brown}},
  \bibinfo {author} {\bibfnamefont {L.~E.}\ \bibnamefont {Kidder}}, \bibinfo
  {author} {\bibfnamefont {A.~H.}\ \bibnamefont {Mroue}}, \bibinfo {author}
  {\bibfnamefont {H.~P.}\ \bibnamefont {Pfeiffer}}, \bibinfo {author}
  {\bibfnamefont {M.~A.}\ \bibnamefont {Scheel}}, \bibinfo {author}
  {\bibfnamefont {G.~B.}\ \bibnamefont {Cook}},\ and\ \bibinfo {author}
  {\bibfnamefont {S.~A.}\ \bibnamefont {Teukolsky}},\ }\bibfield  {title}
  {\bibinfo {title} {{High-accuracy comparison of numerical relativity
  simulations with post-Newtonian expansions}},\ }\href
  {https://doi.org/10.1103/PhysRevD.76.124038} {\bibfield  {journal} {\bibinfo
  {journal} {Phys. Rev. D}\ }\textbf {\bibinfo {volume} {76}},\ \bibinfo
  {pages} {124038} (\bibinfo {year} {2007})},\ \Eprint
  {https://arxiv.org/abs/0710.0158} {arXiv:0710.0158 [gr-qc]} \BibitemShut
  {NoStop}%
\bibitem [{\citenamefont {{Bishop}}\ \emph {et~al.}(1996)\citenamefont
  {{Bishop}}, \citenamefont {{G{\'o}mez}}, \citenamefont {{Lehner}},\ and\
  \citenamefont {{Winicour}}}]{1996PhRvD..54.6153B}%
  \BibitemOpen
  \bibfield  {author} {\bibinfo {author} {\bibfnamefont {N.~T.}\ \bibnamefont
  {{Bishop}}}, \bibinfo {author} {\bibfnamefont {R.}~\bibnamefont
  {{G{\'o}mez}}}, \bibinfo {author} {\bibfnamefont {L.}~\bibnamefont
  {{Lehner}}},\ and\ \bibinfo {author} {\bibfnamefont {J.}~\bibnamefont
  {{Winicour}}},\ }\bibfield  {title} {\bibinfo {title} {{Cauchy-characteristic
  extraction in numerical relativity}},\ }\href
  {https://doi.org/10.1103/PhysRevD.54.6153} {\bibfield  {journal} {\bibinfo
  {journal} {Phys. Rev. D}\ }\textbf {\bibinfo {volume} {54}},\ \bibinfo
  {pages} {6153} (\bibinfo {year} {1996})}\BibitemShut {NoStop}%
\bibitem [{\citenamefont {Bishop}\ and\ \citenamefont
  {Rezzolla}(2016)}]{Bishop:2016lgv}%
  \BibitemOpen
  \bibfield  {author} {\bibinfo {author} {\bibfnamefont {N.~T.}\ \bibnamefont
  {Bishop}}\ and\ \bibinfo {author} {\bibfnamefont {L.}~\bibnamefont
  {Rezzolla}},\ }\bibfield  {title} {\bibinfo {title} {{Extraction of
  Gravitational Waves in Numerical Relativity}},\ }\href
  {https://doi.org/10.1007/s41114-016-0001-9} {\bibfield  {journal} {\bibinfo
  {journal} {Living Rev. Rel.}\ }\textbf {\bibinfo {volume} {19}},\ \bibinfo
  {pages} {2} (\bibinfo {year} {2016})},\ \Eprint
  {https://arxiv.org/abs/1606.02532} {arXiv:1606.02532 [gr-qc]} \BibitemShut
  {NoStop}%
\bibitem [{\citenamefont {{Handmer}}\ \emph {et~al.}(2016)\citenamefont
  {{Handmer}}, \citenamefont {{Szil{\'a}gyi}},\ and\ \citenamefont
  {{Winicour}}}]{2016CQGra..33v5007H}%
  \BibitemOpen
  \bibfield  {author} {\bibinfo {author} {\bibfnamefont {C.~J.}\ \bibnamefont
  {{Handmer}}}, \bibinfo {author} {\bibfnamefont {B.}~\bibnamefont
  {{Szil{\'a}gyi}}},\ and\ \bibinfo {author} {\bibfnamefont {J.}~\bibnamefont
  {{Winicour}}},\ }\bibfield  {title} {\bibinfo {title} {{Spectral Cauchy
  characteristic extraction of strain, news and gravitational radiation
  flux}},\ }\href {https://doi.org/10.1088/0264-9381/33/22/225007} {\bibfield
  {journal} {\bibinfo  {journal} {Class. Quantum Grav.}\ }\textbf {\bibinfo
  {volume} {33}},\ \bibinfo {eid} {225007} (\bibinfo {year} {2016})},\ \Eprint
  {https://arxiv.org/abs/1605.04332} {arXiv:1605.04332 [gr-qc]} \BibitemShut
  {NoStop}%
\bibitem [{\citenamefont {{Barkett}}\ \emph {et~al.}(2020)\citenamefont
  {{Barkett}}, \citenamefont {{Moxon}}, \citenamefont {{Scheel}},\ and\
  \citenamefont {{Szil{\'a}gyi}}}]{2020PhRvD.102b4004B}%
  \BibitemOpen
  \bibfield  {author} {\bibinfo {author} {\bibfnamefont {K.}~\bibnamefont
  {{Barkett}}}, \bibinfo {author} {\bibfnamefont {J.}~\bibnamefont {{Moxon}}},
  \bibinfo {author} {\bibfnamefont {M.~A.}\ \bibnamefont {{Scheel}}},\ and\
  \bibinfo {author} {\bibfnamefont {B.}~\bibnamefont {{Szil{\'a}gyi}}},\
  }\bibfield  {title} {\bibinfo {title} {{Spectral Cauchy-characteristic
  extraction of the gravitational wave news function}},\ }\href
  {https://doi.org/10.1103/PhysRevD.102.024004} {\bibfield  {journal} {\bibinfo
   {journal} {Phys. Rev. D}\ }\textbf {\bibinfo {volume} {102}},\ \bibinfo
  {eid} {024004} (\bibinfo {year} {2020})},\ \Eprint
  {https://arxiv.org/abs/1910.09677} {arXiv:1910.09677 [gr-qc]} \BibitemShut
  {NoStop}%
\bibitem [{\citenamefont {{Moxon}}\ \emph {et~al.}(2020)\citenamefont
  {{Moxon}}, \citenamefont {{Scheel}},\ and\ \citenamefont
  {{Teukolsky}}}]{2020PhRvD.102d4052M}%
  \BibitemOpen
  \bibfield  {author} {\bibinfo {author} {\bibfnamefont {J.}~\bibnamefont
  {{Moxon}}}, \bibinfo {author} {\bibfnamefont {M.~A.}\ \bibnamefont
  {{Scheel}}},\ and\ \bibinfo {author} {\bibfnamefont {S.~A.}\ \bibnamefont
  {{Teukolsky}}},\ }\bibfield  {title} {\bibinfo {title} {{Improved
  Cauchy-characteristic evolution system for high-precision numerical
  relativity waveforms}},\ }\href {https://doi.org/10.1103/PhysRevD.102.044052}
  {\bibfield  {journal} {\bibinfo  {journal} {Phys. Rev. D}\ }\textbf {\bibinfo
  {volume} {102}},\ \bibinfo {eid} {044052} (\bibinfo {year} {2020})},\ \Eprint
  {https://arxiv.org/abs/2007.01339} {arXiv:2007.01339 [gr-qc]} \BibitemShut
  {NoStop}%
\bibitem [{\citenamefont {Moxon}\ \emph {et~al.}(2023)\citenamefont {Moxon},
  \citenamefont {Scheel}, \citenamefont {Teukolsky}, \citenamefont {Deppe},
  \citenamefont {Fischer}, \citenamefont {H\'ebert}, \citenamefont {Kidder},\
  and\ \citenamefont {Throwe}}]{Moxon:2021gbv}%
  \BibitemOpen
  \bibfield  {author} {\bibinfo {author} {\bibfnamefont {J.}~\bibnamefont
  {Moxon}}, \bibinfo {author} {\bibfnamefont {M.~A.}\ \bibnamefont {Scheel}},
  \bibinfo {author} {\bibfnamefont {S.~A.}\ \bibnamefont {Teukolsky}}, \bibinfo
  {author} {\bibfnamefont {N.}~\bibnamefont {Deppe}}, \bibinfo {author}
  {\bibfnamefont {N.}~\bibnamefont {Fischer}}, \bibinfo {author} {\bibfnamefont
  {F.}~\bibnamefont {H\'ebert}}, \bibinfo {author} {\bibfnamefont {L.~E.}\
  \bibnamefont {Kidder}},\ and\ \bibinfo {author} {\bibfnamefont
  {W.}~\bibnamefont {Throwe}},\ }\bibfield  {title} {\bibinfo {title} {{SpECTRE
  Cauchy-characteristic evolution system for rapid, precise waveform
  extraction}},\ }\href {https://doi.org/10.1103/PhysRevD.107.064013}
  {\bibfield  {journal} {\bibinfo  {journal} {Phys. Rev. D}\ }\textbf {\bibinfo
  {volume} {107}},\ \bibinfo {pages} {064013} (\bibinfo {year} {2023})},\
  \Eprint {https://arxiv.org/abs/2110.08635} {arXiv:2110.08635 [gr-qc]}
  \BibitemShut {NoStop}%
\bibitem [{\citenamefont {Mitman}\ \emph {et~al.}(2020)\citenamefont {Mitman},
  \citenamefont {Moxon}, \citenamefont {Scheel}, \citenamefont {Teukolsky},
  \citenamefont {Boyle}, \citenamefont {Deppe}, \citenamefont {Kidder},\ and\
  \citenamefont {Throwe}}]{Mitman:2020pbt}%
  \BibitemOpen
  \bibfield  {author} {\bibinfo {author} {\bibfnamefont {K.}~\bibnamefont
  {Mitman}}, \bibinfo {author} {\bibfnamefont {J.}~\bibnamefont {Moxon}},
  \bibinfo {author} {\bibfnamefont {M.~A.}\ \bibnamefont {Scheel}}, \bibinfo
  {author} {\bibfnamefont {S.~A.}\ \bibnamefont {Teukolsky}}, \bibinfo {author}
  {\bibfnamefont {M.}~\bibnamefont {Boyle}}, \bibinfo {author} {\bibfnamefont
  {N.}~\bibnamefont {Deppe}}, \bibinfo {author} {\bibfnamefont {L.~E.}\
  \bibnamefont {Kidder}},\ and\ \bibinfo {author} {\bibfnamefont
  {W.}~\bibnamefont {Throwe}},\ }\bibfield  {title} {\bibinfo {title}
  {{Computation of displacement and spin gravitational memory in numerical
  relativity}},\ }\href {https://doi.org/10.1103/PhysRevD.102.104007}
  {\bibfield  {journal} {\bibinfo  {journal} {Phys. Rev. D}\ }\textbf {\bibinfo
  {volume} {102}},\ \bibinfo {pages} {104007} (\bibinfo {year} {2020})},\
  \Eprint {https://arxiv.org/abs/2007.11562} {arXiv:2007.11562 [gr-qc]}
  \BibitemShut {NoStop}%
\bibitem [{\citenamefont {Mitman}\ \emph
  {et~al.}(2021{\natexlab{a}})\citenamefont {Mitman} \emph
  {et~al.}}]{Mitman:2020bjf}%
  \BibitemOpen
  \bibfield  {author} {\bibinfo {author} {\bibfnamefont {K.}~\bibnamefont
  {Mitman}} \emph {et~al.},\ }\bibfield  {title} {\bibinfo {title} {{Adding
  gravitational memory to waveform catalogs using BMS balance laws}},\ }\href
  {https://doi.org/10.1103/PhysRevD.103.024031} {\bibfield  {journal} {\bibinfo
   {journal} {Phys. Rev. D}\ }\textbf {\bibinfo {volume} {103}},\ \bibinfo
  {pages} {024031} (\bibinfo {year} {2021}{\natexlab{a}})},\ \Eprint
  {https://arxiv.org/abs/2011.01309} {arXiv:2011.01309 [gr-qc]} \BibitemShut
  {NoStop}%
\bibitem [{\citenamefont {Chu}\ \emph {et~al.}(2016)\citenamefont {Chu},
  \citenamefont {Fong}, \citenamefont {Kumar}, \citenamefont {Pfeiffer},
  \citenamefont {Boyle}, \citenamefont {Hemberger}, \citenamefont {Kidder},
  \citenamefont {Scheel},\ and\ \citenamefont {Szilagyi}}]{Chu:2015kft}%
  \BibitemOpen
  \bibfield  {author} {\bibinfo {author} {\bibfnamefont {T.}~\bibnamefont
  {Chu}}, \bibinfo {author} {\bibfnamefont {H.}~\bibnamefont {Fong}}, \bibinfo
  {author} {\bibfnamefont {P.}~\bibnamefont {Kumar}}, \bibinfo {author}
  {\bibfnamefont {H.~P.}\ \bibnamefont {Pfeiffer}}, \bibinfo {author}
  {\bibfnamefont {M.}~\bibnamefont {Boyle}}, \bibinfo {author} {\bibfnamefont
  {D.~A.}\ \bibnamefont {Hemberger}}, \bibinfo {author} {\bibfnamefont {L.~E.}\
  \bibnamefont {Kidder}}, \bibinfo {author} {\bibfnamefont {M.~A.}\
  \bibnamefont {Scheel}},\ and\ \bibinfo {author} {\bibfnamefont
  {B.}~\bibnamefont {Szilagyi}},\ }\bibfield  {title} {\bibinfo {title} {{On
  the accuracy and precision of numerical waveforms: Effect of waveform
  extraction methodology}},\ }\href
  {https://doi.org/10.1088/0264-9381/33/16/165001} {\bibfield  {journal}
  {\bibinfo  {journal} {Class. Quant. Grav.}\ }\textbf {\bibinfo {volume}
  {33}},\ \bibinfo {pages} {165001} (\bibinfo {year} {2016})},\ \Eprint
  {https://arxiv.org/abs/1512.06800} {arXiv:1512.06800 [gr-qc]} \BibitemShut
  {NoStop}%
\bibitem [{\citenamefont {Pollney}\ and\ \citenamefont
  {Reisswig}(2011)}]{Pollney:2010hs}%
  \BibitemOpen
  \bibfield  {author} {\bibinfo {author} {\bibfnamefont {D.}~\bibnamefont
  {Pollney}}\ and\ \bibinfo {author} {\bibfnamefont {C.}~\bibnamefont
  {Reisswig}},\ }\bibfield  {title} {\bibinfo {title} {{Gravitational memory in
  binary black hole mergers}},\ }\href
  {https://doi.org/10.1088/2041-8205/732/1/L13} {\bibfield  {journal} {\bibinfo
   {journal} {Astrophys. J. Lett.}\ }\textbf {\bibinfo {volume} {732}},\
  \bibinfo {pages} {L13} (\bibinfo {year} {2011})},\ \Eprint
  {https://arxiv.org/abs/1004.4209} {arXiv:1004.4209 [gr-qc]} \BibitemShut
  {NoStop}%
\bibitem [{\citenamefont {Mitman}\ \emph
  {et~al.}(2021{\natexlab{b}})\citenamefont {Mitman} \emph
  {et~al.}}]{Mitman:2021xkq}%
  \BibitemOpen
  \bibfield  {author} {\bibinfo {author} {\bibfnamefont {K.}~\bibnamefont
  {Mitman}} \emph {et~al.},\ }\bibfield  {title} {\bibinfo {title} {{Fixing the
  BMS frame of numerical relativity waveforms}},\ }\href
  {https://doi.org/10.1103/PhysRevD.104.024051} {\bibfield  {journal} {\bibinfo
   {journal} {Phys. Rev. D}\ }\textbf {\bibinfo {volume} {104}},\ \bibinfo
  {pages} {024051} (\bibinfo {year} {2021}{\natexlab{b}})},\ \Eprint
  {https://arxiv.org/abs/2105.02300} {arXiv:2105.02300 [gr-qc]} \BibitemShut
  {NoStop}%
\bibitem [{\citenamefont {Mitman}\ \emph {et~al.}(2022)\citenamefont {Mitman}
  \emph {et~al.}}]{Mitman:2022kwt}%
  \BibitemOpen
  \bibfield  {author} {\bibinfo {author} {\bibfnamefont {K.}~\bibnamefont
  {Mitman}} \emph {et~al.},\ }\bibfield  {title} {\bibinfo {title} {{Fixing the
  BMS frame of numerical relativity waveforms with BMS charges}},\ }\href
  {https://doi.org/10.1103/PhysRevD.106.084029} {\bibfield  {journal} {\bibinfo
   {journal} {Phys. Rev. D}\ }\textbf {\bibinfo {volume} {106}},\ \bibinfo
  {pages} {084029} (\bibinfo {year} {2022})},\ \Eprint
  {https://arxiv.org/abs/2208.04356} {arXiv:2208.04356 [gr-qc]} \BibitemShut
  {NoStop}%
\bibitem [{\citenamefont {P\"urrer}\ and\ \citenamefont
  {Haster}(2020)}]{Purrer:2019jcp}%
  \BibitemOpen
  \bibfield  {author} {\bibinfo {author} {\bibfnamefont {M.}~\bibnamefont
  {P\"urrer}}\ and\ \bibinfo {author} {\bibfnamefont {C.-J.}\ \bibnamefont
  {Haster}},\ }\bibfield  {title} {\bibinfo {title} {{Gravitational waveform
  accuracy requirements for future ground-based detectors}},\ }\href
  {https://doi.org/10.1103/PhysRevResearch.2.023151} {\bibfield  {journal}
  {\bibinfo  {journal} {Phys. Rev. Res.}\ }\textbf {\bibinfo {volume} {2}},\
  \bibinfo {pages} {023151} (\bibinfo {year} {2020})},\ \Eprint
  {https://arxiv.org/abs/1912.10055} {arXiv:1912.10055 [gr-qc]} \BibitemShut
  {NoStop}%
\bibitem [{\citenamefont {Barsotti}\ \emph {et~al.}(2018)\citenamefont
  {Barsotti}, \citenamefont {McCuller}, \citenamefont {Evans},\ and\
  \citenamefont {Fritschel}}]{LIGO-T1800042-v5}%
  \BibitemOpen
  \bibfield  {author} {\bibinfo {author} {\bibfnamefont {L.}~\bibnamefont
  {Barsotti}}, \bibinfo {author} {\bibfnamefont {L.}~\bibnamefont {McCuller}},
  \bibinfo {author} {\bibfnamefont {M.}~\bibnamefont {Evans}},\ and\ \bibinfo
  {author} {\bibfnamefont {P.}~\bibnamefont {Fritschel}},\ }\href
  {https://dcc.ligo.org/LIGO-T1800042/public} {\emph {\bibinfo {title} {The A+
  Design Curve}}},\ \bibinfo {type} {Tech. Rep.}\ \bibinfo {number}
  {LIGO-T1800042-v5}\ (\bibinfo  {institution} {California Institute of
  Technology and Massachusetts Institute of Technology},\ \bibinfo {year}
  {2018})\BibitemShut {NoStop}%
\bibitem [{\citenamefont {Flanagan}\ and\ \citenamefont
  {Hughes}(1998)}]{Flanagan:1997kp}%
  \BibitemOpen
  \bibfield  {author} {\bibinfo {author} {\bibfnamefont {E.~E.}\ \bibnamefont
  {Flanagan}}\ and\ \bibinfo {author} {\bibfnamefont {S.~A.}\ \bibnamefont
  {Hughes}},\ }\bibfield  {title} {\bibinfo {title} {{Measuring gravitational
  waves from binary black hole coalescences: 2. The Waves' information and its
  extraction, with and without templates}},\ }\href
  {https://doi.org/10.1103/PhysRevD.57.4566} {\bibfield  {journal} {\bibinfo
  {journal} {Phys. Rev. D}\ }\textbf {\bibinfo {volume} {57}},\ \bibinfo
  {pages} {4566} (\bibinfo {year} {1998})},\ \Eprint
  {https://arxiv.org/abs/gr-qc/9710129} {arXiv:gr-qc/9710129} \BibitemShut
  {NoStop}%
\bibitem [{\citenamefont {Damour}\ \emph {et~al.}(2011)\citenamefont {Damour},
  \citenamefont {Nagar},\ and\ \citenamefont {Trias}}]{Damour:2010zb}%
  \BibitemOpen
  \bibfield  {author} {\bibinfo {author} {\bibfnamefont {T.}~\bibnamefont
  {Damour}}, \bibinfo {author} {\bibfnamefont {A.}~\bibnamefont {Nagar}},\ and\
  \bibinfo {author} {\bibfnamefont {M.}~\bibnamefont {Trias}},\ }\bibfield
  {title} {\bibinfo {title} {{Accuracy and effectualness of closed-form,
  frequency-domain waveforms for non-spinning black hole binaries}},\ }\href
  {https://doi.org/10.1103/PhysRevD.83.024006} {\bibfield  {journal} {\bibinfo
  {journal} {Phys. Rev. D}\ }\textbf {\bibinfo {volume} {83}},\ \bibinfo
  {pages} {024006} (\bibinfo {year} {2011})},\ \Eprint
  {https://arxiv.org/abs/1009.5998} {arXiv:1009.5998 [gr-qc]} \BibitemShut
  {NoStop}%
\bibitem [{\citenamefont {Chatziioannou}\ \emph
  {et~al.}(2017{\natexlab{b}})\citenamefont {Chatziioannou}, \citenamefont
  {Klein}, \citenamefont {Yunes},\ and\ \citenamefont
  {Cornish}}]{Chatziioannou:2017tdw}%
  \BibitemOpen
  \bibfield  {author} {\bibinfo {author} {\bibfnamefont {K.}~\bibnamefont
  {Chatziioannou}}, \bibinfo {author} {\bibfnamefont {A.}~\bibnamefont
  {Klein}}, \bibinfo {author} {\bibfnamefont {N.}~\bibnamefont {Yunes}},\ and\
  \bibinfo {author} {\bibfnamefont {N.}~\bibnamefont {Cornish}},\ }\bibfield
  {title} {\bibinfo {title} {{Constructing Gravitational Waves from Generic
  Spin-Precessing Compact Binary Inspirals}},\ }\href
  {https://doi.org/10.1103/PhysRevD.95.104004} {\bibfield  {journal} {\bibinfo
  {journal} {Phys. Rev. D}\ }\textbf {\bibinfo {volume} {95}},\ \bibinfo
  {pages} {104004} (\bibinfo {year} {2017}{\natexlab{b}})},\ \Eprint
  {https://arxiv.org/abs/1703.03967} {arXiv:1703.03967 [gr-qc]} \BibitemShut
  {NoStop}%
\bibitem [{\citenamefont {Lovelace}\ \emph {et~al.}(2025)\citenamefont
  {Lovelace} \emph {et~al.}}]{Lovelace:2024wra}%
  \BibitemOpen
  \bibfield  {author} {\bibinfo {author} {\bibfnamefont {G.}~\bibnamefont
  {Lovelace}} \emph {et~al.},\ }\bibfield  {title} {\bibinfo {title}
  {{Simulating binary black hole mergers using discontinuous Galerkin
  methods}},\ }\href {https://doi.org/10.1088/1361-6382/ad9f19} {\bibfield
  {journal} {\bibinfo  {journal} {Class. Quant. Grav.}\ }\textbf {\bibinfo
  {volume} {42}},\ \bibinfo {pages} {035001} (\bibinfo {year} {2025})},\
  \Eprint {https://arxiv.org/abs/2410.00265} {arXiv:2410.00265 [gr-qc]}
  \BibitemShut {NoStop}%
\bibitem [{\citenamefont {Kiuchi}\ \emph {et~al.}(2022)\citenamefont {Kiuchi},
  \citenamefont {Held}, \citenamefont {Sekiguchi},\ and\ \citenamefont
  {Shibata}}]{Kiuchi:2022ubj}%
  \BibitemOpen
  \bibfield  {author} {\bibinfo {author} {\bibfnamefont {K.}~\bibnamefont
  {Kiuchi}}, \bibinfo {author} {\bibfnamefont {L.~E.}\ \bibnamefont {Held}},
  \bibinfo {author} {\bibfnamefont {Y.}~\bibnamefont {Sekiguchi}},\ and\
  \bibinfo {author} {\bibfnamefont {M.}~\bibnamefont {Shibata}},\ }\bibfield
  {title} {\bibinfo {title} {{Implementation of advanced Riemann solvers in a
  neutrino-radiation magnetohydrodynamics code in numerical relativity and its
  application to a binary neutron star merger}},\ }\href
  {https://doi.org/10.1103/PhysRevD.106.124041} {\bibfield  {journal} {\bibinfo
   {journal} {Phys. Rev. D}\ }\textbf {\bibinfo {volume} {106}},\ \bibinfo
  {pages} {124041} (\bibinfo {year} {2022})},\ \Eprint
  {https://arxiv.org/abs/2205.04487} {arXiv:2205.04487 [astro-ph.HE]}
  \BibitemShut {NoStop}%
\bibitem [{\citenamefont {Fields}\ \emph {et~al.}(2025)\citenamefont {Fields},
  \citenamefont {Zhu}, \citenamefont {Radice}, \citenamefont {Stone},
  \citenamefont {Cook}, \citenamefont {Bernuzzi},\ and\ \citenamefont
  {Daszuta}}]{Fields:2024pob}%
  \BibitemOpen
  \bibfield  {author} {\bibinfo {author} {\bibfnamefont {J.}~\bibnamefont
  {Fields}}, \bibinfo {author} {\bibfnamefont {H.}~\bibnamefont {Zhu}},
  \bibinfo {author} {\bibfnamefont {D.}~\bibnamefont {Radice}}, \bibinfo
  {author} {\bibfnamefont {J.~M.}\ \bibnamefont {Stone}}, \bibinfo {author}
  {\bibfnamefont {W.}~\bibnamefont {Cook}}, \bibinfo {author} {\bibfnamefont
  {S.}~\bibnamefont {Bernuzzi}},\ and\ \bibinfo {author} {\bibfnamefont
  {B.}~\bibnamefont {Daszuta}},\ }\bibfield  {title} {\bibinfo {title}
  {{Performance-portable Binary Neutron Star Mergers with AthenaK}},\ }\href
  {https://doi.org/10.3847/1538-4365/ad9687} {\bibfield  {journal} {\bibinfo
  {journal} {Astrophys. J. Suppl.}\ }\textbf {\bibinfo {volume} {276}},\
  \bibinfo {pages} {35} (\bibinfo {year} {2025})},\ \Eprint
  {https://arxiv.org/abs/2409.10384} {arXiv:2409.10384 [astro-ph.HE]}
  \BibitemShut {NoStop}%
\bibitem [{\citenamefont {Fambri}\ \emph {et~al.}(2018)\citenamefont {Fambri},
  \citenamefont {Dumbser}, \citenamefont {K{\"o}ppel}, \citenamefont
  {Rezzolla},\ and\ \citenamefont {Zanotti}}]{Fambri:2018udk}%
  \BibitemOpen
  \bibfield  {author} {\bibinfo {author} {\bibfnamefont {F.}~\bibnamefont
  {Fambri}}, \bibinfo {author} {\bibfnamefont {M.}~\bibnamefont {Dumbser}},
  \bibinfo {author} {\bibfnamefont {S.}~\bibnamefont {K{\"o}ppel}}, \bibinfo
  {author} {\bibfnamefont {L.}~\bibnamefont {Rezzolla}},\ and\ \bibinfo
  {author} {\bibfnamefont {O.}~\bibnamefont {Zanotti}},\ }\bibfield  {title}
  {\bibinfo {title} {{ADER discontinuous Galerkin schemes for
  general-relativistic ideal magnetohydrodynamics}},\ }\href
  {https://doi.org/10.1093/mnras/sty734} {\bibfield  {journal} {\bibinfo
  {journal} {Mon. Not. Roy. Astron. Soc.}\ }\textbf {\bibinfo {volume} {477}},\
  \bibinfo {pages} {4543} (\bibinfo {year} {2018})},\ \Eprint
  {https://arxiv.org/abs/1801.02839} {arXiv:1801.02839 [physics.comp-ph]}
  \BibitemShut {NoStop}%
\bibitem [{\citenamefont {H{\'e}bert}\ \emph {et~al.}(2018)\citenamefont
  {H{\'e}bert}, \citenamefont {Kidder},\ and\ \citenamefont
  {Teukolsky}}]{Hebert:2018xbk}%
  \BibitemOpen
  \bibfield  {author} {\bibinfo {author} {\bibfnamefont {F.}~\bibnamefont
  {H{\'e}bert}}, \bibinfo {author} {\bibfnamefont {L.~E.}\ \bibnamefont
  {Kidder}},\ and\ \bibinfo {author} {\bibfnamefont {S.~A.}\ \bibnamefont
  {Teukolsky}},\ }\bibfield  {title} {\bibinfo {title} {{General-relativistic
  neutron star evolutions with the discontinuous Galerkin method}},\ }\href
  {https://doi.org/10.1103/PhysRevD.98.044041} {\bibfield  {journal} {\bibinfo
  {journal} {Phys. Rev. D}\ }\textbf {\bibinfo {volume} {98}},\ \bibinfo
  {pages} {044041} (\bibinfo {year} {2018})},\ \Eprint
  {https://arxiv.org/abs/1804.02003} {arXiv:1804.02003 [gr-qc]} \BibitemShut
  {NoStop}%
\bibitem [{\citenamefont {Deppe}\ \emph {et~al.}(2022)\citenamefont {Deppe}
  \emph {et~al.}}]{Deppe:2021bhi}%
  \BibitemOpen
  \bibfield  {author} {\bibinfo {author} {\bibfnamefont {N.}~\bibnamefont
  {Deppe}} \emph {et~al.},\ }\bibfield  {title} {\bibinfo {title} {{Simulating
  magnetized neutron stars with discontinuous Galerkin methods}},\ }\href
  {https://doi.org/10.1103/PhysRevD.105.123031} {\bibfield  {journal} {\bibinfo
   {journal} {Phys. Rev. D}\ }\textbf {\bibinfo {volume} {105}},\ \bibinfo
  {pages} {123031} (\bibinfo {year} {2022})},\ \Eprint
  {https://arxiv.org/abs/2109.12033} {arXiv:2109.12033 [gr-qc]} \BibitemShut
  {NoStop}%
\bibitem [{\citenamefont {Jan}\ \emph {et~al.}(2024)\citenamefont {Jan},
  \citenamefont {Ferguson}, \citenamefont {Lange}, \citenamefont {Shoemaker},\
  and\ \citenamefont {Zimmerman}}]{Jan:2023raq}%
  \BibitemOpen
  \bibfield  {author} {\bibinfo {author} {\bibfnamefont {A.}~\bibnamefont
  {Jan}}, \bibinfo {author} {\bibfnamefont {D.}~\bibnamefont {Ferguson}},
  \bibinfo {author} {\bibfnamefont {J.}~\bibnamefont {Lange}}, \bibinfo
  {author} {\bibfnamefont {D.}~\bibnamefont {Shoemaker}},\ and\ \bibinfo
  {author} {\bibfnamefont {A.}~\bibnamefont {Zimmerman}},\ }\bibfield  {title}
  {\bibinfo {title} {{Accuracy limitations of existing numerical relativity
  waveforms on the data analysis of current and future ground-based
  detectors}},\ }\href {https://doi.org/10.1103/PhysRevD.110.024023} {\bibfield
   {journal} {\bibinfo  {journal} {Phys. Rev. D}\ }\textbf {\bibinfo {volume}
  {110}},\ \bibinfo {pages} {024023} (\bibinfo {year} {2024})},\ \Eprint
  {https://arxiv.org/abs/2312.10241} {arXiv:2312.10241 [gr-qc]} \BibitemShut
  {NoStop}%
\end{thebibliography}%

\end{document}